\documentclass[aps,prl,reprint,groupedaddress]{revtex4-1}

\usepackage{graphicx}
\usepackage{dcolumn}
\usepackage{mathtools, nccmath}
\usepackage{bm}
\usepackage{xcolor}
\usepackage{amsmath}
\usepackage[export]{adjustbox}
\usepackage{framed}
\usepackage[]{graphicx}
\usepackage[caption=false]{subfig}
\usepackage{hyperref}
\usepackage[mathlines]{lineno}
\usepackage{textcomp}
\usepackage{gensymb}
\usepackage[super]{nth}
\usepackage{graphicx,color}
\usepackage{hyperref}
\usepackage{babel}
\usepackage{booktabs}
\usepackage{dsfont}
\usepackage[percent]{overpic}
\usepackage{ulem}

\usepackage{stackengine}

\newcommand{\pdag}{{\phantom{\dagger}}}

\newcommand{\bs}[1]{\boldsymbol{#1}}
\newcommand{\pd}{{\phantom{\dagger}}}

\begin{document}

\title{Edge-selective extremal damping from topological heritage of dissipative Chern insulators}

\author{Suraj S. Hegde}
\email{suraj.hegde@tu-dresden.de}
\author{Toni Ehmcke}
\author{Tobias Meng}
\affiliation{Institute of Theoretical Physics and W\"urzburg-Dresden Cluster of Excellence ct.qmat, Technische Universit\"at Dresden, 01062 Dresden, Germany}

\begin{abstract}
One of the most important practical hallmarks of topological matter is the presence of topologically protected, exponentially localized edge states at interfaces of regions characterized by unequal topological invariants. Here, we show that even when driven far from their equilibrium ground state, Chern insulators can inherit topological edge features from their parent Hamiltonian. In particular, we show that the asymptotic long-time approach of the  non-equilibrium steady state, governed by a Lindblad Master equation, can exhibit edge-selective extremal damping. This phenomenon derives from edge states of non-Hermitian extensions of the parent Chern insulator Hamiltonian. The combination of (non-Hermitian) topology and dissipation hence allows to design topologically robust, spatially localized damping patterns.
\end{abstract}

\date{\today}

\maketitle

\normalem
\emph{Introduction.}
The notion of topological phases of matter  has recently been extended to systems with dissipation and driving described by quantum Liouvillian evolution \cite{avron_2012_adiabatic_response,Altland, Bardyn2012,Bardyn_2013,Rivas2013,Albert2016,Linzner2016,Dangel2018,van_Caspel_2019,Wanjura_2020,Lieu2020,Kawasaki22,Gneiting22,niu2022topological, yang2023symmetry,kawasaki2023, Nava23}, and to non-Hermitian models. Liouvillian dynamics has for example been utilized for the purely dissipative preparation of topological non-equilibrium steady states (NESSs) \cite{Bardyn2012,Bardyn_2013,Rivas2013,Albert2016,Linzner2016,Dangel2018,Goldstein2019}. Non-Hermitian Hamiltonians even possess unique topological features such as spectral topology, modified bulk-boundary correspondence, skin effect, and have been realised in multiple experimental settings \cite{Kawabata2019,Ashida_2020,Bergholtz,banerjee2022,Okuma_2023}. Uniting these two branches of research, recent studies have begun to analyze how non-Hermitian topology can emerge in Liouvillian dynamics \cite{van_Caspel_2019,Song2019,Sayyad2021,Yang22,Kawasaki22, yang2023symmetry,kawasaki2023,Pocklington23}. There, topology appears in two important ways \cite{Altland,Lieu2020,Sayyad2021}: one is in the dynamics of approaching the NESS, the other is in the characteristics of the NESS itself. Using  a dissipative Chern insulator as a prototype, we now unveil how the combination of dissipation with the topology can be used to create topologically protected, spatially localized features in the approach towards the NESS.  In particular, we demonstrate the existence of exponentially edge-localized extremal damping channels as explicit physical signatures in the open quantum  dynamics, characterised by topology and exceptional points, and having no counterparts in Hermitian topological phases.
\begin{figure}
\centering
\includegraphics[width=0.9\columnwidth]{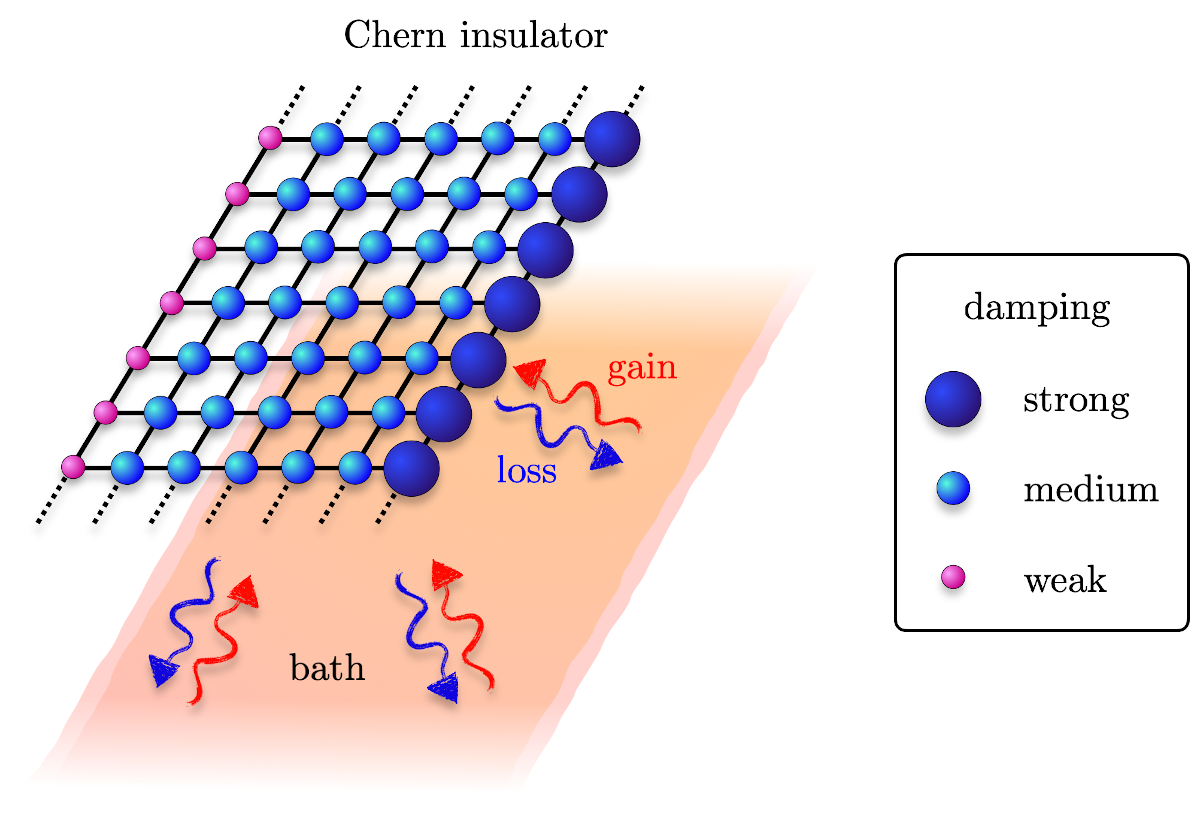}
\caption{A Chern insulator coupled to a bath inducing loss or gain on every site. The topology of the Chern insulator can stabilize edge-selective extremal damping, i.e.~maximal damping on one edge and minimal damping on the opposite edge, as indicated by the size of the lattice site symbols.}
\label{fig:Schematic}
\end{figure}

\normalem
\emph{The model.}
Our starting point is a paradigmatic two-band model for a Chern insulator on a two-dimensional square lattice \cite{bernevig_book}. With periodic boundary conditions (PBC) along $x$ and $y$, the model Hamiltonian reads
\begin{align}
H&=\sum_{\bs{k}}\Psi_{\bs{k}}^\dagger\,\left(\bs{d}\left(\bs{k}\right)\cdot\bs{\sigma}-\mu\,\mathds{1}\right)\,\Psi_{\bs{k}}^\pd,\label{eq:chern_insulator_Hamiltonian}
\end{align}

where $d_x\left(\bs{k}\right)=m-\alpha\,\cos\left(k_x\right)-\alpha\,\cos\left(k_y\right)$, $d_{y}\left(\bs{k}\right)=\beta\,\sin\left(k_{x}\right)$, and $d_{z}\left(\bs{k}\right)=\beta\,\sin\left(k_{y}\right)$,
while $\Psi_{\bs{k}}^T=(c_{\bs{k}\uparrow},c_{\bs{k}\downarrow})$ is the spinor of annihilation operators for electrons with momentum $\bs{k}$ and spin ${s=\uparrow,\downarrow}$. The vector of Pauli matrices is $\bs{\sigma}$, the chemical potential is $\mu$, $m$ denotes an effective mass, and $\alpha,\beta$ quantify hopping amplitudes in the underlying tight-binding model. We furthermore use units such that $e=\hbar=a=1$, where  $e>0$ is the elementary charge, and $a$ the lattice spacing. Without coupling to an environment, $H$ features topological transitions at $m=-2 \alpha$, $m=0$, and  $m=2 \alpha$ with topological phases characterized by Chern numbers $C=-\text{sgn}(m)$ for  $|m|<2 \alpha$ \cite{bernevig_book}.

As sketched in Fig.~\ref{fig:Schematic}, we weakly couple the Chern insulator to a Markovian bath, such that its dynamics is described by a Lindblad  Master equation for the density matrix $\rho$ \cite{manzano_review},
\begin{align}
\dot{\rho}=-i\left[H,\rho\right]+\sum_j\,\left(L_j\,\rho\,L_j^\dagger-\frac{1}{2}\,\left\{L_j^\dagger L_j,\rho\right\}\right).
\end{align}
The jump operators $L_j$ encode loss and gain for $j$-electrons, where $j$ comprises the spin $s$ as well as the lattice site or momentum. We focus on a system preserving translation invariance by using the same jump operators at each site. With PBC, we can transform to momentum space, where the jump operators read
\begin{align}
L_{\bs{k}s}=\begin{cases}\sqrt{2\,\Gamma_s}\,c_{\bs{k}s}^\pd&\text{with }\Gamma_s>0\text{ for loss},\\
\sqrt{-2\,\Gamma_s}\,c_{\bs{k}s}^\dagger&\text{with }\Gamma_s<0\text{ for gain}.\end{cases}
\end{align}
In the remainder, it is convenient to define the net loss or gain as $\Gamma_0=(\Gamma_\uparrow+\Gamma_\downarrow)/2$, and the relative loss or gain as $\Gamma_z=(\Gamma_\uparrow-\Gamma_\downarrow)/2$.

\normalem
\emph{The non-equilibrium steady state and its relation to band topology.}
In the closed limit $L_j=0~\forall j$, dynamics is described by Schr\"odinger's equation. Since every product state of single-particle eigenstates is then a steady state, a closed Chern insulator has infinitely many steady states. If in contrast any finite amount of gain or loss is present, we find that the non-equilibrium steady state (NESS) is unique, as is often the case in dissipative systems \cite{manzano_review}.

\begin{figure}
\centering

\includegraphics[height=0.35\columnwidth]{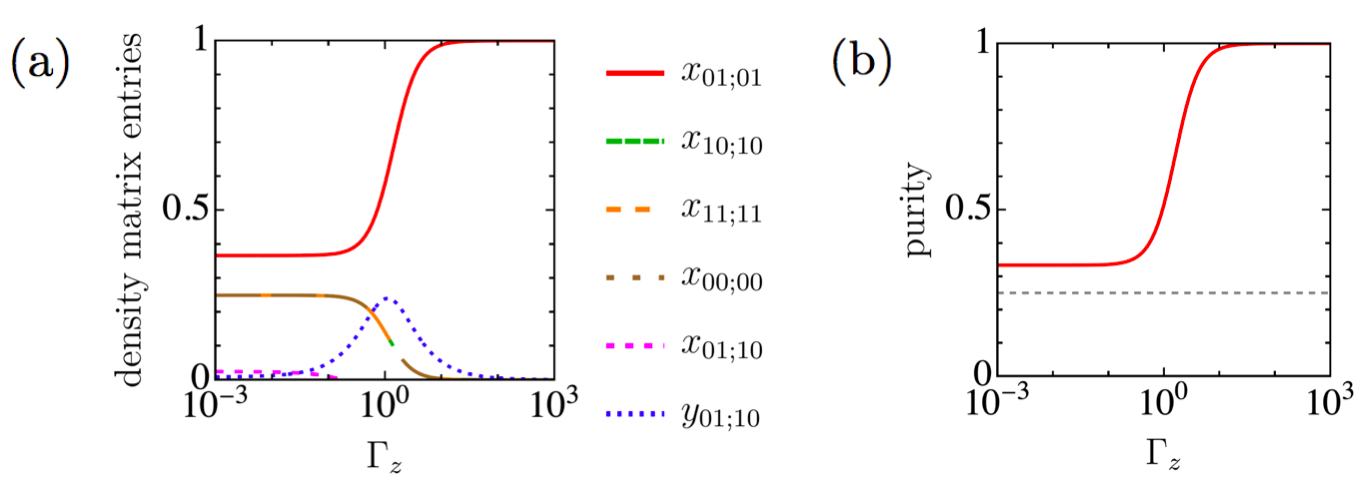}
\caption{Panel (a): Independent non-zero components of the momentum-resolved NESS density matrix for $m=3$, $\alpha=\beta=1$, $\mu=0$, $\Gamma_0=0$, and $\bs{k}^T=(0.38302, -0.05893)$. The density matrix is $\rho_{\bs{k}}^{\text{NESS}}=\sum_{a,b,c,d}(x_{ab;cd}+i\,y_{ab;cd})\,|ab\rangle\langle cd|$ with $x_{ab;cd}=x_{cd;ab}$ and $y_{ab;cd}=-y_{cd;ab}$. 
Panel  (b): corresponding purity $\text{tr}((\rho_{\bs{k}}^{\text{NESS}})^2)$. The purity of a completely mixed state is indicated by the dashed line.}
\label{fig:NESSrhoComponents}
\end{figure}

For $|\Gamma_0|>|\Gamma_z|$, both spin species experience either loss or gain. Since the empty and fully occupied states are also many-particle eigenstates of $H$, the unique NESS is either an entirely empty or a fully occupied pure state. If instead $|\Gamma_z|>|\Gamma_0|$, there is one lossy and one gainy spin species. We then find that the NESS is not a simple dark state, but rather a mixed state determined by the competition of Hamiltonian dynamics and quantum jumps. Only for asymptotically large loss or gain does the NESS approach a pure state determined by the jump operators. For $\Gamma_z\gg|\Gamma_0|$, this state for example corresponds to a fully spin-$\downarrow$-polarized half-filled system. The non-trivial mixed state arising for smaller $|\Gamma_z|$ is most conveniently discussed for PBC, when the NESS density matrix is a tensor product of momentum-resolved density matrices $\rho_{\bs{k}}^{\text{NESS}}$. We write the latter as linear combinations of  $|ab\rangle\langle cd|$, where $a, b,c,d\in\{0,1\}$ label occupancies of electronic states $|n_{\bs{k}\uparrow} n_{\bs{k}\downarrow}\rangle$ with spin $\uparrow,\downarrow$ at the given momentum $\bs{k}$. The Liouvillian can be expressed as a matrix acting on the vector combining the 16 independent real prefactors of $|ab\rangle\langle cd|$. Importantly, the kernel of this Liouvillian matrix can be determined analytically \cite{supplement}, which allows us to deduce closed-form expressions for all steady-state properties of our dissipative Chern insulator with PBC. Fig.~\ref{fig:NESSrhoComponents} for example illustrates the density matrix for a generic momentum, and shows that the NESS is generically strongly mixed.

\begin{figure}
\centering

\includegraphics[height=0.35\columnwidth]{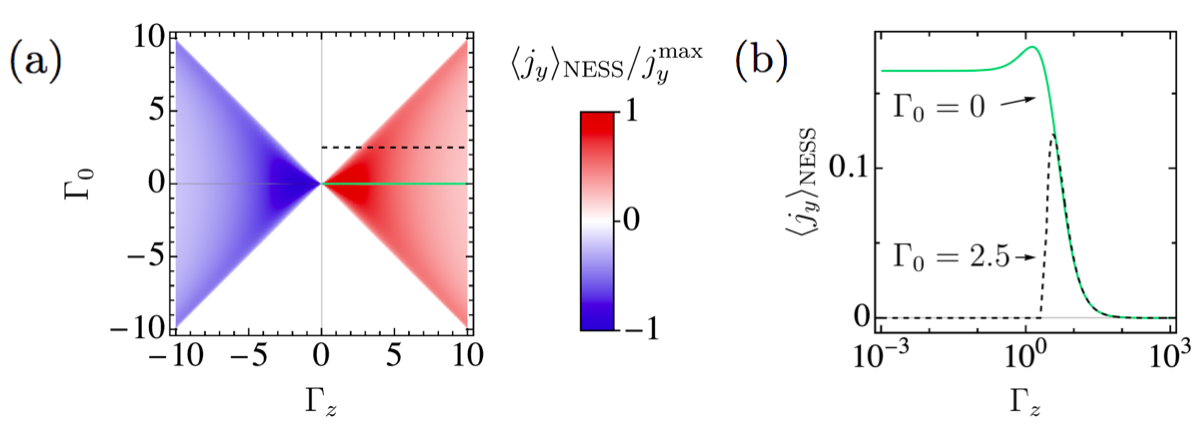}
\caption{The steady state current resulting from the mixed state character. Panel (a): $\langle j_y\rangle_{\text{NESS}}$ as function of $\Gamma_0$ and $\Gamma_z$ in units of its maximal amplitude $j_y^{\text{max}}$. Panel (b): cuts along fixed $\Gamma_0$ as indicated in panel (a). We use $m=3$, $\alpha=\beta=1$, and $\mu=0$, as well as PBC along $x$ and $y$.}
\label{fig:NESScurrentY}
\end{figure}

Given that a fully empty system, a fully occupied system, and (slightly less obviously) a half-filled, spin-polarized system all have zero net current, a finite NESS current density $\langle\bs{j}\rangle_{\text{NESS}}$ provides a convenient experimental proxy of the NESS's mixed state character resulting from the competition between Hamiltonian dynamics and quantum jumps ($\langle\cdot\rangle_{\text{NESS}}$ denotes steady state expectation values). We define the current from the system-internal \cite{avron_2012_adiabatic_response} electron flow $\bs{j}_{\bs{k}} = \Psi_{\bs{k}}^\dagger\,\,[\nabla_{\bs{k}}\,(\bs{d}(\bs{k})\cdot\bs{\sigma})]\,\Psi_{\bs{k}}$ as $\langle\bs{j}\rangle_{\text{NESS}}=-\int d^2k\,\langle \bs{j}_{\bs{k}}\rangle_{\text{NESS}}/(2\pi)^2$. As shown in Fig.~\ref{fig:NESScurrentY}, the largest steady state currents appear when $|\Gamma_0|\lesssim|\Gamma_z|$, i.e.~when the system has just transited from having only loss (or gain) to a regime in which one mode still exhibits loss (or gain), while the other mode has entered its gainy (lossy) regime. This regime maximizes $\langle\bs{j}\rangle_{\text{NESS}}$ because it is associated with a strongly mixed density matrix. 

The topological character of a (closed) Chern insulator is formally defined by the Chern number of its bands, and their occupation in the ground state. This notion of topology clearly cannot carry over unchanged to the dissipative limit. Considering a system in which the coupling to the environment is turned on at a time $t_0$, some of the usual signatures of topology may survive for a limited amount of time. Technically, this can be phrased in the language of a non-Hermitian Hamiltonian $H_{\text{nH}}^{\text{initial}}=H_0-i\sum_jL_j^\dagger L_j^\pdag/2$ that describes the time evolution without quantum jumps at short initial times \cite{Minganti_2019,Roccati}. It has for example been shown that the spectrum of $H_{\text{nH}}^{\text{initial}}$ can still exhibit edge states \cite{Yao2018,Kawabata2018,Bartlett2023}. The Hall conductance, on the contrary, immediately loses its quantization \cite{Chen2018,Philip2018,groenendijk_universal_hall,Hirsbrunner2019}.  As time grows, one could suspect all remaining topological signatures of the original Hamiltonian to be washed out because the NESS is a complicated mixed state that depends strongly  on the jump operators. We for example find that the NESS current does not feature obvious signatures of topological phase transitions \cite{supplement}.

\begin{figure}
\begin{center}
\begin{tabular}{ r l }
\includegraphics[height=0.38\columnwidth]{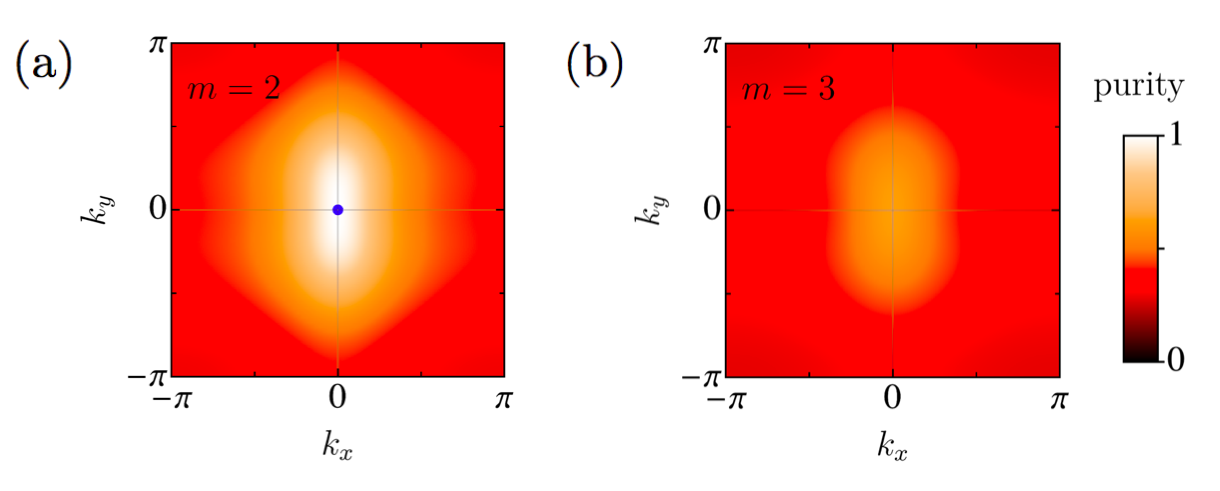}
\end{tabular}
\end{center}
\caption{Momentum-resolved purity $\text{tr}((\rho_{\bs{k}}^{\text{NESS}})^2)$ for $\alpha=\beta=1$, $\Gamma_0=0.08$, and $\Gamma_z=1.2$ for $m=2$ in panel (a), and $m=3$ in panel (b). The blue dot highlights the momentum associated with a pure state.}
\label{fig:NESSpurity}
\end{figure}

Nevertheless, some signatures of the original topology survive at all times. Consider for example topological transitions of the original Hamiltonian. Those are associated with bulk gap closings at specific momenta $\bs{k}_i$, for which the Bloch Hamiltonian is a trivial unit matrix. The Hamiltonian therefore plays no part in determining the NESS at these momenta. The density matrix $\rho_{\bs{k}_i}^{\text{NESS}}$ is then exclusively determined by the jump operators, and the NESS becomes the specific dark state chosen by the jump operators. In the particular model considered here, we even find that  the momentum-resolved purity is a perfect marker for the band topology of the closed Chern insulator: a NESS with unit purity in general arises when a unique pure dark state (annihilated by all jump operators) happens to be an eigenstate of the Hamiltonian. For $|\Gamma_z|>|\Gamma_0|$, that state is a half-filled, spin-polarized state, which is an  eigenstate of $H$ provided $d_x(\bs{k})=d_y(\bs{k})=0$ for some $\bs{k}$. This can be satisfied for all $|m|\leq2\,\alpha$. As shown in Fig.~\ref{fig:NESSpurity}, topological phases of the original Chern Hamiltonian are thus heralded by the existence and location of unit purity points in the momentum-resolved steady-state density matrix. For other models as well, the purity will only peak at topological transitions. Similarly, the jump operators may not have a pure dark state, in which case the purity will only reach a correspondingly reduced value at the momenta associated with gap closings.

\normalem
\emph{Extremal edge-selective damping.}
Having characterized the NESS, we now turn to the main focus of the present study: signatures of the topology of $H$ in the dynamics of approaching the NESS. As our central result, we find that dissipative Chern insulators can feature edge-selective extremal damping as a consequence of topological edge states associated with dynamically-emergent non-Hermitian topology.

To simplify our calculations for finite-size systems, we now switch to the method of ``third quantisation'' \cite{Prosen2008,Prosen_2010}. First, every fermionic operator is converted to a pair of Majorana operators via $c_j=(w_{2j-1}-i\,w_{2j})/2$, where $j$ indicates both lattice site/momentum, and spin. Dropping constant energy offsets, this allows to express the Hamiltonian and jump operators as $H= \sum_{p,q}w_p\,\mathcal{H}_{pq}\,w_q$ and $L_j=\sum_p l_{j,p}\,w_p$, respectively. Second, adjoint creation (annihilation) operators $\hat{c}_p^\dagger$ ($\hat{c}_p^\pdag$) describing the effect of $w_p$ in the Hilbert space of operators are introduced. The Liouvillian can then be cast into the form

\begin{align}
    \hat{\mathcal{L}}=  \frac{1}{2}  \,\sum_{ij}
    \begin{pmatrix} \hat{\bs{c}}^\dagger & \hat{\bs{c}}^{\pdag} \end{pmatrix} \begin{pmatrix}
    -X^{\dagger} & iY \\
    0 & X 
    \end{pmatrix} \begin{pmatrix}
    \hat{\bs{c}}^\dagger\\ \hat{\bs{c}}^{\pdag}
    \end{pmatrix} - \frac{1}{2}\, \text{Tr}(X).
    \end{align}
 Here, $\hat{\bs{c}}^{\dagger}$ ($\hat{\bs{c}}^{\pdag}$) is the vector of adjoint creation (annihilation) operators, while $X=-4i\,\mathcal{H}+\mathcal{M}+\mathcal{M}^{T}$ and $Y=-2i\,(\mathcal{M}-\mathcal{M}^T)$. The components $\mathcal{H}_{pq}$ of $\mathcal{H}$ are defined by the Hamiltonian in Majorana form, while $\mathcal{M}$ follows from the jump operators as $\mathcal{M}_{pq} =\sum_{j}l_{j,p}l_{j,q}^*$. 
 
 The so-called damping matrix $X$ plays a key role for the long-time dynamics approaching the NESS, and allows to define an effective non-Hermitian damping Hamiltonian $H_{\text{nH}}^{\text{damping}}=i\,X\neq H_{\text{nH}}^{\text{initial}}$ \cite{Lieu2020}. For PBC, the non-Hermitian damping Hamiltonian can be block-diagonalized via a unitary $U$ yielding \cite{supplement} $U^\dagger\,H_{\text{nH}}^{\text{damping}}(\bs{k})\,U = \text{diag}({H}_{\text{nH}}^{(\text{+})}(\bs{k}),{H}_{\text{nH}}^{(\text{-})}(\bs{k}))$ with
 
\begin{align}
{H}_{\text{nH}}^{(\pm)}(\bs{k})=& \bs{d}\left(\bs{k}\right)\cdot\bs{\sigma}\pm i\,\xi\,\Gamma_{<}\,\sigma_z+\left(\mp \mu+ i\,|\Gamma_{>}|\right)\,\mathds{1},\label{eq:nh_ham}
\end{align}
where $\Gamma_{<} = \Theta(|\Gamma_0|-|\Gamma_z|)\,\Gamma_z+\Theta(|\Gamma_z|-|\Gamma_0|)\,\Gamma_0$, $\Gamma_{>} = \Theta(|\Gamma_0|-|\Gamma_z|)\,\Gamma_0+\Theta(|\Gamma_z|-|\Gamma_0|)\,\Gamma_z$, and $\xi=\text{sgn}\left(\Gamma_{>}\right)$.

\begin{figure}
\centering
\includegraphics[height=0.4\columnwidth]{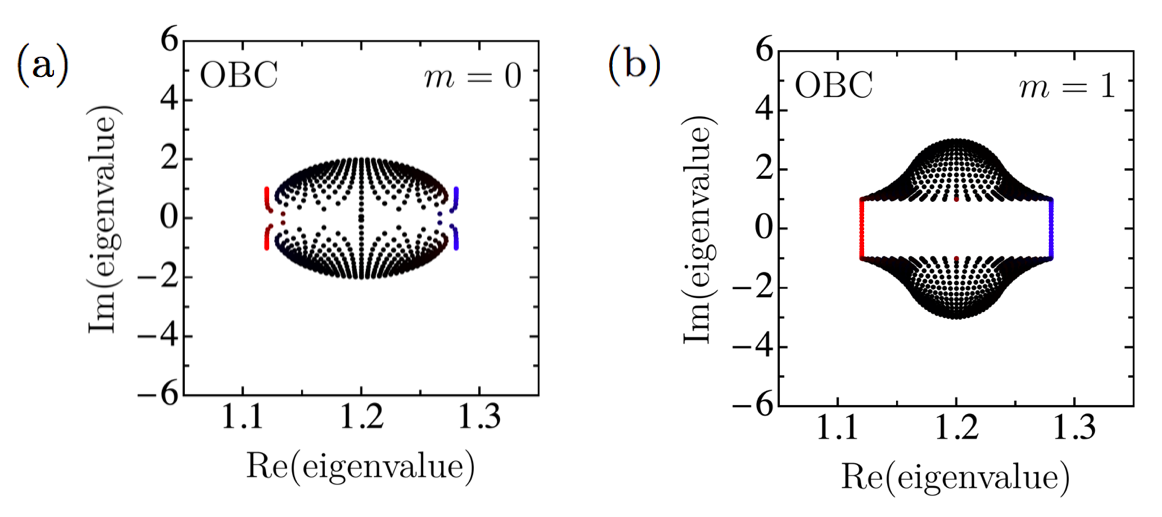}

\caption{Eigenvalues of the damping matrix X for 15 sites and OBC along $x$, combining 50 equidistant values of $k_y$. The dot color indicates the localization of the corresponding eigenstates: bulk states are black, states at the left (right) edge red (blue). We set $m=0$ in panel (a), $m=1$ in panel (b), and use $\Gamma_0=0.08$, $\Gamma_z=1.2$, $\alpha=\beta=1$, and $\mu=0$.}
\label{fig:Xspectrum}
\end{figure}

 In Hermitian topological systems, an important experimental consequence of topological bulk phases are gapless edge states. Non-Hermitian topology also allows for edge states but with contrasting features. For open boundary conditions (OBC) along $x$ and PBC along $y$,  the results of Ref.~\cite{Kawabata2018,supplement} show that edge-states exist for ${H}_{\text{nH}}^{(\pm)}$  in a large window of parameters, both in gapped phases with bulk Chern-invariants as well as in inherently non-Hermitian gapless phases with spectral winding around exceptional points. From the form of Eq. \ref{eq:nh_ham}, these topological edge states can be identified as being inherited from the topology of the non-Hermitian extensions of the parent Hamiltonian $H$.  The edge states of $H_{nH}^+$, for example, are spin-polarized $\downarrow$ ($\uparrow$) on the left (right) edge, and have eigenvalues
$\pm(  \beta \sin k_y + i \xi \Gamma_< )+ i |\Gamma_>| $. Crucially this means that for $X$ (which is $-iH_{nH}^+$), the edge-localised eigenfunctions have largest and smallest real parts of eigenvalues. This  is shown in Fig.~\ref{fig:Xspectrum} both for $m=0$, a phase in which $X$ has bulk exceptional points when subjected to PBC, and $m=1$, a phase with a bulk line gap  \cite{supplement}. We hence dub these states as `extremal edge states'. 

\begin{figure}
\centering

\includegraphics[height=0.4\columnwidth]{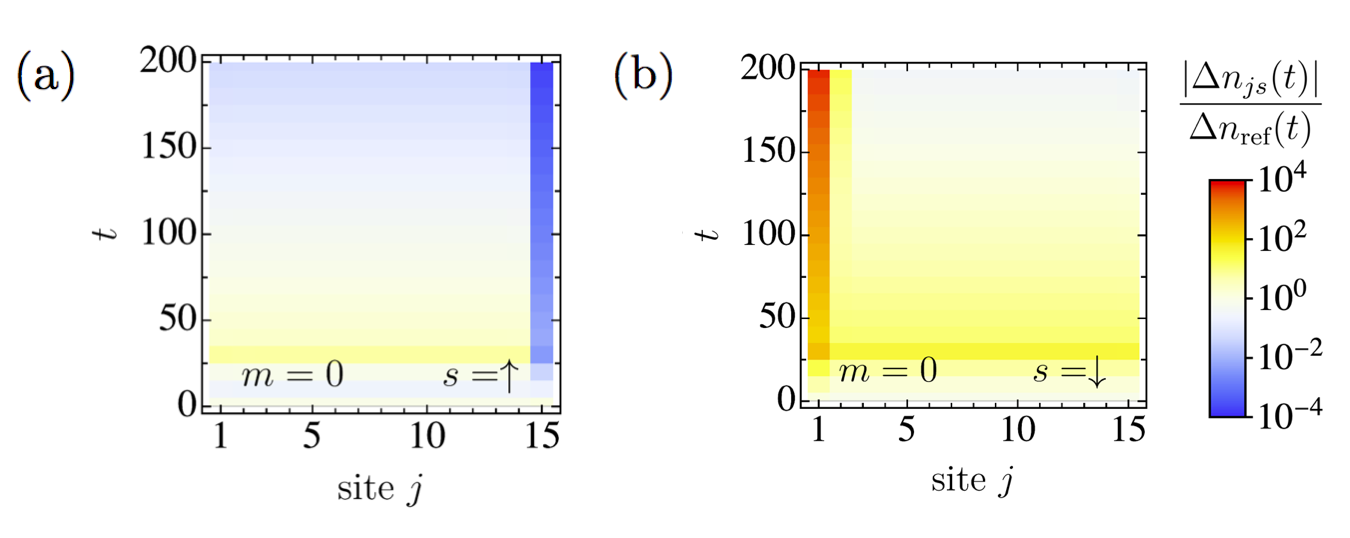}
\caption{Deviation of the electronic density from its steady state value as function of time $t$ and site number $j$ for spin $s$, $\Delta n_{js}(t)$, in a system with 15 sites, OBC along $x$, and PBC along $y$ after initialization in a homogeneous half-filled state at $t=0$. Panel (a) shows $s=\uparrow$, panel (b) corresponds to $s=\downarrow$. We  set all parameters as in Fig.~\ref{fig:Xspectrum} (a), and fix $k_y= 1.629$. The reference density deviation $\Delta n_{\text{ref}}(t)$ is the geometric mean of the largest and smallest density deviations from the steady state value at time $t$ over all sites and spins.}
\label{fig:damping_edge_state}
\end{figure}

The most important experimental implication of these extremal edge states is extremal edge-selective damping. Formally, this follows from the time-evolution of the covariance matrix $\mathcal{C}(t)$ with components $\mathcal{C}_{pq}=\delta_{pq}-\text{tr}(\rho(t)\,w_pw_q)$. The covariance matrix approaches its steady state value $\mathcal{C}_{\text{NESS}}$ according to ${d\Delta\mathcal{C}(t)}/{dt} = -\Delta\mathcal{C}(t)\,X-X^\dagger\,\Delta\mathcal{C}(t)$, where  $\Delta \mathcal{C}(t)=\mathcal{C}(t)-\mathcal{C}_{\text{NESS}}$. This means that $\Delta\mathcal{C}(t)$ can be expressed in terms of the left eigenvectors $\bs{l}_j$ of $X$ and the corresponding eigenvalues $x_j$ as
\begin{align}
\Delta\mathcal{C}(t) = \sum_{j>k}C_{jk}\,e^{-(x_j+x_k)t}\,\left(\bs{l}_j\otimes\bs{l}_k-\bs{l}_k\otimes\bs{l}_j\right),
\end{align}
where $C_{jk}$ are coefficients that depend on the initial density matrix \cite{Song2019,van_Caspel_2019,Yang22}. We find that in the topological regime, the extremal real-parts of eigenvalues with edge-localised eigenfunctions of $X$ dominate  the contributions to  $\Delta \mathcal{C}(t)$ at one edge than in the bulk, and it is analogously suppressed at the opposite edge. This is the edge-selective extremal damping. An experimentally observable consequence is for example the decay of the local on-site densities $n_{js}=c_{js}^\dagger c_{js}^\pdag$ towards their steady state values as measured by $\Delta n_{js}(t) = \langle n_{js}\rangle(t)- \langle n_{js}\rangle_{\text{NESS}}$. As shown in Fig.~\ref{fig:damping_edge_state} for $m=0$ and a generic momentum $k_y$ associated with extremal edge states, edge-selective extremal  damping can also be observed for inherently non-Hermitian phases in which $X$ has exceptional points for PBC and is gapless along imaginary energy \cite{Kawabata2018,supplement}.  This has no counterpart in the closed topological systems and is a completely new manifestation of the dissipative coupling of Chern insulator. (A detailed study of other regimes given in \cite{supplement}).  A system with OBC along $y$ and PBC along $x$, has point-gap topology leading to skin effect and chiral damping \cite{supplement,Song2019,Yang22} in contrast to edge-damping.

\normalem
\emph{Conclusions.}
In this work, we have analyzed the topological heritage of Chern insulators coupled to an environment. Even when driven far from their equilibrium ground states, open Chern insulator preserves remnants of the band topology associated with their Hamiltonians. In particular, the approach of the steady state is governed by an effective non-Hermitian damping Hamiltonian generalizing the Hermitian parent Hamiltonian via the damping matrix.

As a key result, we find that topological edge states in the effective non-Hermitian damping Hamiltonian can be used to engineer damping properties of open Chern insulators also beyond the Liouvillian skin effect. In particular, we find that extremal edge states, i.e.~edge states of the damping matrix with extremal real parts, allow to focus damping on one edge, and to similarly suppress it on the opposite edge. This paves the way for using non-Hermitian topology to design topologically protected damping landscapes in electronic systems. Practical implementations  will benefit from the spin polarization of the extremal edge states. If for example spin-$\uparrow$ electrons leak much more strongly to the environment than spin-$\downarrow$-electrons, $\Gamma_\uparrow\gg\Gamma_\downarrow$, tuning the system to a regime with extremal edge states will result in a non-equilibrium state that after a short initial phase is essentially empty except for one edge, where an exponentially localized stripe of spin-$\downarrow$ electrons remains up to a time that diverges as  $\Gamma_\downarrow\to0$ \cite{supplement}. Finally, given the impressive theoretical and experimental progress in implementing non-Hermitian Hamiltonians in dissipative photonic systems, these platforms might provide alternative avenues for the physics discussed here \cite{RMPOzawa,Lan_2022,YanZhao2023}.

\normalem
\emph{Note:} While preparing this manuscript we became aware of a related work \cite{cherifi2023nonhermitian} that implements closely related physics in lossy classical waveguides. A related work \cite{FYang23} appeared after our pre-print with overlapping results and also discussing edge-selective damping in other models.

\begin{acknowledgments}
The authors are grateful to Jan Budich and Emil Bergholtz for valuable discussions. All authors acknowledge funding by the Deutsche Forschungsgemeinschaft (DFG) via the Emmy Noether Programme (Quantum Design grant, ME4844/1, project-id 327807255), project A04 of the Collaborative Research Center SFB 1143 (project-id 247310070), and the Cluster of Excellence on Complexity and Topology in Quantum Matter ct.qmat (EXC 2147, project-id 390858490).
\end{acknowledgments}

S.H.~ and T.E.~contributed equally to this work.

\bibliography{main2}

\providecommand{\noopsort}[1]{}\providecommand{\singleletter}[1]{#1}%
\begin{thebibliography}{48}%
\makeatletter
\providecommand \@ifxundefined [1]{%
 \@ifx{#1\undefined}
}%
\providecommand \@ifnum [1]{%
 \ifnum #1\expandafter \@firstoftwo
 \else \expandafter \@secondoftwo
 \fi
}%
\providecommand \@ifx [1]{%
 \ifx #1\expandafter \@firstoftwo
 \else \expandafter \@secondoftwo
 \fi
}%
\providecommand \natexlab [1]{#1}%
\providecommand \enquote  [1]{``#1''}%
\providecommand \bibnamefont  [1]{#1}%
\providecommand \bibfnamefont [1]{#1}%
\providecommand \citenamefont [1]{#1}%
\providecommand \href@noop [0]{\@secondoftwo}%
\providecommand \href [0]{\begingroup \@sanitize@url \@href}%
\providecommand \@href[1]{\@@startlink{#1}\@@href}%
\providecommand \@@href[1]{\endgroup#1\@@endlink}%
\providecommand \@sanitize@url [0]{\catcode `\\12\catcode `\$12\catcode
  `\&12\catcode `\#12\catcode `\^12\catcode `\_12\catcode `\%12\relax}%
\providecommand \@@startlink[1]{}%
\providecommand \@@endlink[0]{}%
\providecommand \url  [0]{\begingroup\@sanitize@url \@url }%
\providecommand \@url [1]{\endgroup\@href {#1}{\urlprefix }}%
\providecommand \urlprefix  [0]{URL }%
\providecommand \Eprint [0]{\href }%
\providecommand \doibase [0]{http://dx.doi.org/}%
\providecommand \selectlanguage [0]{\@gobble}%
\providecommand \bibinfo  [0]{\@secondoftwo}%
\providecommand \bibfield  [0]{\@secondoftwo}%
\providecommand \translation [1]{[#1]}%
\providecommand \BibitemOpen [0]{}%
\providecommand \bibitemStop [0]{}%
\providecommand \bibitemNoStop [0]{.\EOS\space}%
\providecommand \EOS [0]{\spacefactor3000\relax}%
\providecommand \BibitemShut  [1]{\csname bibitem#1\endcsname}%
\let\auto@bib@innerbib\@empty
\bibitem [{\citenamefont {Avron}\ \emph {et~al.}(2012)\citenamefont {Avron},
  \citenamefont {Fraas},\ and\ \citenamefont
  {Graf}}]{avron_2012_adiabatic_response}%
  \BibitemOpen
  \bibfield  {author} {\bibinfo {author} {\bibfnamefont {J.~E.}\ \bibnamefont
  {Avron}}, \bibinfo {author} {\bibfnamefont {M.}~\bibnamefont {Fraas}}, \ and\
  \bibinfo {author} {\bibfnamefont {G.~M.}\ \bibnamefont {Graf}},\ }\href
  {\doibase 10.1007/s10955-012-0550-6} {\bibfield  {journal} {\bibinfo
  {journal} {J. Stat. Phys.}\ }\textbf {\bibinfo {volume} {148}},\ \bibinfo
  {pages} {800} (\bibinfo {year} {2012})}\BibitemShut {NoStop}%
\bibitem [{\citenamefont {Altland}\ \emph {et~al.}(2021)\citenamefont
  {Altland}, \citenamefont {Fleischhauer},\ and\ \citenamefont
  {Diehl}}]{Altland}%
  \BibitemOpen
  \bibfield  {author} {\bibinfo {author} {\bibfnamefont {A.}~\bibnamefont
  {Altland}}, \bibinfo {author} {\bibfnamefont {M.}~\bibnamefont
  {Fleischhauer}}, \ and\ \bibinfo {author} {\bibfnamefont {S.}~\bibnamefont
  {Diehl}},\ }\href {\doibase 10.1103/PhysRevX.11.021037} {\bibfield  {journal}
  {\bibinfo  {journal} {Phys. Rev. X}\ }\textbf {\bibinfo {volume} {11}},\
  \bibinfo {pages} {021037} (\bibinfo {year} {2021})}\BibitemShut {NoStop}%
\bibitem [{\citenamefont {Bardyn}\ \emph {et~al.}(2012)\citenamefont {Bardyn},
  \citenamefont {Baranov}, \citenamefont {Rico}, \citenamefont {\ifmmode
  \dot{I}\else \.{I}\fi{}mamo\ifmmode~\breve{g}\else \u{g}\fi{}lu},
  \citenamefont {Zoller},\ and\ \citenamefont {Diehl}}]{Bardyn2012}%
  \BibitemOpen
  \bibfield  {author} {\bibinfo {author} {\bibfnamefont {C.-E.}\ \bibnamefont
  {Bardyn}}, \bibinfo {author} {\bibfnamefont {M.~A.}\ \bibnamefont {Baranov}},
  \bibinfo {author} {\bibfnamefont {E.}~\bibnamefont {Rico}}, \bibinfo {author}
  {\bibfnamefont {A.}~\bibnamefont {\ifmmode \dot{I}\else
  \.{I}\fi{}mamo\ifmmode~\breve{g}\else \u{g}\fi{}lu}}, \bibinfo {author}
  {\bibfnamefont {P.}~\bibnamefont {Zoller}}, \ and\ \bibinfo {author}
  {\bibfnamefont {S.}~\bibnamefont {Diehl}},\ }\href {\doibase
  10.1103/PhysRevLett.109.130402} {\bibfield  {journal} {\bibinfo  {journal}
  {Phys. Rev. Lett.}\ }\textbf {\bibinfo {volume} {109}},\ \bibinfo {pages}
  {130402} (\bibinfo {year} {2012})}\BibitemShut {NoStop}%
\bibitem [{\citenamefont {Bardyn}\ \emph {et~al.}(2013)\citenamefont {Bardyn},
  \citenamefont {Baranov}, \citenamefont {Kraus}, \citenamefont {Rico},
  \citenamefont {{\.{I} }mamo{\u{g}}lu}, \citenamefont {Zoller},\ and\
  \citenamefont {Diehl}}]{Bardyn_2013}%
  \BibitemOpen
  \bibfield  {author} {\bibinfo {author} {\bibfnamefont {C.-E.}\ \bibnamefont
  {Bardyn}}, \bibinfo {author} {\bibfnamefont {M.~A.}\ \bibnamefont {Baranov}},
  \bibinfo {author} {\bibfnamefont {C.~V.}\ \bibnamefont {Kraus}}, \bibinfo
  {author} {\bibfnamefont {E.}~\bibnamefont {Rico}}, \bibinfo {author}
  {\bibfnamefont {A.}~\bibnamefont {{\.{I} }mamo{\u{g}}lu}}, \bibinfo {author}
  {\bibfnamefont {P.}~\bibnamefont {Zoller}}, \ and\ \bibinfo {author}
  {\bibfnamefont {S.}~\bibnamefont {Diehl}},\ }\href {\doibase
  10.1088/1367-2630/15/8/085001} {\bibfield  {journal} {\bibinfo  {journal}
  {New Journal of Physics}\ }\textbf {\bibinfo {volume} {15}},\ \bibinfo
  {pages} {085001} (\bibinfo {year} {2013})}\BibitemShut {NoStop}%
\bibitem [{\citenamefont {Rivas}\ \emph {et~al.}(2013)\citenamefont {Rivas},
  \citenamefont {Viyuela},\ and\ \citenamefont {Martin-Delgado}}]{Rivas2013}%
  \BibitemOpen
  \bibfield  {author} {\bibinfo {author} {\bibfnamefont {A.}~\bibnamefont
  {Rivas}}, \bibinfo {author} {\bibfnamefont {O.}~\bibnamefont {Viyuela}}, \
  and\ \bibinfo {author} {\bibfnamefont {M.~A.}\ \bibnamefont
  {Martin-Delgado}},\ }\href {\doibase 10.1103/PhysRevB.88.155141} {\bibfield
  {journal} {\bibinfo  {journal} {Phys. Rev. B}\ }\textbf {\bibinfo {volume}
  {88}},\ \bibinfo {pages} {155141} (\bibinfo {year} {2013})}\BibitemShut
  {NoStop}%
\bibitem [{\citenamefont {Albert}\ \emph {et~al.}(2016)\citenamefont {Albert},
  \citenamefont {Bradlyn}, \citenamefont {Fraas},\ and\ \citenamefont
  {Jiang}}]{Albert2016}%
  \BibitemOpen
  \bibfield  {author} {\bibinfo {author} {\bibfnamefont {V.~V.}\ \bibnamefont
  {Albert}}, \bibinfo {author} {\bibfnamefont {B.}~\bibnamefont {Bradlyn}},
  \bibinfo {author} {\bibfnamefont {M.}~\bibnamefont {Fraas}}, \ and\ \bibinfo
  {author} {\bibfnamefont {L.}~\bibnamefont {Jiang}},\ }\href {\doibase
  10.1103/PhysRevX.6.041031} {\bibfield  {journal} {\bibinfo  {journal} {Phys.
  Rev. X}\ }\textbf {\bibinfo {volume} {6}},\ \bibinfo {pages} {041031}
  (\bibinfo {year} {2016})}\BibitemShut {NoStop}%
\bibitem [{\citenamefont {Linzner}\ \emph {et~al.}(2016)\citenamefont
  {Linzner}, \citenamefont {Wawer}, \citenamefont {Grusdt},\ and\ \citenamefont
  {Fleischhauer}}]{Linzner2016}%
  \BibitemOpen
  \bibfield  {author} {\bibinfo {author} {\bibfnamefont {D.}~\bibnamefont
  {Linzner}}, \bibinfo {author} {\bibfnamefont {L.}~\bibnamefont {Wawer}},
  \bibinfo {author} {\bibfnamefont {F.}~\bibnamefont {Grusdt}}, \ and\ \bibinfo
  {author} {\bibfnamefont {M.}~\bibnamefont {Fleischhauer}},\ }\href {\doibase
  10.1103/PhysRevB.94.201105} {\bibfield  {journal} {\bibinfo  {journal} {Phys.
  Rev. B}\ }\textbf {\bibinfo {volume} {94}},\ \bibinfo {pages} {201105}
  (\bibinfo {year} {2016})}\BibitemShut {NoStop}%
\bibitem [{\citenamefont {Dangel}\ \emph {et~al.}(2018)\citenamefont {Dangel},
  \citenamefont {Wagner}, \citenamefont {Cartarius}, \citenamefont {Main},\
  and\ \citenamefont {Wunner}}]{Dangel2018}%
  \BibitemOpen
  \bibfield  {author} {\bibinfo {author} {\bibfnamefont {F.}~\bibnamefont
  {Dangel}}, \bibinfo {author} {\bibfnamefont {M.}~\bibnamefont {Wagner}},
  \bibinfo {author} {\bibfnamefont {H.}~\bibnamefont {Cartarius}}, \bibinfo
  {author} {\bibfnamefont {J.}~\bibnamefont {Main}}, \ and\ \bibinfo {author}
  {\bibfnamefont {G.}~\bibnamefont {Wunner}},\ }\href {\doibase
  10.1103/PhysRevA.98.013628} {\bibfield  {journal} {\bibinfo  {journal} {Phys.
  Rev. A}\ }\textbf {\bibinfo {volume} {98}},\ \bibinfo {pages} {013628}
  (\bibinfo {year} {2018})}\BibitemShut {NoStop}%
\bibitem [{\citenamefont {van Caspel}\ \emph {et~al.}(2019)\citenamefont {van
  Caspel}, \citenamefont {Arze},\ and\ \citenamefont
  {Castillo}}]{van_Caspel_2019}%
  \BibitemOpen
  \bibfield  {author} {\bibinfo {author} {\bibfnamefont {M.}~\bibnamefont {van
  Caspel}}, \bibinfo {author} {\bibfnamefont {S.~E.~T.}\ \bibnamefont {Arze}},
  \ and\ \bibinfo {author} {\bibfnamefont {I.~P.}\ \bibnamefont {Castillo}},\
  }\href {\doibase 10.21468/scipostphys.6.2.026} {\bibfield  {journal}
  {\bibinfo  {journal} {{SciPost} Physics}\ }\textbf {\bibinfo {volume} {6}}
  (\bibinfo {year} {2019}),\ 10.21468/scipostphys.6.2.026}\BibitemShut
  {NoStop}%
\bibitem [{\citenamefont {Wanjura}\ \emph {et~al.}(2020)\citenamefont
  {Wanjura}, \citenamefont {Brunelli},\ and\ \citenamefont
  {Nunnenkamp}}]{Wanjura_2020}%
  \BibitemOpen
  \bibfield  {author} {\bibinfo {author} {\bibfnamefont {C.~C.}\ \bibnamefont
  {Wanjura}}, \bibinfo {author} {\bibfnamefont {M.}~\bibnamefont {Brunelli}}, \
  and\ \bibinfo {author} {\bibfnamefont {A.}~\bibnamefont {Nunnenkamp}},\
  }\href {\doibase 10.1038/s41467-020-16863-9} {\bibfield  {journal} {\bibinfo
  {journal} {Nature Communications}\ }\textbf {\bibinfo {volume} {11}}
  (\bibinfo {year} {2020}),\ 10.1038/s41467-020-16863-9}\BibitemShut {NoStop}%
\bibitem [{\citenamefont {Lieu}\ \emph {et~al.}(2020)\citenamefont {Lieu},
  \citenamefont {McGinley},\ and\ \citenamefont {Cooper}}]{Lieu2020}%
  \BibitemOpen
  \bibfield  {author} {\bibinfo {author} {\bibfnamefont {S.}~\bibnamefont
  {Lieu}}, \bibinfo {author} {\bibfnamefont {M.}~\bibnamefont {McGinley}}, \
  and\ \bibinfo {author} {\bibfnamefont {N.~R.}\ \bibnamefont {Cooper}},\
  }\href {\doibase 10.1103/PhysRevLett.124.040401} {\bibfield  {journal}
  {\bibinfo  {journal} {Phys. Rev. Lett.}\ }\textbf {\bibinfo {volume} {124}},\
  \bibinfo {pages} {040401} (\bibinfo {year} {2020})}\BibitemShut {NoStop}%
\bibitem [{\citenamefont {Kawasaki}\ \emph {et~al.}(2022)\citenamefont
  {Kawasaki}, \citenamefont {Mochizuki},\ and\ \citenamefont
  {Obuse}}]{Kawasaki22}%
  \BibitemOpen
  \bibfield  {author} {\bibinfo {author} {\bibfnamefont {M.}~\bibnamefont
  {Kawasaki}}, \bibinfo {author} {\bibfnamefont {K.}~\bibnamefont {Mochizuki}},
  \ and\ \bibinfo {author} {\bibfnamefont {H.}~\bibnamefont {Obuse}},\ }\href
  {\doibase 10.1103/PhysRevB.106.035408} {\bibfield  {journal} {\bibinfo
  {journal} {Phys. Rev. B}\ }\textbf {\bibinfo {volume} {106}},\ \bibinfo
  {pages} {035408} (\bibinfo {year} {2022})}\BibitemShut {NoStop}%
\bibitem [{\citenamefont {Gneiting}\ \emph {et~al.}(2022)\citenamefont
  {Gneiting}, \citenamefont {Koottandavida}, \citenamefont {Rozhkov},\ and\
  \citenamefont {Nori}}]{Gneiting22}%
  \BibitemOpen
  \bibfield  {author} {\bibinfo {author} {\bibfnamefont {C.}~\bibnamefont
  {Gneiting}}, \bibinfo {author} {\bibfnamefont {A.}~\bibnamefont
  {Koottandavida}}, \bibinfo {author} {\bibfnamefont {A.~V.}\ \bibnamefont
  {Rozhkov}}, \ and\ \bibinfo {author} {\bibfnamefont {F.}~\bibnamefont
  {Nori}},\ }\href {\doibase 10.1103/PhysRevResearch.4.023036} {\bibfield
  {journal} {\bibinfo  {journal} {Phys. Rev. Res.}\ }\textbf {\bibinfo {volume}
  {4}},\ \bibinfo {pages} {023036} (\bibinfo {year} {2022})}\BibitemShut
  {NoStop}%
\bibitem [{\citenamefont {Niu}\ and\ \citenamefont
  {Wang}(2022)}]{niu2022topological}%
  \BibitemOpen
  \bibfield  {author} {\bibinfo {author} {\bibfnamefont {X.}~\bibnamefont
  {Niu}}\ and\ \bibinfo {author} {\bibfnamefont {J.}~\bibnamefont {Wang}},\
  }\href {https://arxiv.org/abs/2211.04233} {\enquote {\bibinfo {title}
  {{Topological extension including quantum jump}},}\ } (\bibinfo {year}
  {2022}),\ \Eprint {http://arxiv.org/abs/2211.04233} {2211.04233} \BibitemShut
  {NoStop}%
\bibitem [{\citenamefont {Yang}\ \emph
  {et~al.}(2023{\natexlab{a}})\citenamefont {Yang}, \citenamefont {Wei},
  \citenamefont {Tong}, \citenamefont {Cao},\ and\ \citenamefont
  {Kou}}]{yang2023symmetry}%
  \BibitemOpen
  \bibfield  {author} {\bibinfo {author} {\bibfnamefont {F.}~\bibnamefont
  {Yang}}, \bibinfo {author} {\bibfnamefont {Z.}~\bibnamefont {Wei}}, \bibinfo
  {author} {\bibfnamefont {X.}~\bibnamefont {Tong}}, \bibinfo {author}
  {\bibfnamefont {K.}~\bibnamefont {Cao}}, \ and\ \bibinfo {author}
  {\bibfnamefont {S.-P.}\ \bibnamefont {Kou}},\ }\href
  {https://arxiv.org/abs/2301.03208} {\enquote {\bibinfo {title} {{Symmetry
  classes of dissipative topological insulators with edge dark state}},}\ }
  (\bibinfo {year} {2023}{\natexlab{a}}),\ \Eprint
  {http://arxiv.org/abs/2301.03208} {2301.03208} \BibitemShut {NoStop}%
\bibitem [{\citenamefont {Kawasaki}\ and\ \citenamefont
  {Obuse}(2023)}]{kawasaki2023}%
  \BibitemOpen
  \bibfield  {author} {\bibinfo {author} {\bibfnamefont {M.}~\bibnamefont
  {Kawasaki}}\ and\ \bibinfo {author} {\bibfnamefont {H.}~\bibnamefont
  {Obuse}},\ }\href {https://arxiv.org/abs/2301.08446} {\enquote {\bibinfo
  {title} {{Topological Phases in a PT-Symmetric Dissipative Kitaev Chain}},}\
  } (\bibinfo {year} {2023}),\ \Eprint {http://arxiv.org/abs/2301.08446}
  {2301.08446} \BibitemShut {NoStop}%
\bibitem [{\citenamefont {Nava}\ \emph {et~al.}(2023)\citenamefont {Nava},
  \citenamefont {Campagnano}, \citenamefont {Sodano},\ and\ \citenamefont
  {Giuliano}}]{Nava23}%
  \BibitemOpen
  \bibfield  {author} {\bibinfo {author} {\bibfnamefont {A.}~\bibnamefont
  {Nava}}, \bibinfo {author} {\bibfnamefont {G.}~\bibnamefont {Campagnano}},
  \bibinfo {author} {\bibfnamefont {P.}~\bibnamefont {Sodano}}, \ and\ \bibinfo
  {author} {\bibfnamefont {D.}~\bibnamefont {Giuliano}},\ }\href {\doibase
  10.1103/PhysRevB.107.035113} {\bibfield  {journal} {\bibinfo  {journal}
  {Phys. Rev. B}\ }\textbf {\bibinfo {volume} {107}},\ \bibinfo {pages}
  {035113} (\bibinfo {year} {2023})}\BibitemShut {NoStop}%
\bibitem [{\citenamefont {Goldstein}(2019)}]{Goldstein2019}%
  \BibitemOpen
  \bibfield  {author} {\bibinfo {author} {\bibfnamefont {M.}~\bibnamefont
  {Goldstein}},\ }\href {\doibase 10.21468/SciPostPhys.7.5.067} {\bibfield
  {journal} {\bibinfo  {journal} {SciPost Phys.}\ }\textbf {\bibinfo {volume}
  {7}},\ \bibinfo {pages} {067} (\bibinfo {year} {2019})}\BibitemShut {NoStop}%
\bibitem [{\citenamefont {Kawabata}\ \emph {et~al.}(2019)\citenamefont
  {Kawabata}, \citenamefont {Shiozaki}, \citenamefont {Ueda},\ and\
  \citenamefont {Sato}}]{Kawabata2019}%
  \BibitemOpen
  \bibfield  {author} {\bibinfo {author} {\bibfnamefont {K.}~\bibnamefont
  {Kawabata}}, \bibinfo {author} {\bibfnamefont {K.}~\bibnamefont {Shiozaki}},
  \bibinfo {author} {\bibfnamefont {M.}~\bibnamefont {Ueda}}, \ and\ \bibinfo
  {author} {\bibfnamefont {M.}~\bibnamefont {Sato}},\ }\href {\doibase
  10.1103/PhysRevX.9.041015} {\bibfield  {journal} {\bibinfo  {journal} {Phys.
  Rev. X}\ }\textbf {\bibinfo {volume} {9}},\ \bibinfo {pages} {041015}
  (\bibinfo {year} {2019})}\BibitemShut {NoStop}%
\bibitem [{\citenamefont {Ashida}\ \emph {et~al.}(2020)\citenamefont {Ashida},
  \citenamefont {Gong},\ and\ \citenamefont {Ueda}}]{Ashida_2020}%
  \BibitemOpen
  \bibfield  {author} {\bibinfo {author} {\bibfnamefont {Y.}~\bibnamefont
  {Ashida}}, \bibinfo {author} {\bibfnamefont {Z.}~\bibnamefont {Gong}}, \ and\
  \bibinfo {author} {\bibfnamefont {M.}~\bibnamefont {Ueda}},\ }\href {\doibase
  10.1080/00018732.2021.1876991} {\bibfield  {journal} {\bibinfo  {journal}
  {Advances in Physics}\ }\textbf {\bibinfo {volume} {69}},\ \bibinfo {pages}
  {249} (\bibinfo {year} {2020})}\BibitemShut {NoStop}%
\bibitem [{\citenamefont {Bergholtz}\ \emph {et~al.}(2021)\citenamefont
  {Bergholtz}, \citenamefont {Budich},\ and\ \citenamefont
  {Kunst}}]{Bergholtz}%
  \BibitemOpen
  \bibfield  {author} {\bibinfo {author} {\bibfnamefont {E.~J.}\ \bibnamefont
  {Bergholtz}}, \bibinfo {author} {\bibfnamefont {J.~C.}\ \bibnamefont
  {Budich}}, \ and\ \bibinfo {author} {\bibfnamefont {F.~K.}\ \bibnamefont
  {Kunst}},\ }\href {\doibase 10.1103/RevModPhys.93.015005} {\bibfield
  {journal} {\bibinfo  {journal} {Rev. Mod. Phys.}\ }\textbf {\bibinfo {volume}
  {93}},\ \bibinfo {pages} {015005} (\bibinfo {year} {2021})}\BibitemShut
  {NoStop}%
\bibitem [{\citenamefont {Banerjee}\ \emph {et~al.}(2022)\citenamefont
  {Banerjee}, \citenamefont {Sarkar}, \citenamefont {Dey},\ and\ \citenamefont
  {Narayan}}]{banerjee2022}%
  \BibitemOpen
  \bibfield  {author} {\bibinfo {author} {\bibfnamefont {A.}~\bibnamefont
  {Banerjee}}, \bibinfo {author} {\bibfnamefont {R.}~\bibnamefont {Sarkar}},
  \bibinfo {author} {\bibfnamefont {S.}~\bibnamefont {Dey}}, \ and\ \bibinfo
  {author} {\bibfnamefont {A.}~\bibnamefont {Narayan}},\ }\href
  {https://arxiv.org/abs/2212.06478} {\enquote {\bibinfo {title}
  {{Non-Hermitian Topological Phases: Principles and Prospects}},}\ } (\bibinfo
  {year} {2022}),\ \Eprint {http://arxiv.org/abs/2212.06478} {2212.06478}
  \BibitemShut {NoStop}%
\bibitem [{\citenamefont {Okuma}\ and\ \citenamefont
  {Sato}(2023)}]{Okuma_2023}%
  \BibitemOpen
  \bibfield  {author} {\bibinfo {author} {\bibfnamefont {N.}~\bibnamefont
  {Okuma}}\ and\ \bibinfo {author} {\bibfnamefont {M.}~\bibnamefont {Sato}},\
  }\href {\doibase 10.1146/annurev-conmatphys-040521-033133} {\bibfield
  {journal} {\bibinfo  {journal} {Annual Review of Condensed Matter Physics}\
  }\textbf {\bibinfo {volume} {14}},\ \bibinfo {pages} {83} (\bibinfo {year}
  {2023})}\BibitemShut {NoStop}%
\bibitem [{\citenamefont {Song}\ \emph {et~al.}(2019)\citenamefont {Song},
  \citenamefont {Yao},\ and\ \citenamefont {Wang}}]{Song2019}%
  \BibitemOpen
  \bibfield  {author} {\bibinfo {author} {\bibfnamefont {F.}~\bibnamefont
  {Song}}, \bibinfo {author} {\bibfnamefont {S.}~\bibnamefont {Yao}}, \ and\
  \bibinfo {author} {\bibfnamefont {Z.}~\bibnamefont {Wang}},\ }\href {\doibase
  10.1103/PhysRevLett.123.170401} {\bibfield  {journal} {\bibinfo  {journal}
  {Phys. Rev. Lett.}\ }\textbf {\bibinfo {volume} {123}},\ \bibinfo {pages}
  {170401} (\bibinfo {year} {2019})}\BibitemShut {NoStop}%
\bibitem [{\citenamefont {Sayyad}\ \emph {et~al.}(2021)\citenamefont {Sayyad},
  \citenamefont {Yu}, \citenamefont {Grushin},\ and\ \citenamefont
  {Sieberer}}]{Sayyad2021}%
  \BibitemOpen
  \bibfield  {author} {\bibinfo {author} {\bibfnamefont {S.}~\bibnamefont
  {Sayyad}}, \bibinfo {author} {\bibfnamefont {J.}~\bibnamefont {Yu}}, \bibinfo
  {author} {\bibfnamefont {A.~G.}\ \bibnamefont {Grushin}}, \ and\ \bibinfo
  {author} {\bibfnamefont {L.~M.}\ \bibnamefont {Sieberer}},\ }\href {\doibase
  10.1103/PhysRevResearch.3.033022} {\bibfield  {journal} {\bibinfo  {journal}
  {Phys. Rev. Res.}\ }\textbf {\bibinfo {volume} {3}},\ \bibinfo {pages}
  {033022} (\bibinfo {year} {2021})}\BibitemShut {NoStop}%
\bibitem [{\citenamefont {Yang}\ \emph {et~al.}(2022)\citenamefont {Yang},
  \citenamefont {Jiang},\ and\ \citenamefont {Bergholtz}}]{Yang22}%
  \BibitemOpen
  \bibfield  {author} {\bibinfo {author} {\bibfnamefont {F.}~\bibnamefont
  {Yang}}, \bibinfo {author} {\bibfnamefont {Q.-D.}\ \bibnamefont {Jiang}}, \
  and\ \bibinfo {author} {\bibfnamefont {E.~J.}\ \bibnamefont {Bergholtz}},\
  }\href {\doibase 10.1103/PhysRevResearch.4.023160} {\bibfield  {journal}
  {\bibinfo  {journal} {Phys. Rev. Res.}\ }\textbf {\bibinfo {volume} {4}},\
  \bibinfo {pages} {023160} (\bibinfo {year} {2022})}\BibitemShut {NoStop}%
\bibitem [{\citenamefont {Pocklington}\ \emph {et~al.}(2023)\citenamefont
  {Pocklington}, \citenamefont {Wang},\ and\ \citenamefont
  {Clerk}}]{Pocklington23}%
  \BibitemOpen
  \bibfield  {author} {\bibinfo {author} {\bibfnamefont {A.}~\bibnamefont
  {Pocklington}}, \bibinfo {author} {\bibfnamefont {Y.-X.}\ \bibnamefont
  {Wang}}, \ and\ \bibinfo {author} {\bibfnamefont {A.~A.}\ \bibnamefont
  {Clerk}},\ }\href {\doibase 10.1103/PhysRevLett.130.123602} {\bibfield
  {journal} {\bibinfo  {journal} {Phys. Rev. Lett.}\ }\textbf {\bibinfo
  {volume} {130}},\ \bibinfo {pages} {123602} (\bibinfo {year}
  {2023})}\BibitemShut {NoStop}%
\bibitem [{\citenamefont {Bernevig}\ and\ \citenamefont
  {Hughes}(2013)}]{bernevig_book}%
  \BibitemOpen
  \bibfield  {author} {\bibinfo {author} {\bibfnamefont {B.~A.}\ \bibnamefont
  {Bernevig}}\ and\ \bibinfo {author} {\bibfnamefont {T.~L.}\ \bibnamefont
  {Hughes}},\ }\href {http://www.jstor.org/stable/j.ctt19cc2gc} {\emph
  {\bibinfo {title} {{Topological Insulators and Topological
  Superconductors}}}}\ (\bibinfo  {publisher} {Princeton University Press},\
  \bibinfo {year} {2013})\BibitemShut {NoStop}%
\bibitem [{\citenamefont {Manzano}(2020)}]{manzano_review}%
  \BibitemOpen
  \bibfield  {author} {\bibinfo {author} {\bibfnamefont {D.}~\bibnamefont
  {Manzano}},\ }\href {\doibase 10.1063/1.5115323} {\bibfield  {journal}
  {\bibinfo  {journal} {AIP Advances}\ }\textbf {\bibinfo {volume} {10}},\
  \bibinfo {pages} {025106} (\bibinfo {year} {2020})}\BibitemShut {NoStop}%
\bibitem [{sup()}]{supplement}%
  \BibitemOpen
  \href@noop {} {}\bibinfo {note} {See Supplemental Material for detailed
  analytical calculation of NESS under PBCs, derivation of non-Hermitian
  Hamiltonian associated with the damping matrix, the symmetry analysis and
  spectra of the damping matrix, extensive numerical results on the
  edge-selective extremal damping for a range of system parameters and also
  containing which also includes Refs.\cite{Ruter2010,Zhang2018}}\BibitemShut
  {NoStop}%
\bibitem [{\citenamefont {Minganti}\ \emph {et~al.}(2019)\citenamefont
  {Minganti}, \citenamefont {Miranowicz}, \citenamefont {Chhajlany},\ and\
  \citenamefont {Nori}}]{Minganti_2019}%
  \BibitemOpen
  \bibfield  {author} {\bibinfo {author} {\bibfnamefont {F.}~\bibnamefont
  {Minganti}}, \bibinfo {author} {\bibfnamefont {A.}~\bibnamefont
  {Miranowicz}}, \bibinfo {author} {\bibfnamefont {R.~W.}\ \bibnamefont
  {Chhajlany}}, \ and\ \bibinfo {author} {\bibfnamefont {F.}~\bibnamefont
  {Nori}},\ }\href {\doibase 10.1103/physreva.100.062131} {\bibfield  {journal}
  {\bibinfo  {journal} {Physical Review A}\ }\textbf {\bibinfo {volume} {100}}
  (\bibinfo {year} {2019}),\ 10.1103/physreva.100.062131}\BibitemShut {NoStop}%
\bibitem [{\citenamefont {Roccati}\ \emph {et~al.}(2022)\citenamefont
  {Roccati}, \citenamefont {Palma}, \citenamefont {Ciccarello},\ and\
  \citenamefont {Bagarello}}]{Roccati}%
  \BibitemOpen
  \bibfield  {author} {\bibinfo {author} {\bibfnamefont {F.}~\bibnamefont
  {Roccati}}, \bibinfo {author} {\bibfnamefont {G.~M.}\ \bibnamefont {Palma}},
  \bibinfo {author} {\bibfnamefont {F.}~\bibnamefont {Ciccarello}}, \ and\
  \bibinfo {author} {\bibfnamefont {F.}~\bibnamefont {Bagarello}},\ }\href
  {\doibase 10.1142/S1230161222500044} {\bibfield  {journal} {\bibinfo
  {journal} {Open Systems \& Information Dynamics}\ }\textbf {\bibinfo {volume}
  {29}},\ \bibinfo {pages} {2250004} (\bibinfo {year} {2022})}\BibitemShut
  {NoStop}%
\bibitem [{\citenamefont {{Yao, Shunyu and Song, Fei and Wang,
  Zhong}}(2018)}]{Yao2018}%
  \BibitemOpen
  \bibfield  {author} {\bibinfo {author} {\bibnamefont {{Yao, Shunyu and Song,
  Fei and Wang, Zhong}}},\ }\href {\doibase 10.1103/PhysRevLett.121.136802}
  {\bibfield  {journal} {\bibinfo  {journal} {Phys. Rev. Lett.}\ }\textbf
  {\bibinfo {volume} {121}},\ \bibinfo {pages} {136802} (\bibinfo {year}
  {2018})}\BibitemShut {NoStop}%
\bibitem [{\citenamefont {Kawabata}\ \emph {et~al.}(2018)\citenamefont
  {Kawabata}, \citenamefont {Shiozaki},\ and\ \citenamefont
  {Ueda}}]{Kawabata2018}%
  \BibitemOpen
  \bibfield  {author} {\bibinfo {author} {\bibfnamefont {K.}~\bibnamefont
  {Kawabata}}, \bibinfo {author} {\bibfnamefont {K.}~\bibnamefont {Shiozaki}},
  \ and\ \bibinfo {author} {\bibfnamefont {M.}~\bibnamefont {Ueda}},\ }\href
  {\doibase 10.1103/PhysRevB.98.165148} {\bibfield  {journal} {\bibinfo
  {journal} {Phys. Rev. B}\ }\textbf {\bibinfo {volume} {98}},\ \bibinfo
  {pages} {165148} (\bibinfo {year} {2018})}\BibitemShut {NoStop}%
\bibitem [{\citenamefont {Bartlett}\ and\ \citenamefont
  {Zhao}(2023)}]{Bartlett2023}%
  \BibitemOpen
  \bibfield  {author} {\bibinfo {author} {\bibfnamefont {J.}~\bibnamefont
  {Bartlett}}\ and\ \bibinfo {author} {\bibfnamefont {E.}~\bibnamefont
  {Zhao}},\ }\href {\doibase 10.1103/PhysRevB.107.035101} {\bibfield  {journal}
  {\bibinfo  {journal} {Phys. Rev. B}\ }\textbf {\bibinfo {volume} {107}},\
  \bibinfo {pages} {035101} (\bibinfo {year} {2023})}\BibitemShut {NoStop}%
\bibitem [{\citenamefont {Chen}\ and\ \citenamefont {Zhai}(2018)}]{Chen2018}%
  \BibitemOpen
  \bibfield  {author} {\bibinfo {author} {\bibfnamefont {Y.}~\bibnamefont
  {Chen}}\ and\ \bibinfo {author} {\bibfnamefont {H.}~\bibnamefont {Zhai}},\
  }\href {\doibase 10.1103/PhysRevB.98.245130} {\bibfield  {journal} {\bibinfo
  {journal} {Phys. Rev. B}\ }\textbf {\bibinfo {volume} {98}},\ \bibinfo
  {pages} {245130} (\bibinfo {year} {2018})}\BibitemShut {NoStop}%
\bibitem [{\citenamefont {Philip}\ \emph {et~al.}(2018)\citenamefont {Philip},
  \citenamefont {Hirsbrunner},\ and\ \citenamefont {Gilbert}}]{Philip2018}%
  \BibitemOpen
  \bibfield  {author} {\bibinfo {author} {\bibfnamefont {T.~M.}\ \bibnamefont
  {Philip}}, \bibinfo {author} {\bibfnamefont {M.~R.}\ \bibnamefont
  {Hirsbrunner}}, \ and\ \bibinfo {author} {\bibfnamefont {M.~J.}\ \bibnamefont
  {Gilbert}},\ }\href {\doibase 10.1103/PhysRevB.98.155430} {\bibfield
  {journal} {\bibinfo  {journal} {Phys. Rev. B}\ }\textbf {\bibinfo {volume}
  {98}},\ \bibinfo {pages} {155430} (\bibinfo {year} {2018})}\BibitemShut
  {NoStop}%
\bibitem [{\citenamefont {Groenendijk}\ \emph {et~al.}(2021)\citenamefont
  {Groenendijk}, \citenamefont {Schmidt},\ and\ \citenamefont
  {Meng}}]{groenendijk_universal_hall}%
  \BibitemOpen
  \bibfield  {author} {\bibinfo {author} {\bibfnamefont {S.}~\bibnamefont
  {Groenendijk}}, \bibinfo {author} {\bibfnamefont {T.~L.}\ \bibnamefont
  {Schmidt}}, \ and\ \bibinfo {author} {\bibfnamefont {T.}~\bibnamefont
  {Meng}},\ }\href {\doibase 10.1103/PhysRevResearch.3.023001} {\bibfield
  {journal} {\bibinfo  {journal} {Phys. Rev. Res.}\ }\textbf {\bibinfo {volume}
  {3}},\ \bibinfo {pages} {023001} (\bibinfo {year} {2021})}\BibitemShut
  {NoStop}%
\bibitem [{\citenamefont {Hirsbrunner}\ \emph {et~al.}(2019)\citenamefont
  {Hirsbrunner}, \citenamefont {Philip},\ and\ \citenamefont
  {Gilbert}}]{Hirsbrunner2019}%
  \BibitemOpen
  \bibfield  {author} {\bibinfo {author} {\bibfnamefont {M.~R.}\ \bibnamefont
  {Hirsbrunner}}, \bibinfo {author} {\bibfnamefont {T.~M.}\ \bibnamefont
  {Philip}}, \ and\ \bibinfo {author} {\bibfnamefont {M.~J.}\ \bibnamefont
  {Gilbert}},\ }\href {\doibase 10.1103/PhysRevB.100.081104} {\bibfield
  {journal} {\bibinfo  {journal} {Phys. Rev. B}\ }\textbf {\bibinfo {volume}
  {100}},\ \bibinfo {pages} {081104} (\bibinfo {year} {2019})}\BibitemShut
  {NoStop}%
\bibitem [{\citenamefont {Prosen}(2008)}]{Prosen2008}%
  \BibitemOpen
  \bibfield  {author} {\bibinfo {author} {\bibfnamefont {T.}~\bibnamefont
  {Prosen}},\ }\href {\doibase 10.1088/1367-2630/10/4/043026} {\bibfield
  {journal} {\bibinfo  {journal} {New Journal of Physics}\ }\textbf {\bibinfo
  {volume} {10}},\ \bibinfo {pages} {043026} (\bibinfo {year}
  {2008})}\BibitemShut {NoStop}%
\bibitem [{\citenamefont {Prosen}(2010)}]{Prosen_2010}%
  \BibitemOpen
  \bibfield  {author} {\bibinfo {author} {\bibfnamefont {T.}~\bibnamefont
  {Prosen}},\ }\href {\doibase 10.1088/1742-5468/2010/07/p07020} {\bibfield
  {journal} {\bibinfo  {journal} {Journal of Statistical Mechanics: Theory and
  Experiment}\ }\textbf {\bibinfo {volume} {2010}},\ \bibinfo {pages} {P07020}
  (\bibinfo {year} {2010})}\BibitemShut {NoStop}%
\bibitem [{\citenamefont {Ozawa}\ \emph {et~al.}(2019)\citenamefont {Ozawa},
  \citenamefont {Price}, \citenamefont {Amo}, \citenamefont {Goldman},
  \citenamefont {Hafezi}, \citenamefont {Lu}, \citenamefont {Rechtsman},
  \citenamefont {Schuster}, \citenamefont {Simon}, \citenamefont {Zilberberg},\
  and\ \citenamefont {Carusotto}}]{RMPOzawa}%
  \BibitemOpen
  \bibfield  {author} {\bibinfo {author} {\bibfnamefont {T.}~\bibnamefont
  {Ozawa}}, \bibinfo {author} {\bibfnamefont {H.~M.}\ \bibnamefont {Price}},
  \bibinfo {author} {\bibfnamefont {A.}~\bibnamefont {Amo}}, \bibinfo {author}
  {\bibfnamefont {N.}~\bibnamefont {Goldman}}, \bibinfo {author} {\bibfnamefont
  {M.}~\bibnamefont {Hafezi}}, \bibinfo {author} {\bibfnamefont
  {L.}~\bibnamefont {Lu}}, \bibinfo {author} {\bibfnamefont {M.~C.}\
  \bibnamefont {Rechtsman}}, \bibinfo {author} {\bibfnamefont {D.}~\bibnamefont
  {Schuster}}, \bibinfo {author} {\bibfnamefont {J.}~\bibnamefont {Simon}},
  \bibinfo {author} {\bibfnamefont {O.}~\bibnamefont {Zilberberg}}, \ and\
  \bibinfo {author} {\bibfnamefont {I.}~\bibnamefont {Carusotto}},\ }\href
  {\doibase 10.1103/RevModPhys.91.015006} {\bibfield  {journal} {\bibinfo
  {journal} {Rev. Mod. Phys.}\ }\textbf {\bibinfo {volume} {91}},\ \bibinfo
  {pages} {015006} (\bibinfo {year} {2019})}\BibitemShut {NoStop}%
\bibitem [{\citenamefont {Lan}\ \emph {et~al.}(2022)\citenamefont {Lan},
  \citenamefont {Chen}, \citenamefont {Gao}, \citenamefont {Zhang},\ and\
  \citenamefont {Sha}}]{Lan_2022}%
  \BibitemOpen
  \bibfield  {author} {\bibinfo {author} {\bibfnamefont {Z.}~\bibnamefont
  {Lan}}, \bibinfo {author} {\bibfnamefont {M.~L.}\ \bibnamefont {Chen}},
  \bibinfo {author} {\bibfnamefont {F.}~\bibnamefont {Gao}}, \bibinfo {author}
  {\bibfnamefont {S.}~\bibnamefont {Zhang}}, \ and\ \bibinfo {author}
  {\bibfnamefont {W.~E.}\ \bibnamefont {Sha}},\ }\href {\doibase
  10.1016/j.revip.2022.100076} {\bibfield  {journal} {\bibinfo  {journal}
  {Reviews in Physics}\ }\textbf {\bibinfo {volume} {9}},\ \bibinfo {pages}
  {100076} (\bibinfo {year} {2022})}\BibitemShut {NoStop}%
\bibitem [{\citenamefont {Yan}\ \emph {et~al.}(2023)\citenamefont {Yan},
  \citenamefont {Zhao}, \citenamefont {Zhou}, \citenamefont {Ma}, \citenamefont
  {Lyu}, \citenamefont {Chu}, \citenamefont {Hu},\ and\ \citenamefont
  {Gong}}]{YanZhao2023}%
  \BibitemOpen
  \bibfield  {author} {\bibinfo {author} {\bibfnamefont {Q.}~\bibnamefont
  {Yan}}, \bibinfo {author} {\bibfnamefont {B.}~\bibnamefont {Zhao}}, \bibinfo
  {author} {\bibfnamefont {R.}~\bibnamefont {Zhou}}, \bibinfo {author}
  {\bibfnamefont {R.}~\bibnamefont {Ma}}, \bibinfo {author} {\bibfnamefont
  {Q.}~\bibnamefont {Lyu}}, \bibinfo {author} {\bibfnamefont {S.}~\bibnamefont
  {Chu}}, \bibinfo {author} {\bibfnamefont {X.}~\bibnamefont {Hu}}, \ and\
  \bibinfo {author} {\bibfnamefont {Q.}~\bibnamefont {Gong}},\ }\href {\doibase
  doi:10.1515/nanoph-2022-0775} {\bibfield  {journal} {\bibinfo  {journal}
  {Nanophotonics}\ } (\bibinfo {year} {2023}),\
  doi:10.1515/nanoph-2022-0775}\BibitemShut {NoStop}%
\bibitem [{\citenamefont {Cherifi}\ \emph {et~al.}(2023)\citenamefont
  {Cherifi}, \citenamefont {Carlstr\"om}, \citenamefont {Bourennane},\ and\
  \citenamefont {Bergholtz}}]{cherifi2023nonhermitian}%
  \BibitemOpen
  \bibfield  {author} {\bibinfo {author} {\bibfnamefont {W.}~\bibnamefont
  {Cherifi}}, \bibinfo {author} {\bibfnamefont {J.}~\bibnamefont
  {Carlstr\"om}}, \bibinfo {author} {\bibfnamefont {M.}~\bibnamefont
  {Bourennane}}, \ and\ \bibinfo {author} {\bibfnamefont {E.~J.}\ \bibnamefont
  {Bergholtz}},\ }\href {https://arxiv.org/abs/2304.03016} {\bibfield
  {journal} {\bibinfo  {journal} {arXiv:2304.03016}\ } (\bibinfo {year}
  {2023})}\BibitemShut {NoStop}%
\bibitem [{\citenamefont {Yang}\ \emph
  {et~al.}(2023{\natexlab{b}})\citenamefont {Yang}, \citenamefont {Molignini},\
  and\ \citenamefont {Bergholtz}}]{FYang23}%
  \BibitemOpen
  \bibfield  {author} {\bibinfo {author} {\bibfnamefont {F.}~\bibnamefont
  {Yang}}, \bibinfo {author} {\bibfnamefont {P.}~\bibnamefont {Molignini}}, \
  and\ \bibinfo {author} {\bibfnamefont {E.~J.}\ \bibnamefont {Bergholtz}},\
  }\href@noop {} {\enquote {\bibinfo {title} {Dissipative boundary state
  preparation},}\ } (\bibinfo {year} {2023}{\natexlab{b}}),\ \Eprint
  {http://arxiv.org/abs/2305.00031} {arXiv:2305.00031 [cond-mat.mes-hall]}
  \BibitemShut {NoStop}%
\bibitem [{\citenamefont {R{\"u}ter}\ \emph {et~al.}(2010)\citenamefont
  {R{\"u}ter}, \citenamefont {Makris}, \citenamefont {El-Ganainy},
  \citenamefont {Christodoulides}, \citenamefont {Segev},\ and\ \citenamefont
  {Kip}}]{Ruter2010}%
  \BibitemOpen
  \bibfield  {author} {\bibinfo {author} {\bibfnamefont {C.~E.}\ \bibnamefont
  {R{\"u}ter}}, \bibinfo {author} {\bibfnamefont {K.~G.}\ \bibnamefont
  {Makris}}, \bibinfo {author} {\bibfnamefont {R.}~\bibnamefont {El-Ganainy}},
  \bibinfo {author} {\bibfnamefont {D.~N.}\ \bibnamefont {Christodoulides}},
  \bibinfo {author} {\bibfnamefont {M.}~\bibnamefont {Segev}}, \ and\ \bibinfo
  {author} {\bibfnamefont {D.}~\bibnamefont {Kip}},\ }\href {\doibase
  10.1038/nphys1515} {\bibfield  {journal} {\bibinfo  {journal} {Nature
  Physics}\ }\textbf {\bibinfo {volume} {6}},\ \bibinfo {pages} {192} (\bibinfo
  {year} {2010})}\BibitemShut {NoStop}%
\bibitem [{\citenamefont {Zhang}\ \emph {et~al.}(2018)\citenamefont {Zhang},
  \citenamefont {Wang}, \citenamefont {Zhang},\ and\ \citenamefont
  {Song}}]{Zhang2018}%
  \BibitemOpen
  \bibfield  {author} {\bibinfo {author} {\bibfnamefont {K.~L.}\ \bibnamefont
  {Zhang}}, \bibinfo {author} {\bibfnamefont {P.}~\bibnamefont {Wang}},
  \bibinfo {author} {\bibfnamefont {G.}~\bibnamefont {Zhang}}, \ and\ \bibinfo
  {author} {\bibfnamefont {Z.}~\bibnamefont {Song}},\ }\href {\doibase
  10.1103/PhysRevA.98.022128} {\bibfield  {journal} {\bibinfo  {journal} {Phys.
  Rev. A}\ }\textbf {\bibinfo {volume} {98}},\ \bibinfo {pages} {022128}
  (\bibinfo {year} {2018})}\BibitemShut {NoStop}%
\end{thebibliography}%


\providecommand{\noopsort}[1]{}\providecommand{\singleletter}[1]{#1}%
\begin{thebibliography}{8}%
\makeatletter
\providecommand \@ifxundefined [1]{%
 \@ifx{#1\undefined}
}%
\providecommand \@ifnum [1]{%
 \ifnum #1\expandafter \@firstoftwo
 \else \expandafter \@secondoftwo
 \fi
}%
\providecommand \@ifx [1]{%
 \ifx #1\expandafter \@firstoftwo
 \else \expandafter \@secondoftwo
 \fi
}%
\providecommand \natexlab [1]{#1}%
\providecommand \enquote  [1]{``#1''}%
\providecommand \bibnamefont  [1]{#1}%
\providecommand \bibfnamefont [1]{#1}%
\providecommand \citenamefont [1]{#1}%
\providecommand \href@noop [0]{\@secondoftwo}%
\providecommand \href [0]{\begingroup \@sanitize@url \@href}%
\providecommand \@href[1]{\@@startlink{#1}\@@href}%
\providecommand \@@href[1]{\endgroup#1\@@endlink}%
\providecommand \@sanitize@url [0]{\catcode `\\12\catcode `\$12\catcode
  `\&12\catcode `\#12\catcode `\^12\catcode `\_12\catcode `\%12\relax}%
\providecommand \@@startlink[1]{}%
\providecommand \@@endlink[0]{}%
\providecommand \url  [0]{\begingroup\@sanitize@url \@url }%
\providecommand \@url [1]{\endgroup\@href {#1}{\urlprefix }}%
\providecommand \urlprefix  [0]{URL }%
\providecommand \Eprint [0]{\href }%
\providecommand \doibase [0]{https://doi.org/}%
\providecommand \selectlanguage [0]{\@gobble}%
\providecommand \bibinfo  [0]{\@secondoftwo}%
\providecommand \bibfield  [0]{\@secondoftwo}%
\providecommand \translation [1]{[#1]}%
\providecommand \BibitemOpen [0]{}%
\providecommand \bibitemStop [0]{}%
\providecommand \bibitemNoStop [0]{.\EOS\space}%
\providecommand \EOS [0]{\spacefactor3000\relax}%
\providecommand \BibitemShut  [1]{\csname bibitem#1\endcsname}%
\let\auto@bib@innerbib\@empty
\bibitem [{\citenamefont {Kawabata}\ \emph {et~al.}(2018)\citenamefont
  {Kawabata}, \citenamefont {Shiozaki},\ and\ \citenamefont
  {Ueda}}]{Kawabata2018}%
  \BibitemOpen
  \bibfield  {author} {\bibinfo {author} {\bibfnamefont {K.}~\bibnamefont
  {Kawabata}}, \bibinfo {author} {\bibfnamefont {K.}~\bibnamefont {Shiozaki}},\
  and\ \bibinfo {author} {\bibfnamefont {M.}~\bibnamefont {Ueda}},\ }\bibfield
  {title} {\bibinfo {title} {{Anomalous helical edge states in a non-Hermitian
  Chern insulator}},\ }\href {https://doi.org/10.1103/PhysRevB.98.165148}
  {\bibfield  {journal} {\bibinfo  {journal} {Phys. Rev. B}\ }\textbf {\bibinfo
  {volume} {98}},\ \bibinfo {pages} {165148} (\bibinfo {year}
  {2018})}\BibitemShut {NoStop}%
\bibitem [{\citenamefont {Kawasaki}\ \emph {et~al.}(2022)\citenamefont
  {Kawasaki}, \citenamefont {Mochizuki},\ and\ \citenamefont
  {Obuse}}]{Kawasaki22}%
  \BibitemOpen
  \bibfield  {author} {\bibinfo {author} {\bibfnamefont {M.}~\bibnamefont
  {Kawasaki}}, \bibinfo {author} {\bibfnamefont {K.}~\bibnamefont
  {Mochizuki}},\ and\ \bibinfo {author} {\bibfnamefont {H.}~\bibnamefont
  {Obuse}},\ }\bibfield  {title} {\bibinfo {title} {{Topological phases
  protected by shifted sublattice symmetry in dissipative quantum systems}},\
  }\href {https://doi.org/10.1103/PhysRevB.106.035408} {\bibfield  {journal}
  {\bibinfo  {journal} {Phys. Rev. B}\ }\textbf {\bibinfo {volume} {106}},\
  \bibinfo {pages} {035408} (\bibinfo {year} {2022})}\BibitemShut {NoStop}%
\bibitem [{\citenamefont {Kawabata}\ \emph {et~al.}(2019)\citenamefont
  {Kawabata}, \citenamefont {Shiozaki}, \citenamefont {Ueda},\ and\
  \citenamefont {Sato}}]{Kawabata2019}%
  \BibitemOpen
  \bibfield  {author} {\bibinfo {author} {\bibfnamefont {K.}~\bibnamefont
  {Kawabata}}, \bibinfo {author} {\bibfnamefont {K.}~\bibnamefont {Shiozaki}},
  \bibinfo {author} {\bibfnamefont {M.}~\bibnamefont {Ueda}},\ and\ \bibinfo
  {author} {\bibfnamefont {M.}~\bibnamefont {Sato}},\ }\bibfield  {title}
  {\bibinfo {title} {{Symmetry and Topology in Non-Hermitian Physics}},\ }\href
  {https://doi.org/10.1103/PhysRevX.9.041015} {\bibfield  {journal} {\bibinfo
  {journal} {Phys. Rev. X}\ }\textbf {\bibinfo {volume} {9}},\ \bibinfo {pages}
  {041015} (\bibinfo {year} {2019})}\BibitemShut {NoStop}%
\bibitem [{\citenamefont {Bergholtz}\ \emph {et~al.}(2021)\citenamefont
  {Bergholtz}, \citenamefont {Budich},\ and\ \citenamefont
  {Kunst}}]{Bergholtz}%
  \BibitemOpen
  \bibfield  {author} {\bibinfo {author} {\bibfnamefont {E.~J.}\ \bibnamefont
  {Bergholtz}}, \bibinfo {author} {\bibfnamefont {J.~C.}\ \bibnamefont
  {Budich}},\ and\ \bibinfo {author} {\bibfnamefont {F.~K.}\ \bibnamefont
  {Kunst}},\ }\bibfield  {title} {\bibinfo {title} {{Exceptional topology of
  non-Hermitian systems}},\ }\href
  {https://doi.org/10.1103/RevModPhys.93.015005} {\bibfield  {journal}
  {\bibinfo  {journal} {Rev. Mod. Phys.}\ }\textbf {\bibinfo {volume} {93}},\
  \bibinfo {pages} {015005} (\bibinfo {year} {2021})}\BibitemShut {NoStop}%
\bibitem [{\citenamefont {Banerjee}\ \emph {et~al.}(2022)\citenamefont
  {Banerjee}, \citenamefont {Sarkar}, \citenamefont {Dey},\ and\ \citenamefont
  {Narayan}}]{banerjee2022}%
  \BibitemOpen
  \bibfield  {author} {\bibinfo {author} {\bibfnamefont {A.}~\bibnamefont
  {Banerjee}}, \bibinfo {author} {\bibfnamefont {R.}~\bibnamefont {Sarkar}},
  \bibinfo {author} {\bibfnamefont {S.}~\bibnamefont {Dey}},\ and\ \bibinfo
  {author} {\bibfnamefont {A.}~\bibnamefont {Narayan}},\ }\href
  {https://arxiv.org/abs/2212.06478} {\bibinfo {title} {{Non-Hermitian
  Topological Phases: Principles and Prospects}}} (\bibinfo {year} {2022}),\
  \Eprint {https://arxiv.org/abs/2212.06478} {2212.06478} \BibitemShut
  {NoStop}%
\bibitem [{\citenamefont {Okuma}\ and\ \citenamefont
  {Sato}(2023)}]{Okuma_2023}%
  \BibitemOpen
  \bibfield  {author} {\bibinfo {author} {\bibfnamefont {N.}~\bibnamefont
  {Okuma}}\ and\ \bibinfo {author} {\bibfnamefont {M.}~\bibnamefont {Sato}},\
  }\bibfield  {title} {\bibinfo {title} {{Non-Hermitian Topological Phenomena:
  A Review}},\ }\href
  {https://doi.org/10.1146/annurev-conmatphys-040521-033133} {\bibfield
  {journal} {\bibinfo  {journal} {Annual Review of Condensed Matter Physics}\
  }\textbf {\bibinfo {volume} {14}},\ \bibinfo {pages} {83} (\bibinfo {year}
  {2023})}\BibitemShut {NoStop}%
\bibitem [{\citenamefont {R{\"u}ter}\ \emph {et~al.}(2010)\citenamefont
  {R{\"u}ter}, \citenamefont {Makris}, \citenamefont {El-Ganainy},
  \citenamefont {Christodoulides}, \citenamefont {Segev},\ and\ \citenamefont
  {Kip}}]{Ruter2010}%
  \BibitemOpen
  \bibfield  {author} {\bibinfo {author} {\bibfnamefont {C.~E.}\ \bibnamefont
  {R{\"u}ter}}, \bibinfo {author} {\bibfnamefont {K.~G.}\ \bibnamefont
  {Makris}}, \bibinfo {author} {\bibfnamefont {R.}~\bibnamefont {El-Ganainy}},
  \bibinfo {author} {\bibfnamefont {D.~N.}\ \bibnamefont {Christodoulides}},
  \bibinfo {author} {\bibfnamefont {M.}~\bibnamefont {Segev}},\ and\ \bibinfo
  {author} {\bibfnamefont {D.}~\bibnamefont {Kip}},\ }\bibfield  {title}
  {\bibinfo {title} {{Observation of parity--time symmetry in optics}},\ }\href
  {https://doi.org/10.1038/nphys1515} {\bibfield  {journal} {\bibinfo
  {journal} {Nature Physics}\ }\textbf {\bibinfo {volume} {6}},\ \bibinfo
  {pages} {192} (\bibinfo {year} {2010})}\BibitemShut {NoStop}%
\bibitem [{\citenamefont {Zhang}\ \emph {et~al.}(2018)\citenamefont {Zhang},
  \citenamefont {Wang}, \citenamefont {Zhang},\ and\ \citenamefont
  {Song}}]{Zhang2018}%
  \BibitemOpen
  \bibfield  {author} {\bibinfo {author} {\bibfnamefont {K.~L.}\ \bibnamefont
  {Zhang}}, \bibinfo {author} {\bibfnamefont {P.}~\bibnamefont {Wang}},
  \bibinfo {author} {\bibfnamefont {G.}~\bibnamefont {Zhang}},\ and\ \bibinfo
  {author} {\bibfnamefont {Z.}~\bibnamefont {Song}},\ }\bibfield  {title}
  {\bibinfo {title} {{Simple harmonic oscillation in a non-Hermitian
  Su-Schrieffer-Heeger chain at the exceptional point}},\ }\href
  {https://doi.org/10.1103/PhysRevA.98.022128} {\bibfield  {journal} {\bibinfo
  {journal} {Phys. Rev. A}\ }\textbf {\bibinfo {volume} {98}},\ \bibinfo
  {pages} {022128} (\bibinfo {year} {2018})}\BibitemShut {NoStop}%
\end{thebibliography}%

\end{document}


\title{Supplemental Material: Edge-selective extremal damping from topological heritage of dissipative Chern insulators}

\author{Suraj S. Hegde}
\author{Toni Ehmcke}
\author{Tobias Meng}
\affiliation{Institute of Theoretical Physics and W\"urzburg-Dresden Cluster of Excellence ct.qmat, Technische Universit\"at Dresden, 01062 Dresden, Germany}


\date{\today}

\maketitle
\section{Analytics for the non-equilibrium steady state with periodic boundary conditions}
We analyze a Chern insulator that is dissipatively coupled to a Markovian bath and described by a Lindblad Master equation, and here specialize to a translation invariant setting. There are four independent many-body states per momentum $\bs{k}$. We use the basis $|n_{\bs{k}\uparrow} n_{\bs{k}\downarrow}\rangle$, where $n_\sigma\in\{0,1\}$ is the occupancy of the electronic state with spin $\sigma=\uparrow,\downarrow$ at $\bs{k}$. The momentum-resolved density matrix $\rho_{\bs{k}}$ is thus a $(4\times4)$-matrix. Using the basis $|ab\rangle\langle cd|$ with $a, b,c,d\in\{0,1\}$, we write this matrix as $\rho_{\bs{k}}=\sum_{a,b,c,d}(x_{ab;cd}+i\,y_{ab;cd})\,|ab\rangle\langle cd|$. Hermiticity of the density matrix imposes that $x_{ab;cd}=x_{cd;ab}\in\mathds{R}$ and $y_{ab;cd}=-y_{cd;ab}\in\mathds{R}$, which implies that there are only 16 independent real parameters characterizing the momentum-resolved density matrix $\rho_{\bs{k}}$. The Liouvillian superoperator can consequently be written as as $(16\times16)$-matrix acting on the 16-dimensional vector of the independent, non-zero coefficients $x_{ab;cd}$ and $y_{ab;cd}$.

Writing the Hamiltonian as in the main text in the form $H=\sum_{\bs{k}}\Psi_{\bs{k}}^\dagger\,\left(\bs{d}\left(\bs{k}\right)\cdot\bs{\sigma}-\mu\,\mathds{1}\right)\,\Psi_{\bs{k}}^\pd$, we find that the Liouvillian can be represented as

\begin{align}
\dot{\tilde{\bs{w}}}_{\bs{k}}(t) &=\tilde{{B}}_{\bs{k}}\,\tilde{\bs{w}}_{\bs{k}}(t)\quad\text{with}\quad\tilde{\bs{w}}_{\bs{k}}(t) = \begin{pmatrix}\bs{w}_{\bs{k}}(t)\\\bs{v}_{\bs{k}}(t)\\\bs{z}_{\bs{k}}(t)\end{pmatrix}\quad\text{and}\quad\tilde{{B}}_{\bs{k}}=\begin{pmatrix}{B}_{\bs{k}}&0&0\\0&{C}_{\bs{k}}&0\\0&0&{D}_{\bs{k}}\end{pmatrix}\label{eq:wTilde}
\end{align}
where
\begin{align}
\bs{w}_{\bs{k}}(t)&=\begin{pmatrix}x_{00;00}(\bs{k},t)\\x_{01;01}(\bs{k},t)\\x_{01;10}(\bs{k},t)\\y_{01;10}(\bs{k},t)\\x_{10;10}(\bs{k},t)\\x_{11;11}(\bs{k},t)\end{pmatrix}\quad\text{with}\quad\dot{\bs{w}}_{\bs{k}}(t)={B}_{\bs{k}}\,\bs{w}_{\bs{k}}(t),\label{eq:w}\\[0.5cm]
\bs{v}_{\bs{k}}(t)&=\begin{pmatrix}x_{00;01}(\bs{k},t)\\y_{00;01}(\bs{k},t)\\x_{00;10}(\bs{k},t)\\y_{00;10}(\bs{k},t)\\x_{01;11}(\bs{k},t)\\y_{01;11}(\bs{k},t)\\x_{10;11}(\bs{k},t)\\y_{10;11}(\bs{k},t)\end{pmatrix}\quad\text{with}\quad\dot{\bs{v}}_{\bs{k}}(t)={C}_{\bs{k}}\,\bs{v}_{\bs{k}}(t),\label{eq:v}\\[0.5cm]
\bs{z}_{\bs{k}}(t)&=\begin{pmatrix}x_{00;11}(\bs{k},t)\\y_{00;11}(\bs{k},t)\end{pmatrix}\quad\text{with}\quad\dot{\bs{z}}_{\bs{k}}(t)={D}_{\bs{k}}\,\bs{z}_{\bs{k}}(t).\label{eq:z}
\end{align}
A non-equilibrium steady state (NESS) corresponds an eigenvector $\tilde{\bs{w}}_{\bs{k}}(t)$ of $\tilde{\bs{B}}_{\bs{k}}$ with eigenvalue zero. We can further simplify the calculation by noting that the unit trace condition $\text{tr}(\rho_{\bs{k}}(t))=1$ implies $x_{00;00}(t)+x_{01;01}(t)+x_{10;10}(t)+x_{11;11}(t)=1$. This means that $\bs{w}_{\bs{k}}(t)\neq0$ must always be satisfied. $\bs{v}_{\bs{k}}$ and $\bs{z}_{\bs{k}}$, on the other hand, are allowed to vanish, in which case Eqs.~\eqref{eq:v} and \eqref{eq:z} are trivially satisfied. Given that also any NESS has $\bs{w}_{\bs{k}}\neq0$, it is in fact easy to see that a unique NESS must have $\bs{v}_{\bs{k}}=\bs{z}_{\bs{k}}=0$. Finding a unique NESS therefore boils down to finding the eigenvector of the $(6\times6)$-matrix $B_{\bs{k}}$ with eigenvalue zero.

\subsection{Loss for both spin species}
If $\Gamma_0>|\Gamma_z|$, both spin species experience loss. We find that the Liouvillian can then be represented by

\begin{align}
B_{\bs{k}}&=\begin{pmatrix}0&2\,(\Gamma_0-\Gamma_z)&0&0&2\,(\Gamma_0+\Gamma_z)&0\\0&2\,(\Gamma_z-\Gamma_0)&2\,d_y&-2\,d_x&0&2\,(\Gamma_0+\Gamma_z)\\0&-d_y&-2\Gamma_0&-2d_z&d_y&0\\0&d_x&2\,d_z&-2\,\Gamma_0&-d_x&0\\0&0&-2\,d_y&2\,d_x&-2\,(\Gamma_0+\Gamma_z)&2\,(\Gamma_0-\Gamma_z)\\0&0&0&0&0&-4\,\Gamma_0\end{pmatrix},\\[0.5cm]
C_{\bs{k}}&=\begin{pmatrix}\Gamma_z-\Gamma_0&d_z&d_y&-d_x&0&0&2\,(\Gamma_0+\Gamma_z)&0\\-d_z&\Gamma_z-\Gamma_0&d_x&d_y&0&0&0&2\,(\Gamma_0+\Gamma_z)\\-d_y&-d_x&-\Gamma_0-\Gamma_z&-d_z&2\,(\Gamma_0-\Gamma_z)&0&0&0\\d_x&-d_y&d_z&-\Gamma_0-\Gamma_z&0&2\,(\Gamma_z-\Gamma_0)&0&0\\0&0&0&0&\Gamma_z-3\,\Gamma_0&-d_z&d_y&d_x\\0&0&0&0&d_z&\Gamma_z-3\,\Gamma_0&-d_x&d_y\\0&0&0&0&-d_y&d_x&-3\,\Gamma_0-\Gamma_z&d_z\\0&0&0&0&-d_x&-d_y&-d_z&-3\,\Gamma_0-\Gamma_z\end{pmatrix},\\[0.5cm]
D_{\bs{k}}&=\begin{pmatrix}-2\,\Gamma_0&0\\0&-2\,\Gamma_0\end{pmatrix}.
\end{align}
One can check that both $C_{\bs{k}}$ and $D_{\bs{k}}$ generically have only non-zero eigenvalues. Similarly, $B_{\bs{k}}$ generically has only a single eigenvector with eigenvalue zero, namely $\bs{w}_{\bs{k}}=(1,0,0,0,0,0)$. This means that $|n_{\bs{k}\uparrow} n_{\bs{k}\downarrow}\rangle=|0 0\rangle$, the completely empty system, is the unique NESS. 

\subsection{Gain for both spin species}
An analysis along the same lines shows that for $-\Gamma_0>|\Gamma_z|$, there is again a unique steady state. This steady state corresponds to a fully filled empty system, i.e.~to the state $|n_{\bs{k}\uparrow} n_{\bs{k}\downarrow}\rangle=|1 1 \rangle$.

\subsection{Loss for spin-$\uparrow$ electrons, gain for spin-$\uparrow$ electrons}
If $\Gamma_z>|\Gamma_0|$, spin-$\uparrow$ electrons experience loss, while spin-$\downarrow$ electrons experience gain. We have again checked that there is generically only a single NESS, specified by

\begin{align}
x_{00;00}^{\text{NESS}}(\bs{k})=\frac{\left(\Gamma_0+\Gamma_z\right)^2\,\left(d_x^2(\bs{k})+d_y^2(\bs{k})\right)}{4\,\left(\left[d_x^2(\bs{k})+d_y^2(\bs{k})\right]\,\Gamma_z^2+\left[d_z^2(\bs{k})+\Gamma_z^2\right]\,\left[\Gamma_z^2-\Gamma_0^2\right]\right)},\\
x_{01;01}^{\text{NESS}}(\bs{k})=\frac{\left(\Gamma_z^2-\Gamma_0^2\right)\,\left(d_x^2(\bs{k})+d_y^2(\bs{k})+4\,\left[d_z^2(\bs{k})+\Gamma_z^2\right]\right)}{4\,\left(\left[d_x^2(\bs{k})+d_y^2(\bs{k})\right]\,\Gamma_z^2+\left[d_z^2(\bs{k})+\Gamma_z^2\right]\,\left[\Gamma_z^2-\Gamma_0^2\right]\right)},\\
x_{01;10}^{\text{NESS}}(\bs{k})=\frac{\left(\Gamma_z^2-\Gamma_0^2\right)\,\left(-d_x(\bs{k})\,d_z(\bs{k})-d_y(\bs{k})\,\Gamma_z\right)}{2\,\left(\left[d_x^2(\bs{k})+d_y^2(\bs{k})\right]\,\Gamma_z^2+\left[d_z^2(\bs{k})+\Gamma_z^2\right]\,\left[\Gamma_z^2-\Gamma_0^2\right]\right)},\\
y_{01;10}^{\text{NESS}}(\bs{k})=\frac{\left(\Gamma_z^2-\Gamma_0^2\right)\,\left(-d_y(\bs{k})\,d_z(\bs{k})+d_x(\bs{k})\,\Gamma_z\right)}{2\,\left(\left[d_x^2(\bs{k})+d_y^2(\bs{k})\right]\,\Gamma_z^2+\left[d_z^2(\bs{k})+\Gamma_z^2\right]\,\left[\Gamma_z^2-\Gamma_0^2\right]\right)},\\
x_{10;10}^{\text{NESS}}(\bs{k})=\frac{\left(\Gamma_z^2-\Gamma_0^2\right)\,\left(d_x^2(\bs{k})+d_y^2(\bs{k})\right)}{4\,\left(\left[d_x^2(\bs{k})+d_y^2(\bs{k})\right]\,\Gamma_z^2+\left[d_z^2(\bs{k})+\Gamma_z^2\right]\,\left[\Gamma_z^2-\Gamma_0^2\right]\right)},\\
x_{11;11}^{\text{NESS}}(\bs{k})=\frac{\left(\Gamma_0-\Gamma_z\right)^2\,\left(d_x^2(\bs{k})+d_y^2(\bs{k})\right)}{4\,\left(\left[d_x^2(\bs{k})+d_y^2(\bs{k})\right]\,\Gamma_z^2+\left[d_z^2(\bs{k})+\Gamma_z^2\right]\,\left[\Gamma_z^2-\Gamma_0^2\right]\right)},
\end{align}
while all other (independent) $x_{ab;cd}^{\text{NESS}}$ and $y_{ab;cd}^{\text{NESS}}$ vanish.

\subsection{Gain for spin-$\uparrow$ electrons, loss for spin-$\uparrow$ electrons}
If $-\Gamma_z>|\Gamma_0|$, spin-$\uparrow$ electrons experience gain, while spin-$\downarrow$ electrons experience loss. We again obtain a unique NESS, specified by

\begin{align}
x_{00;00}^{\text{NESS}}(\bs{k})=\frac{\left(\Gamma_0-\Gamma_z\right)^2\,\left(d_x^2(\bs{k})+d_y^2(\bs{k})\right)}{4\,\left(\left[d_x^2(\bs{k})+d_y^2(\bs{k})\right]\,\Gamma_z^2+\left[d_z^2(\bs{k})+\Gamma_z^2\right]\,\left[\Gamma_z^2-\Gamma_0^2\right]\right)},\\
x_{01;01}^{\text{NESS}}(\bs{k})=\frac{\left(\Gamma_z^2-\Gamma_0^2\right)\,\left(d_x^2(\bs{k})+d_y^2(\bs{k})\right)}{4\,\left(\left[d_x^2(\bs{k})+d_y^2(\bs{k})\right]\,\Gamma_z^2+\left[d_z^2(\bs{k})+\Gamma_z^2\right]\,\left[\Gamma_z^2-\Gamma_0^2\right]\right)},\\
x_{01;10}^{\text{NESS}}(\bs{k})=\frac{\left(\Gamma_z^2-\Gamma_0^2\right)\,\left(d_x(\bs{k})\,d_z(\bs{k})-d_y(\bs{k})\,\Gamma_z\right)}{2\,\left(\left[d_x^2(\bs{k})+d_y^2(\bs{k})\right]\,\Gamma_z^2+\left[d_z^2(\bs{k})+\Gamma_z^2\right]\,\left[\Gamma_z^2-\Gamma_0^2\right]\right)},\\
y_{01;10}^{\text{NESS}}(\bs{k})=\frac{\left(\Gamma_z^2-\Gamma_0^2\right)\,\left(d_y(\bs{k})\,d_z(\bs{k})+d_x(\bs{k})\,\Gamma_z\right)}{2\,\left(\left[d_x^2(\bs{k})+d_y^2(\bs{k})\right]\,\Gamma_z^2+\left[d_z^2(\bs{k})+\Gamma_z^2\right]\,\left[\Gamma_z^2-\Gamma_0^2\right]\right)},\\
x_{10;10}^{\text{NESS}}(\bs{k})=\frac{\left(\Gamma_z^2-\Gamma_0^2\right)\,\left(d_x^2(\bs{k})+d_y^2(\bs{k})+4\,\left[d_z^2(\bs{k})+\Gamma_z^2\right]\right)}{4\,\left(\left[d_x^2(\bs{k})+d_y^2(\bs{k})\right]\,\Gamma_z^2+\left[d_z^2(\bs{k})+\Gamma_z^2\right]\,\left[\Gamma_z^2-\Gamma_0^2\right]\right)},\\
x_{11;11}^{\text{NESS}}(\bs{k})=\frac{\left(\Gamma_0+\Gamma_z\right)^2\,\left(d_x^2(\bs{k})+d_y^2(\bs{k})\right)}{4\,\left(\left[d_x^2(\bs{k})+d_y^2(\bs{k})\right]\,\Gamma_z^2+\left[d_z^2(\bs{k})+\Gamma_z^2\right]\,\left[\Gamma_z^2-\Gamma_0^2\right]\right)},
\end{align}
while all other (independent) $x_{ab;cd}^{\text{NESS}}$ and $y_{ab;cd}^{\text{NESS}}$ vanish.

\subsection{NESS purity}
As shown above, the NESS is always a pure state if  $|\Gamma_0|>|\Gamma_z|$. If instead  $|\Gamma_0|<|\Gamma_z|$, the purity $\text{tr}((\rho_{\bs{k}}^{\text{NESS}})^2)$ can easily be calculated from the above equations. In general, this yields lengthy and not very enlightening expressions. For $\Gamma_0=0$, however, we find the compact formula
\begin{align}
\left.\text{tr}\left((\rho_{\bs{k}}^{\text{NESS}})^2\right)\right|_{\Gamma_0=0,\,\Gamma_z\neq0} = \frac{\Gamma_z^2+\bs{d}^2(\bs{k})-\left(d_x^2(\bs{k})+d_y^2(\bs{k})\right)/2}{\Gamma_z^2+\bs{d}^2(\bs{k})}\leq1.
\end{align}
For $\Gamma_0=0$, unit purity is hence obtained for $d_x(\bs{k})=d_y(\bs{k})=0$. As discussed in the main text, this condition for unit purity also carries over to finite $\Gamma_0$. Using it allows to analytically determine the momenta with unit purity as

\begin{align}
\bs{k}^T= \begin{cases}
(\pi,\pi)&\text{for}~m=-2\alpha,\\
(\pi,\pm\arccos\left(\frac{m}{\alpha}+1)\right)&\text{for}~-2\alpha<m<0,\\
(0,\pi)~\text{and}~(\pi,0)&\text{for}~m=0,\\
(0,\pm\arccos\left(\frac{m}{\alpha}-1)\right)&\text{for}~0<m<2\alpha,\\
(0,0)&\text{for}~m=2 \alpha.
\end{cases}\label{eq:pure_momenta}
\end{align}

\subsection{NESS current and results for $\langle j_x\rangle_{\text{NESS}}$}
\begin{figure}
\centering
\raisebox{4.75cm}{(a)}
\includegraphics[height=0.3\columnwidth]{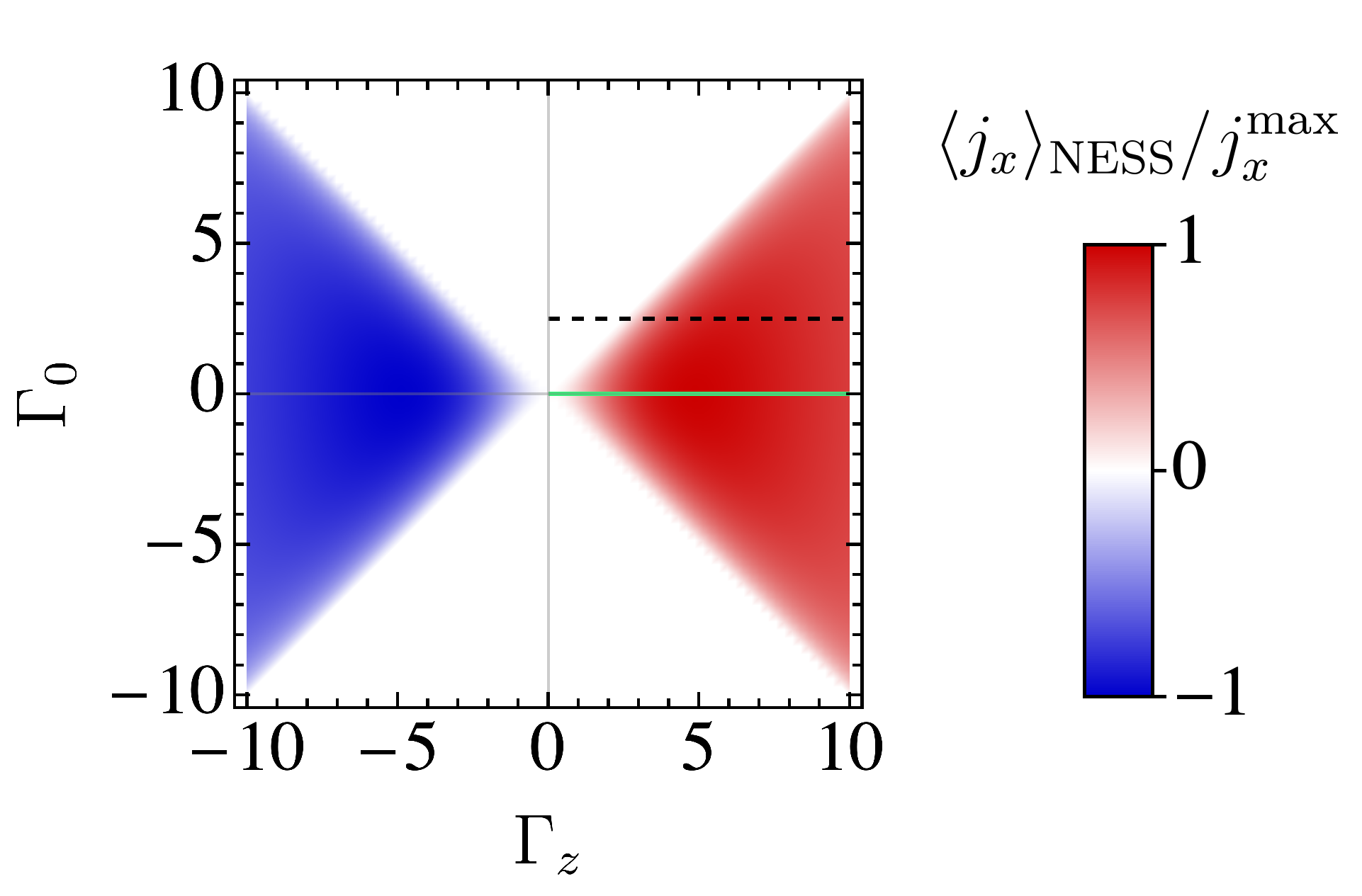}\hspace*{1cm}\raisebox{4.75cm}{(b)}\includegraphics[height=0.3\columnwidth]{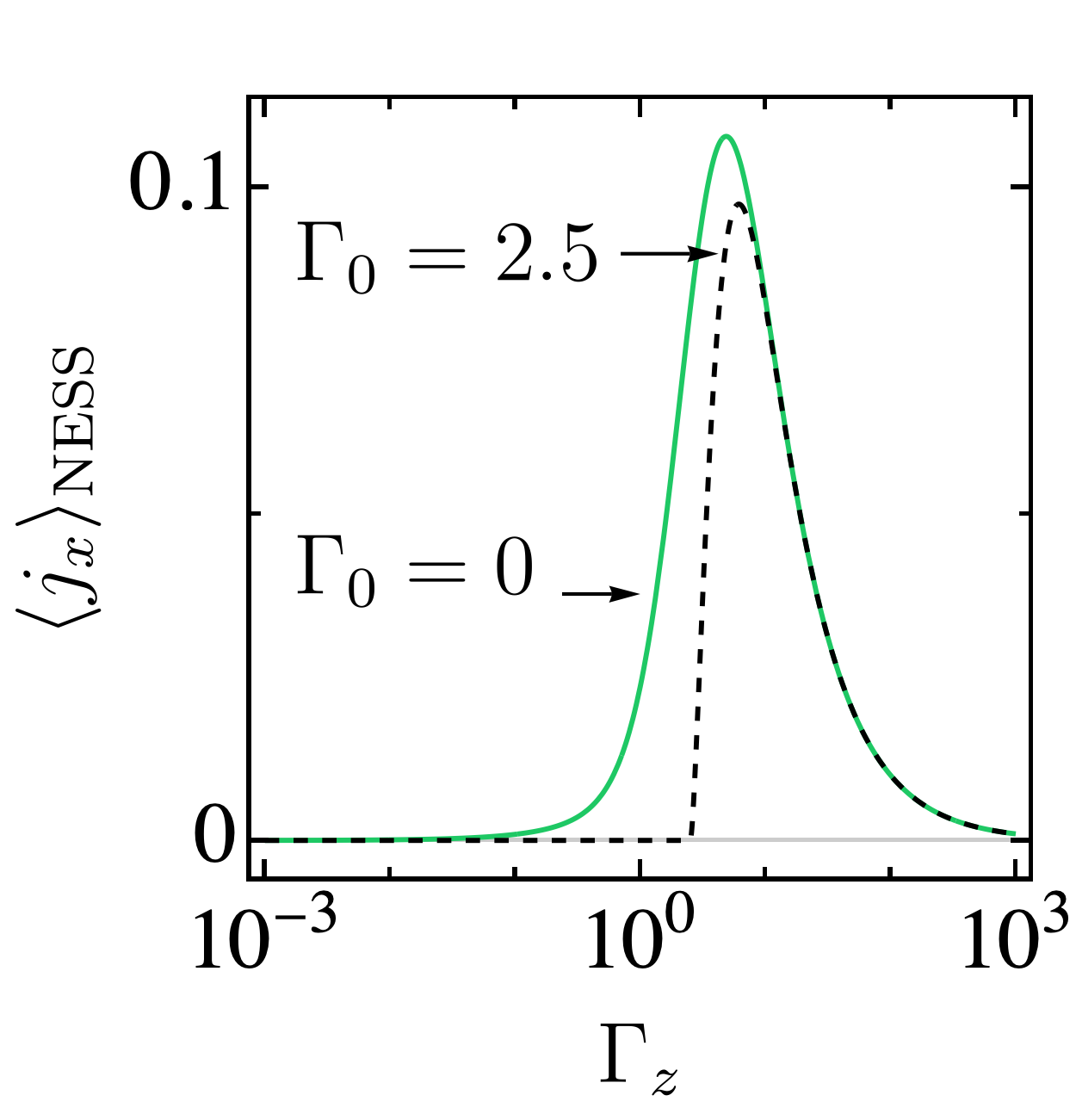}
\caption{The steady state current in $x$-direction. Panel (a): $\langle j_x\rangle_{\text{NESS}}$ as function of $\Gamma_0$ and $\Gamma_z$ in units of the maximal value $j_x^{\text{max}}$. Panel (b): cuts along fixed $\Gamma_0$ as indicated in the left panel. The remaining parameters are $m=3$, $\alpha=\beta=1$, and $\mu=0$.}
\label{fig:NESScurrentX}
\end{figure}

Using the steady state density matrix, we can calculate the net current
\begin{align}
\langle\bs{j}\rangle_{\text{NESS}}=\text{tr}\left(\rho^{\text{NESS}}\,\bs{j}\right) = \int \frac{d^2k}{(2\pi)^2} \text{tr}\left(\rho_{\bs{k}}^{\text{NESS}}\,\bs{j}_{\bs{k}}\right).
\end{align}
The main text already presents the results for $\langle j_y\rangle_{\text{NESS}}$. As shown in Fig.~\ref{fig:NESScurrentX}, similar results are obtained for $\langle j_x\rangle_{\text{NESS}}$.

\subsection{NESS current as a function of $m$}
\begin{figure}
\centering
\raisebox{4.75cm}{(a)}
\includegraphics[height=0.3\columnwidth]{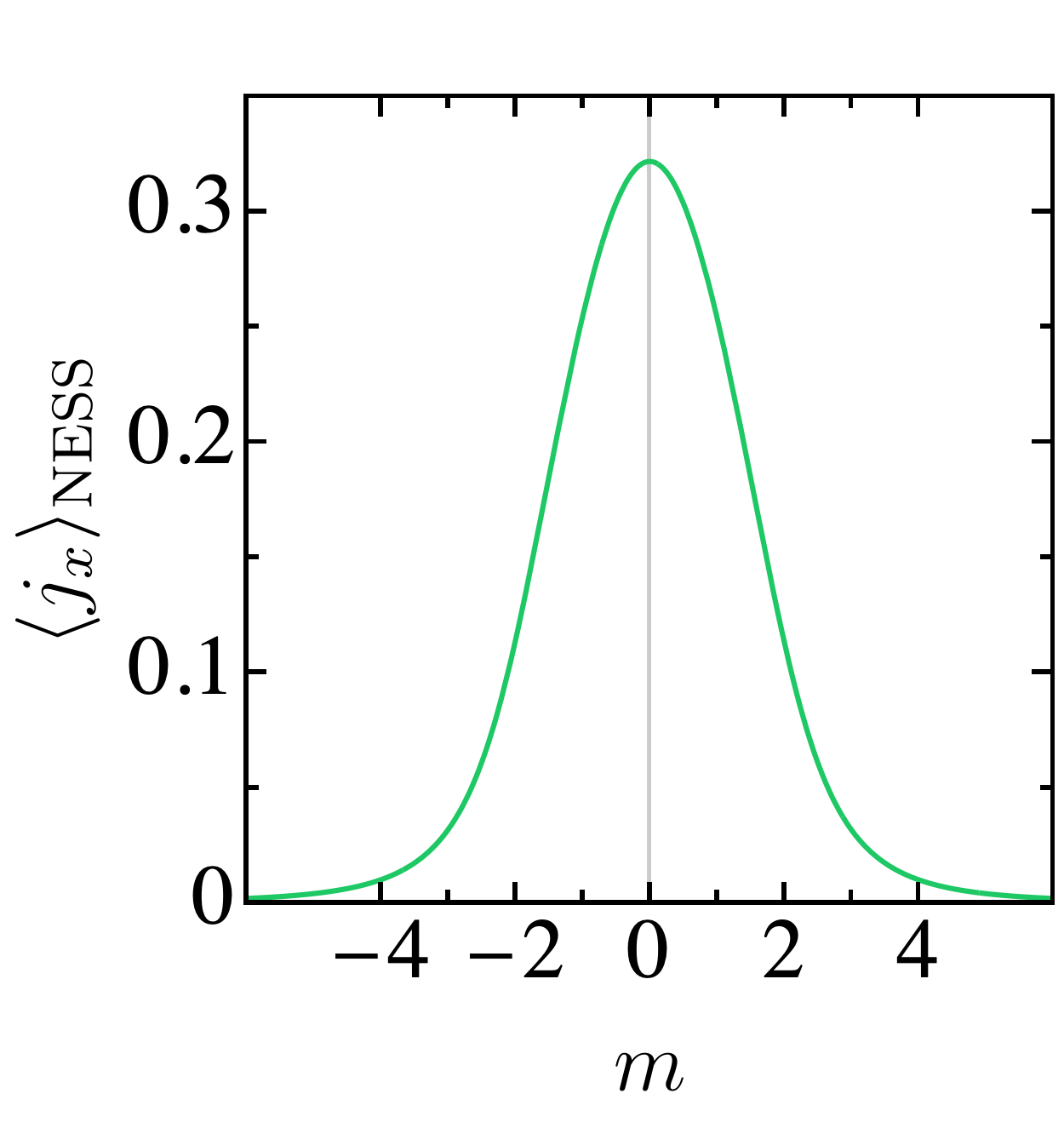}\hspace*{1cm}\raisebox{4.75cm}{(b)}\includegraphics[height=0.3\columnwidth]{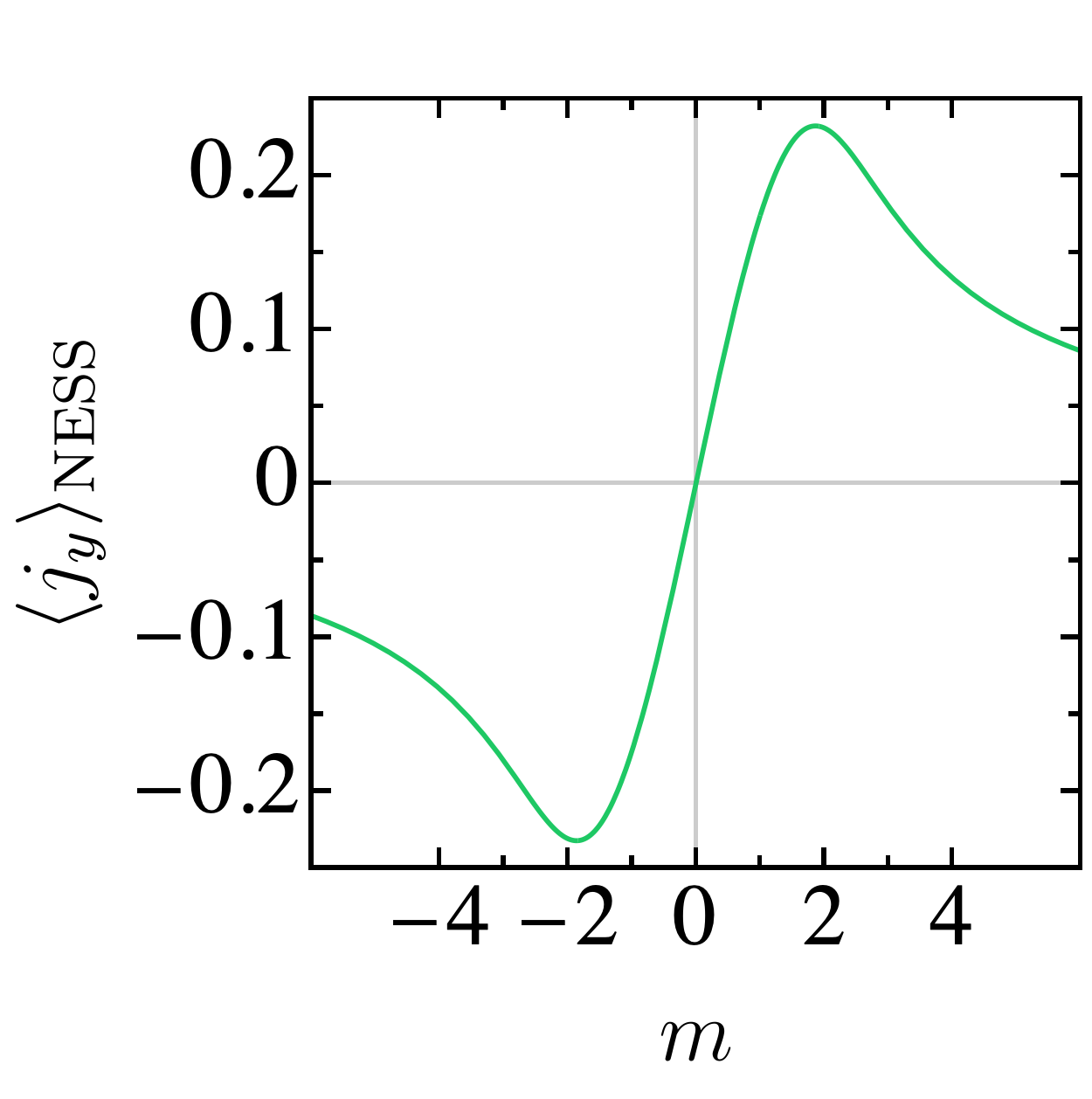}
\caption{$m$-dependence of the steady state current in $x$-direction in panel (a), and along the $y$-direction in panel (b).  We fix  $\alpha=\beta=1$, $\mu=0$, $\Gamma_0=0.08$, and $\Gamma_z=1.2$.}
\label{fig:NESScurrentXm}
\end{figure}

Finally, we can analyze the current as a function of the mass parameter $m$. As declared in the main text, the current does not show any noticeable features as a function of $m$ (except for an (anti-)symmetry with respect to $m=0$), neither at the locations of the topological transitions of the system Hamiltonians $H$, not at any other value of $m$. This is shown in Fig.~\ref{fig:NESScurrentXm}

\section{From the damping matrix to non-Hermitian Hamiltonians}
As explained in the main text, we use the formalism of third quantization to analyze the system with open boundary conditions. The Hamiltonian is then first represented using Majorana operators

\begin{align}
w_{2j-1}=c_j^\pdag+c_j^\dagger\quad\text{and}\quad w_{2j}=i\,\left(c_j^\pdag-c_j^\dagger\right).
\end{align}
Dropping constant global energy shifts, we can rewrite the Hamiltonian in terms of Majorana operators as

\begin{align}
H=\sum_{\bs{k}}\Psi_{\bs{k}}^\dagger\,\left(\bs{d}\left(\bs{k}\right)\cdot\bs{\sigma}-\mu\,\mathds{1}\right)\,\Psi_{\bs{k}}^\pd \equiv \sum_{p,q}w_p\,\mathcal{H}_{pq}\,w_q,
\end{align}
where for open boundary conditions (OBC) the indices $p$ and $q$ encode both lattice site (or momentum), and spin. In addition, we denote the Bloch Hamiltonian as
\begin{align}
h(\mu,\bs{k})=\bs{d}\left(\bs{k}\right)\cdot\bs{\sigma}-\mu\,\mathds{1}.
\end{align}
Similarly, the jump operators can be expressed as

\begin{align}
L_j\equiv\sum_p l_{j,p}\,w_p.
\end{align}
Introducing adjoint creation operators $\hat{c}_j^\pdag$ and adjoint annihilation operators $\hat{c}_j^\dagger$, the Liouvillian is then written as

\begin{align}
    \hat{\mathcal{L}}=  \frac{1}{2}  \,\sum_{ij}
    \begin{pmatrix} \hat{\bs{c}}^\dagger & \hat{\bs{c}}^{\pdag} \end{pmatrix} \begin{pmatrix}
    -X^{\dagger} & iY \\
    0 & X 
    \end{pmatrix} \begin{pmatrix}
    \hat{\bs{c}}^\dagger\\ \hat{\bs{c}}^{\pdag}
    \end{pmatrix} - \frac{1}{2}\, \text{Tr}(X),
    \end{align}
where the vectors of adjoint operators are defined via their components $\left.\hat{\bs{c}}\right|_j = \hat{c}_j^\pdag$  and $\left.\hat{\bs{c}}^\dagger\right|_j = \hat{c}_j^\dagger$, while $X=-4i\,\mathcal{H}+\mathcal{M}+\mathcal{M}^{T}$ and $Y=-2i\,(\mathcal{M}-\mathcal{M}^T)$ with

\begin{align}
\left.\mathcal{H}\right|_{pq}=\mathcal{H}_{pq} \quad\text{and}\quad \left.\mathcal{M}\right|_{pq} =\sum_{j}l_{j,p}l_{j,q}^*.
\end{align}
The matrix $X$ is known as the damping matrix, and as explained in the main text governs the time evolution (damping) of covariance matrix towards its steady-state form. To connect this time evolution to non-Hermitian Hamiltonians, we define

\begin{align}
H_{\text{nH}}^{\text{damping}}=i\,X.
\end{align}
Given that $H_{\text{nH}}^{\text{damping}}$ is defined from an expression involving adjoint Majorana operators, and thus particle-hole-mixed operators, $H_{\text{nH}}^{\text{damping}}$ can be understood as an effective non-Hermitian Bogoliubov-de Gennes Hamiltonian. To make this more explicit, we focus on the case of PBC, where momentum remains a good quantum number. There is then one damping matrix $X_{\bs{k}}$ per momentum, and we use the convention $c_1\equiv c_{\bs{k}\uparrow}$ and $c_2\equiv c_{\bs{k}\downarrow}$. To block-diagonalize the damping matrix, we use the transformation

\begin{align}
\tilde{U}^{}=\frac{1}{\sqrt{2}}\,\begin{pmatrix}1&0&0&-1\\i&0&0&i\\0&1&-1&0\\0&i&i&0\end{pmatrix}.
\end{align}
We find that this transformation brings ($i$ times) the damping matrix to a block-diagonal form reminiscent of a Bogoliubov-de Gennes Hamiltonian, namely
\begin{align}
\tilde{U}^\dagger\,(i\,X_{\bs{k}})\,\tilde{U}= \begin{pmatrix}\tilde{H}_{\text{nH}}^{(\text{+})}(\bs{k})&0\\0&\tilde{H}_{\text{nH}}^{(\text{-})}(\bs{k})\end{pmatrix},
\end{align}
where

\begin{align}
\tilde{H}_{\text{nH}}^{(\text{+})}(\bs{k})&=h(\mu,\bs{k})+i\,\text{sgn}\left(\Gamma_{>}\right)\,\Gamma_{<}\,\sigma_z+i\,|\Gamma_{>}|\,\mathds{1},\\
\tilde{H}_{\text{nH}}^{(\text{-})}(\bs{k})&=-h(\mu,-\bs{k})^T-i\,\text{sgn}\left(\Gamma_{>}\right)\,\Gamma_{<}\,\sigma_z+i\,|\Gamma_{>}|\,\mathds{1}.
\end{align}
Here, $\Gamma_{<}$ and $\Gamma_{>}$ are either $\Gamma_z$ or $\Gamma_0$, $\Gamma_{<}$ being the one with the smaller absolute value, and $\Gamma_{>}$  the one with the larger absolute value. Going beyond the connection to Bogoliubov-de Gennes Hamiltonians and connecting to the discussion of the main text, it is furthermore convenient to define a second unitary transformation

\begin{align}
{U}^{}=\frac{1}{\sqrt{2}}\,\begin{pmatrix}1&0&0&1\\i&0&0&- i\\0&1&-1&0\\0&i&i&0\end{pmatrix},
\end{align}
which yields

\begin{align}
{U}^\dagger\,(i\,X_{\bs{k}})\,{U}= \begin{pmatrix}{H}_{\text{nH}}^{(\text{+})}(\bs{k})&0\\0&{H}_{\text{nH}}^{(\text{-})}(\bs{k})\end{pmatrix},
\end{align}
with

\begin{align}
H_{\text{nH}}^{(\text{+})}(\bs{k})&=h(\mu,\bs{k})+i\,\text{sgn}\left(\Gamma_{>}\right)\,\Gamma_{<}\,\sigma_z+i\,|\Gamma_{>}|\,\mathds{1},\\
H_{\text{nH}}^{(\text{-})}(\bs{k})&=h(-\mu,\bs{k})-i\,\text{sgn}\left(\Gamma_{>}\right)\,\Gamma_{<}\,\sigma_z+i\,|\Gamma_{>}|\,\mathds{1}.
\end{align}

\section{Slab models and symmetries}
When opening up the system along one direction ($x$ or $y$), the momentum along the orthogonal direction remains a good quantum number. Fixing that quantum number defines an effective one-dimensional model. Open boundary conditions along $x$ and $y$ show dramatically different spectra \cite{Kawabata2018}, which can be connected to the fact that slab models along $x$ and $y$ have distinct symmetry properties.

Let us first consider a system of size $L_y$ and with PBC along the $y$-direction, while the system can either be taken with OBC or PBC along $x$ -- this is the case of interest in the main text. The $y$-momentum then has the finite-size quantization $k_y={2\pi n}/{L_y}$ with $n=0,1,\ldots,n-1$. For each such momentum, we obtain an effective one-dimensional Hamiltonian along the $x$-direction. The symmetry analysis is easiest for $k_y=0=\mu$ and PBC along $x$, the case on which we focus now.  The traceless part of the effective one-dimensional Hamiltonian in either the upper or lower blocks then takes the form of a non-Hermitian SSH-model with on-site gain and loss,
\begin{align}
    \mathcal{H}_{\text{nH-SSH}}^{\pm}(k_x)=H_{\text{nH}}^{(\pm)}(k_x) - i\,|\Gamma_{>}|\,\mathds{1} = \begin{pmatrix}m-\alpha-\alpha\,\cos(k_x)\\\beta\,\sin(k_x) \\\pm i\,\Gamma_{<}\end{pmatrix} \cdot \bs{\sigma}.
\end{align}
It is this traceless part of the Hamiltonian on which the symmetry operations will be performed as indicated in Ref.~\cite{Kawasaki22}.
We find that $\mathcal{H}_{\text{nH-SSH}}^{\pm}(k_x)$ has a chiral symmetry $\sigma_z  \,\mathcal{H}_{\text{nH-SSH}}^{(\pm)}(k_x)^{\dagger} \, \sigma_z^{-1}=-\mathcal{H}_{\text{nH-SSH}}^{(\pm)}(k_x)$. As for example explained in Ref.~\cite{Kawabata2019}, this implies that it belongs to class AIII under real Altland-Zirnbauer (AZ) classification, characterised by integer winding numbers with a line-gap in the real part of the spectrum. The non-zero winding implies topologically protected modes whose eigenvalues have a vanishing real part (i.e.~are purely imaginary). Such modes are localised at the boundaries or defects of the system. The system also has a PT-symmetry $\sigma_x\, \mathcal{H}(k_x)^*\, \sigma_x^{-1}=\mathcal{H}(k_x)$. Importantly, it does not exhibit a non-Hermitian skin effect. 

For PBC along $x$ and setting $k_x=0=\mu$, the effective traceless one-dimensional Hamiltonians become

\begin{align}
    \tilde{\mathcal{H}}_{\text{nH-SSH}}^{\pm}(k_y)=H_{\text{nH}}^{(\pm)}(k_y) - i\,|\Gamma_{>}|\,\mathds{1} = \begin{pmatrix}m-\alpha-\alpha\,\cos(k_y)\\0 \\\beta\,\sin(k_y)\pm i\,\Gamma_{<}\end{pmatrix} \cdot \bs{\sigma}.
\end{align}
This Hamiltonian has a sub-lattice symmetry $\sigma_y \, \tilde{\mathcal{H}}_{\text{nH-SSH}}^{(\pm)}(k_y) \, \sigma_y^{-1}=- \tilde{\mathcal{H}}_{\text{nH-SSH}}^{(\pm)}(k_y)$ and belongs to class $\mathcal{S}$-AIII under complex AZ classification with additional sub-lattice symmetry \cite{Kawabata2019}. This difference in symmetry classification as compared to $\mathcal{H}_{\text{nH-SSH}}^{\pm}(k_x)$ entails important differences in physics: it is for example well-known that $\tilde{\mathcal{H}}_{\text{nH-SSH}}^{\pm}(k_y)$ exhibits a non-Hermitian skin-effect instead of extremal edge states \cite{Bergholtz,banerjee2022,Okuma_2023}. 

Imposing OBC along both directions, finally, is known to stabilize edge states only along one of the edges of the two-dimensional system \cite{Kawabata2018}.

\section{Spectrum of the damping matrix}
\begin{figure}
\begin{center}
\begin{tabular}{ ccccc }
\raisebox{4.985cm}{(a)}\raisebox{0.325cm}{\includegraphics[width=0.27\columnwidth]{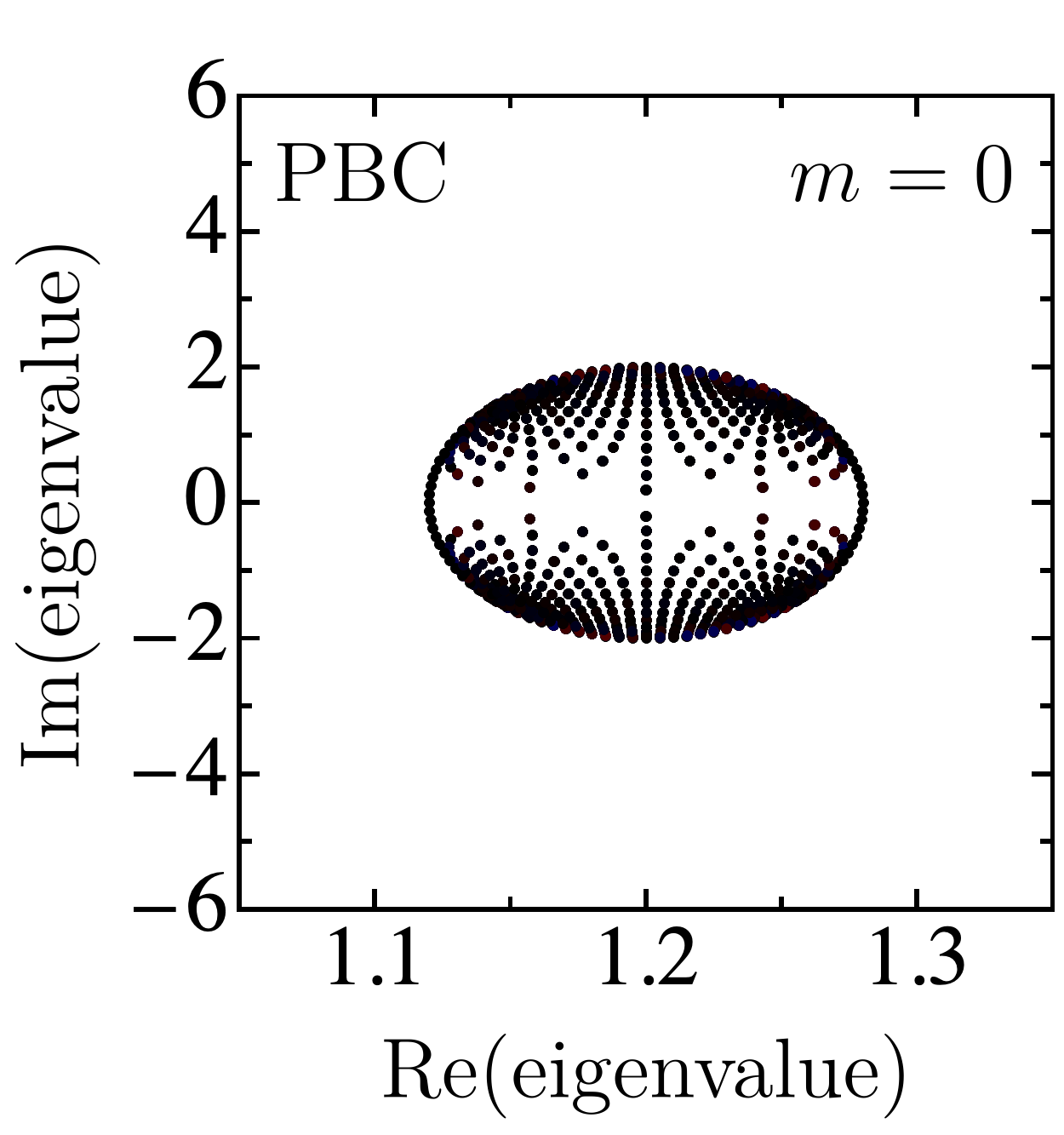}}&\hspace*{0.2cm}&\raisebox{4.985cm}{(b)}\includegraphics[width=0.275\columnwidth]{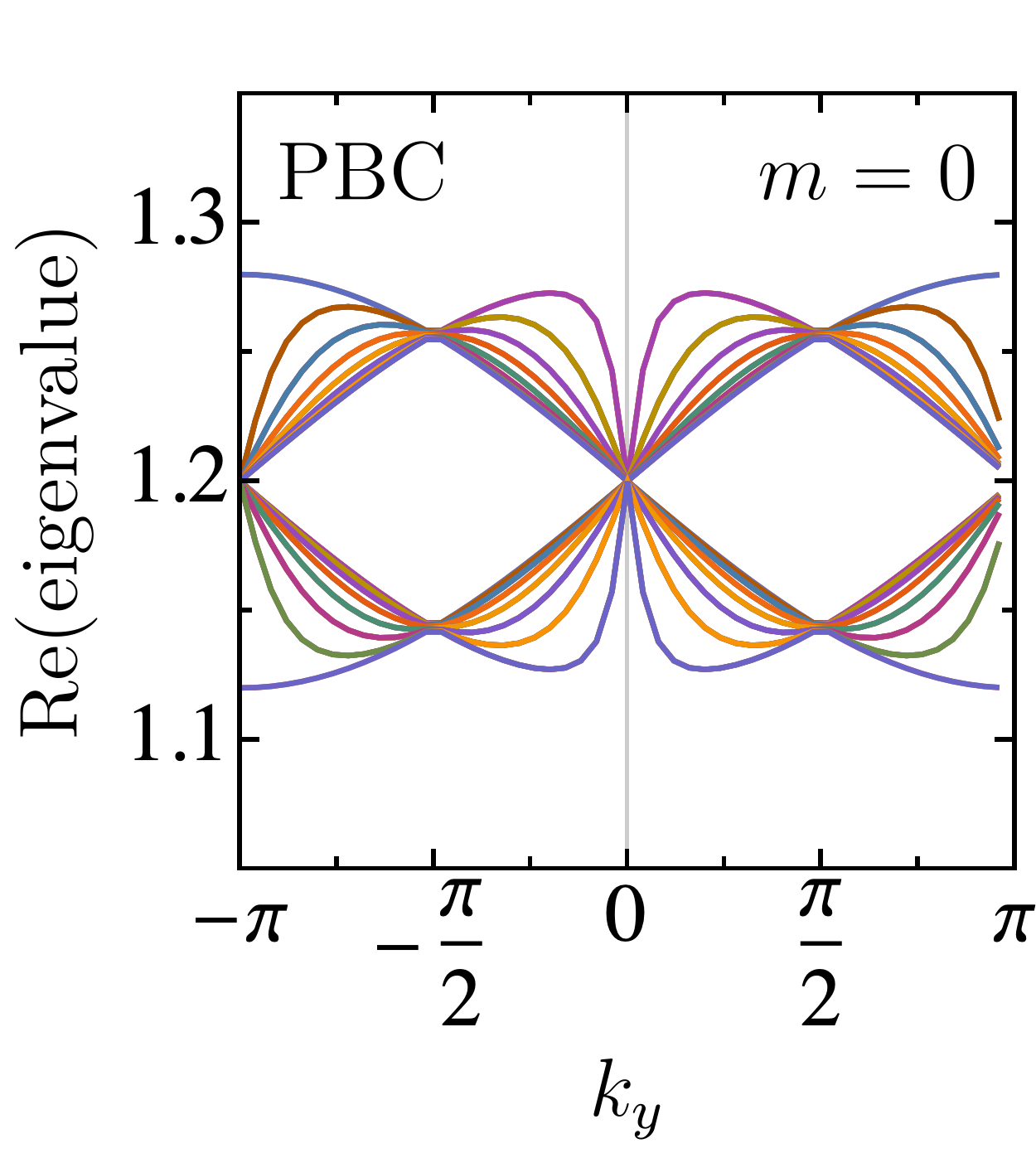}&\hspace*{0.2cm}&\raisebox{4.985cm}{(c)}\includegraphics[width=0.275\columnwidth]{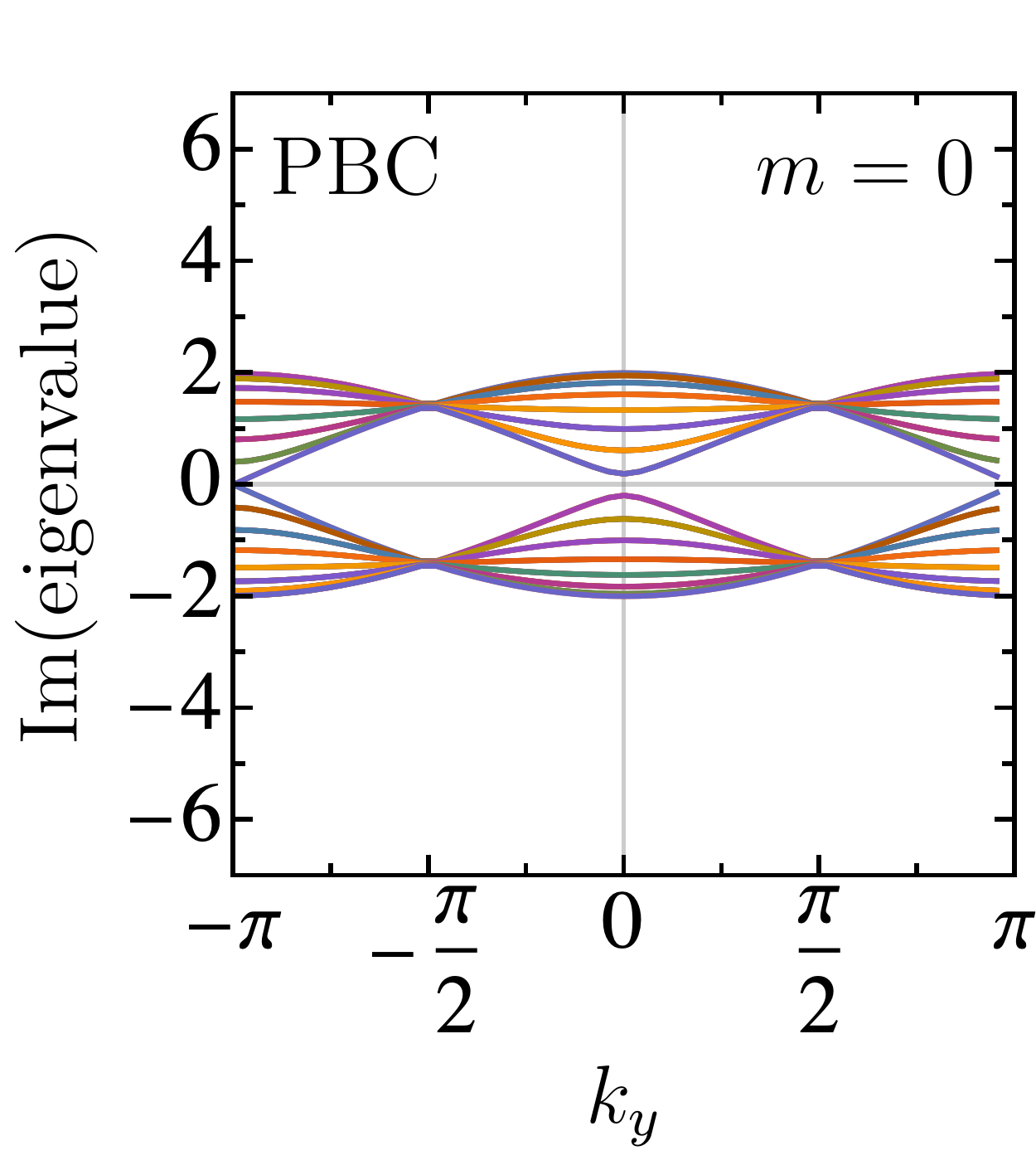}\\
\raisebox{4.985cm}{(d)}\raisebox{0.325cm}{\includegraphics[width=0.27\columnwidth]{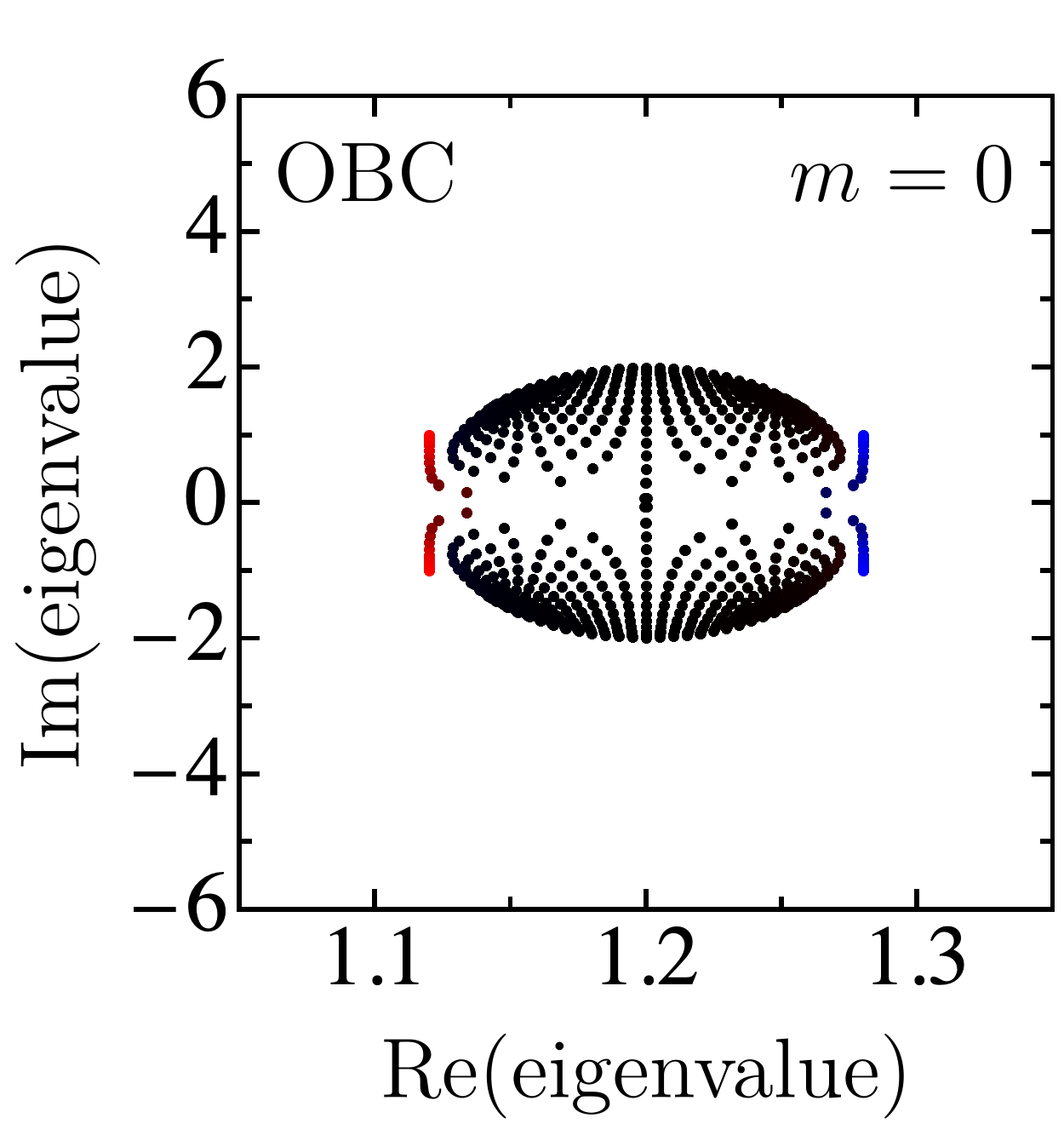}}&\hspace*{0.2cm}&\raisebox{4.985cm}{(e)}\includegraphics[width=0.275\columnwidth]{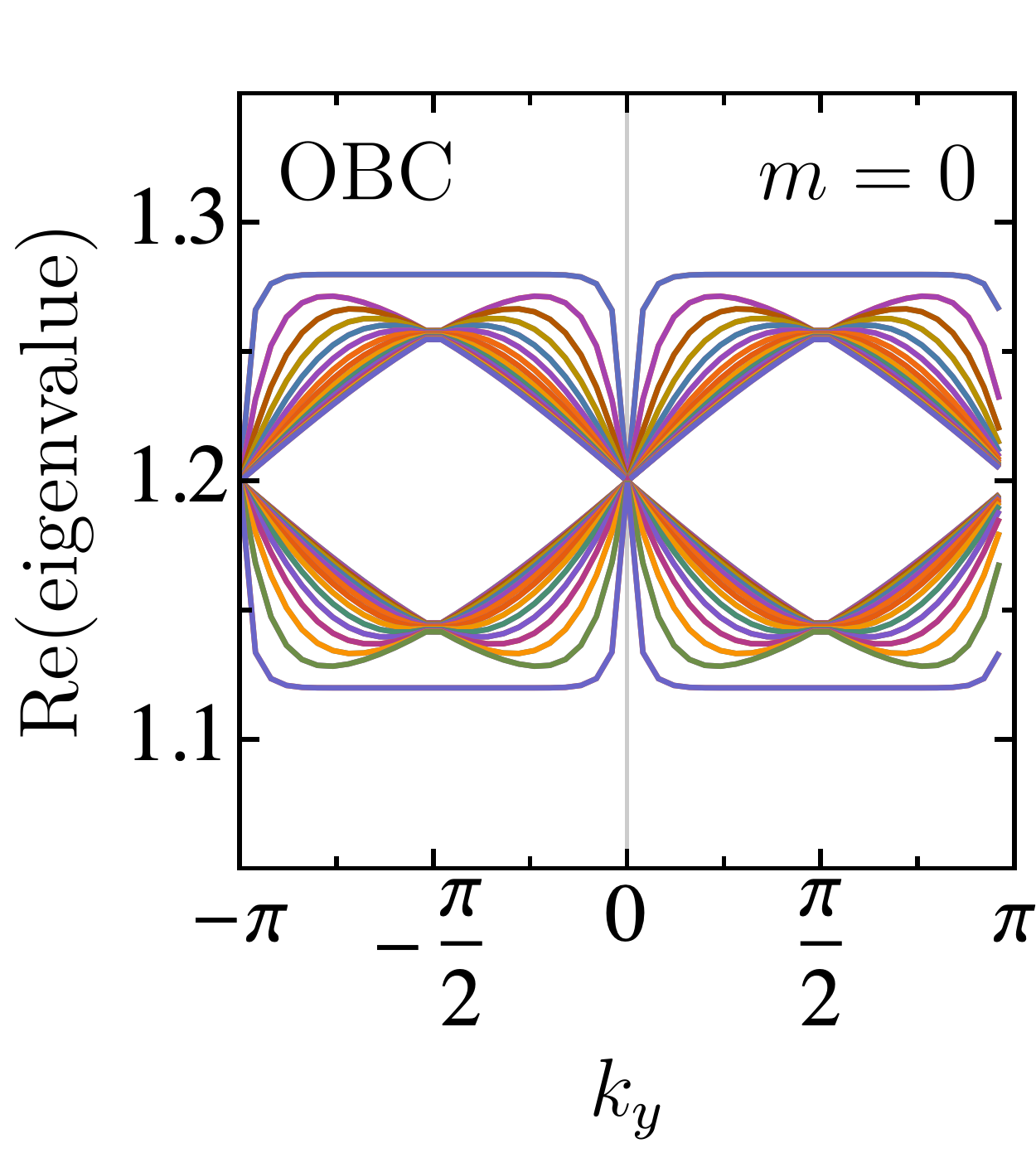}&\hspace*{0.2cm}&\raisebox{4.985cm}{(f)}\includegraphics[width=0.275\columnwidth]{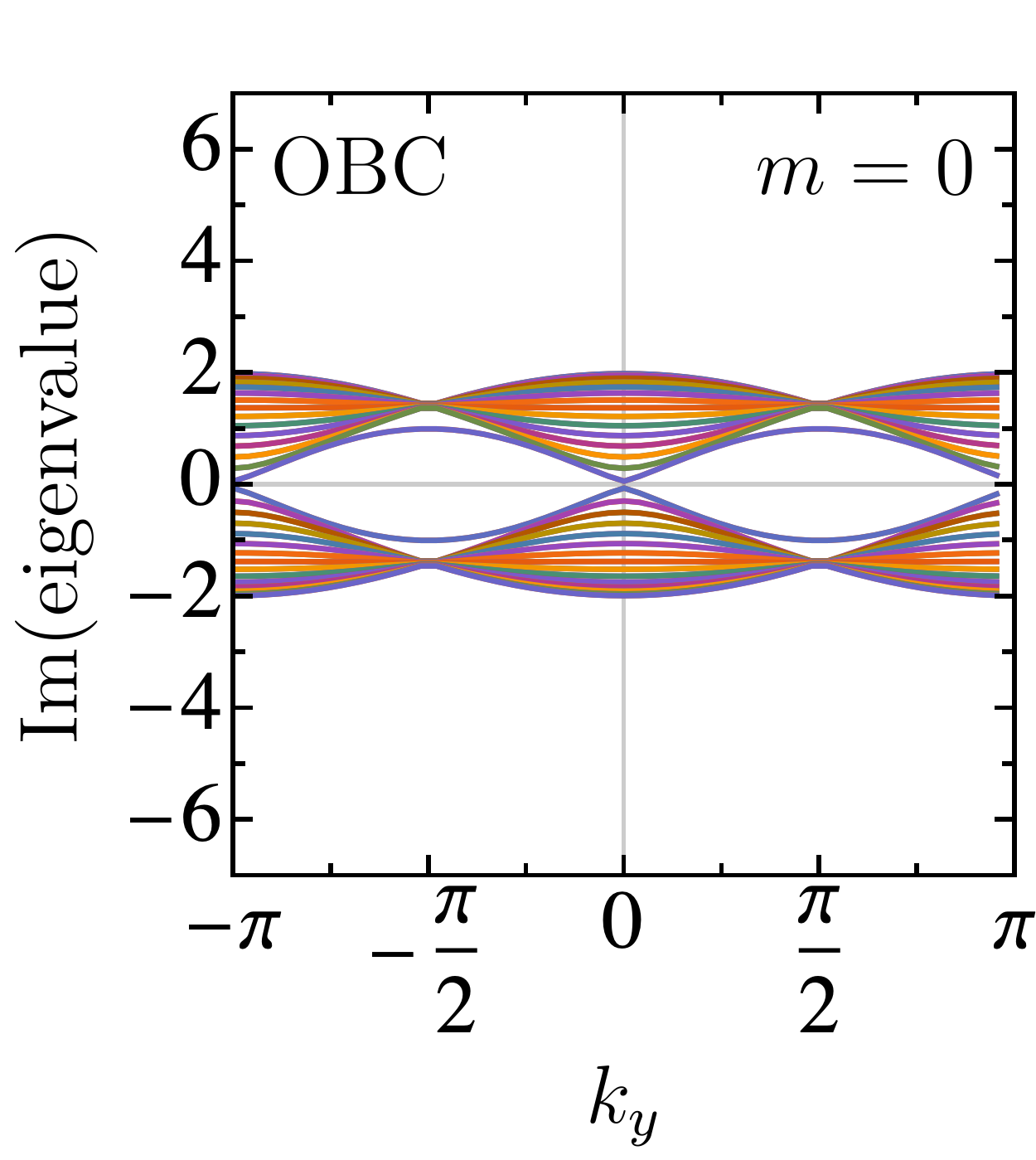}\\
\end{tabular}
\end{center}
\caption{Complex spectrum of the damping matrix $X$ for $m=0$, when $H$ is  in a topological regime with topological edge states, but without a bulk spectral gap. We compare periodic boundary conditions (PBC, top row) and open boundary conditions (OBC, bottom row). We always choose $\alpha=\beta=1$, $\mu=0$, $\Gamma_0=0.08$ and $\Gamma_z=1.2$, consider 15 sites along $x$, and use periodic boundary conditions along $y$. Panels (a) and (d) depict the complex eigenvalues combining 50 equidistant values of $k_y$ with PBC and OBC, respectively. The color indicates the localization of the corresponding eigenstates: bulk states are shown in black, states at the left (right) edge in red (blue). The dependence of the eigenvalues of $X$ on $k_y$ is illustrated panels (b) (PBC, real part), (c) (PBC, imaginary part),  (e) (OBC, real part), and (f) (OBC, imaginary part).}
\label{fig:damping_spectrum_m0}
\end{figure}

The complex eigenvalues of damping matrix $X$ are depicted in Fig.~\ref{fig:damping_spectrum_m2} for the case discussed in the main text, i.e.~$m=1$ and other chosen parameters as mentioned in the caption (here, the closed-system Chern insulator Hamiltonian is in a topological regime with Chern number $C=1$, and the damping matrix has a bulk spectral gap as well as edge states). In addition, we show the spectrum for $m=0$ (corresponding to a topological transition of the Hamiltonian, the damping matrix has no bulk spectral gap, but does feature  edge states) in Fig.~\ref{fig:damping_spectrum_m0}, while Fig.~\ref{fig:damping_spectrum_m2} shows the same for $m=2$ (corresponding to a topological transition of the Hamiltonian, the damping matrix has no bulk spectral gap and no edge states), and Fig.~\ref{fig:damping_spectrum_m3} for $m=3$ (corresponding to a topologically trivial regime of the Hamiltonian, the damping matrix has a spectral gap but no edge states).

\begin{figure}
\begin{center}
\begin{tabular}{ ccccc }
\raisebox{4.985cm}{(a)}\raisebox{0.325cm}{\includegraphics[width=0.27\columnwidth]{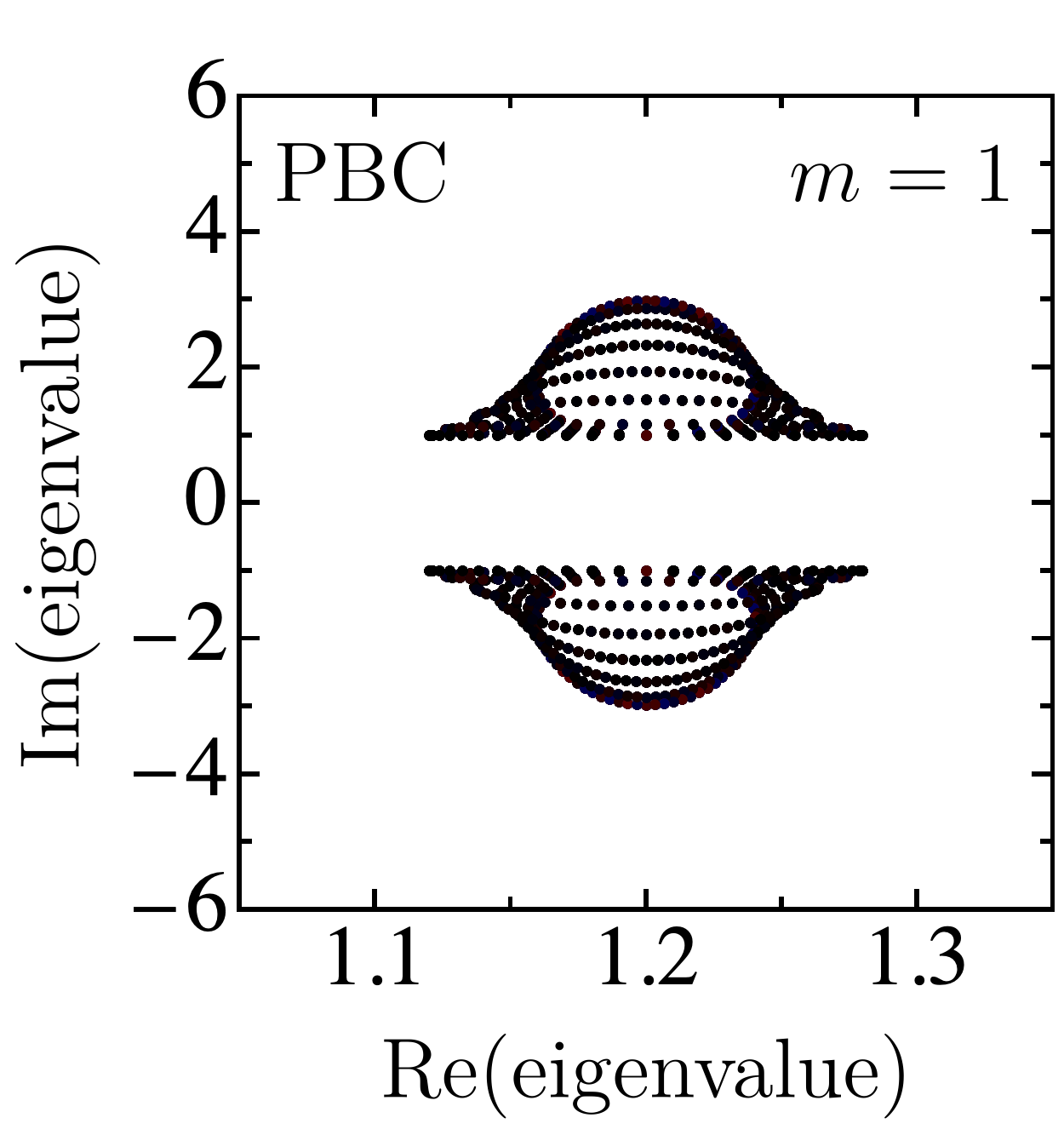}}&\hspace*{0.2cm}&\raisebox{4.985cm}{(b)}\includegraphics[width=0.275\columnwidth]{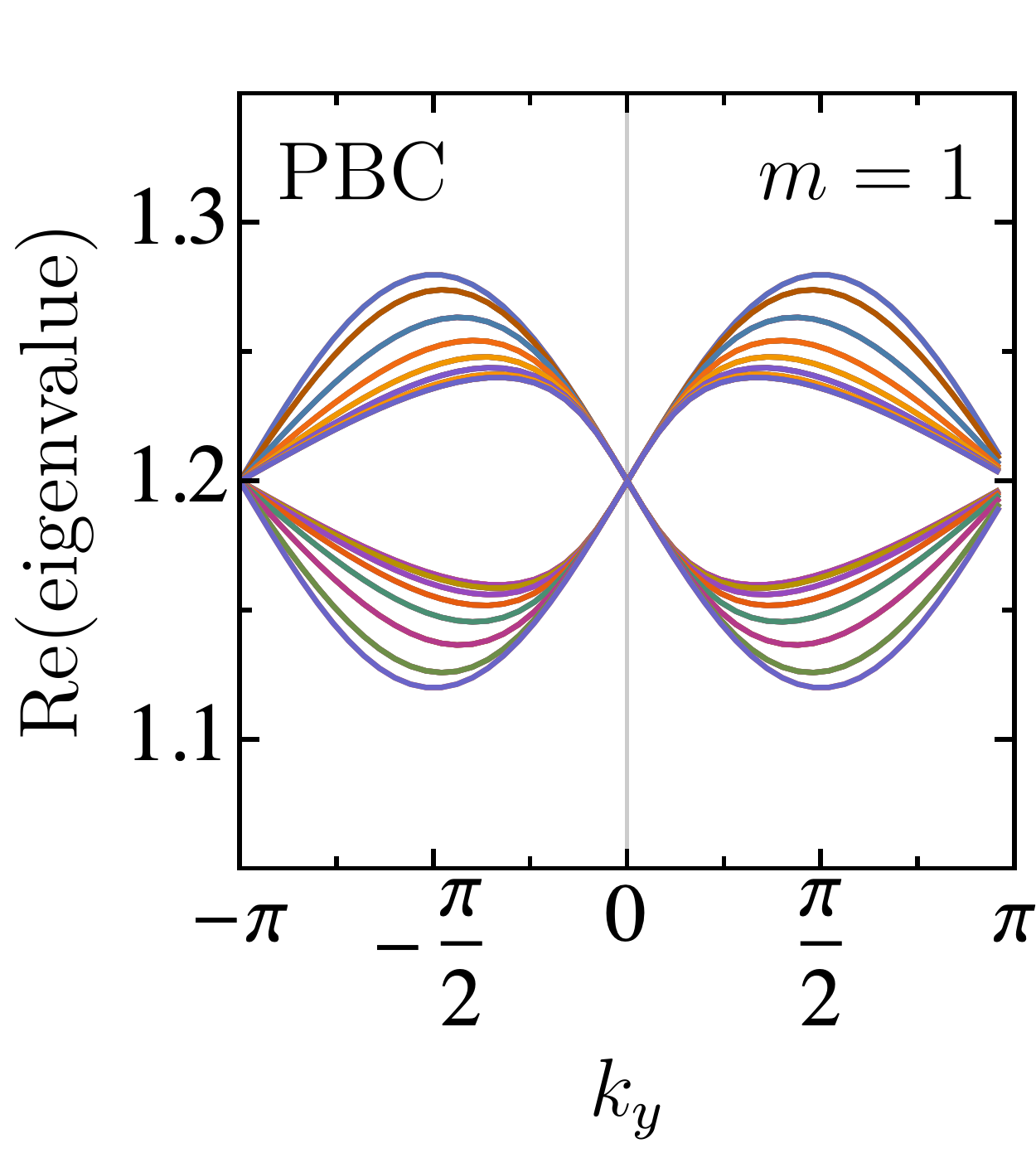}&\hspace*{0.2cm}&\raisebox{4.985cm}{(c)}\includegraphics[width=0.275\columnwidth]{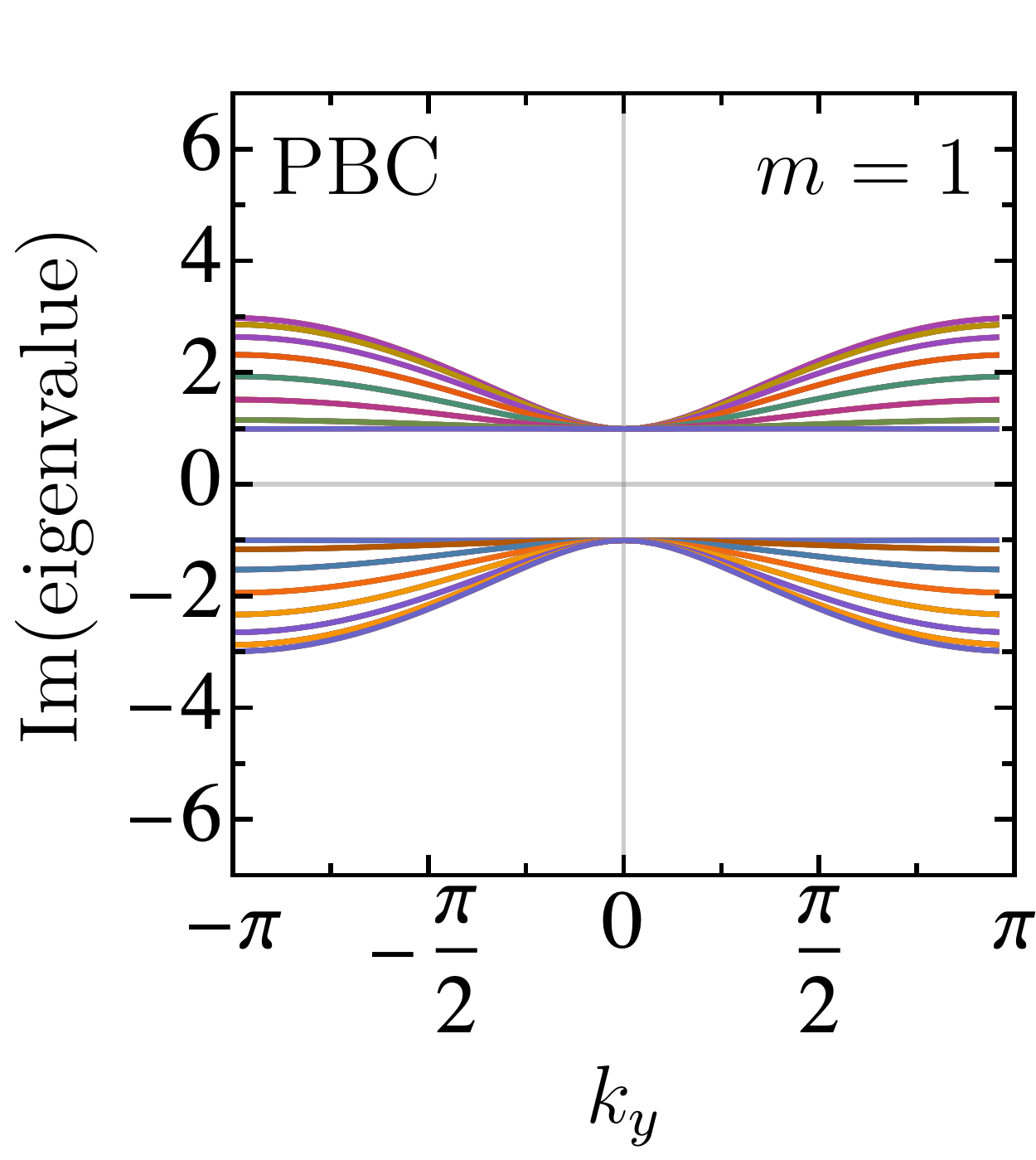}\\
\raisebox{4.985cm}{(d)}\raisebox{0.325cm}{\includegraphics[width=0.27\columnwidth]{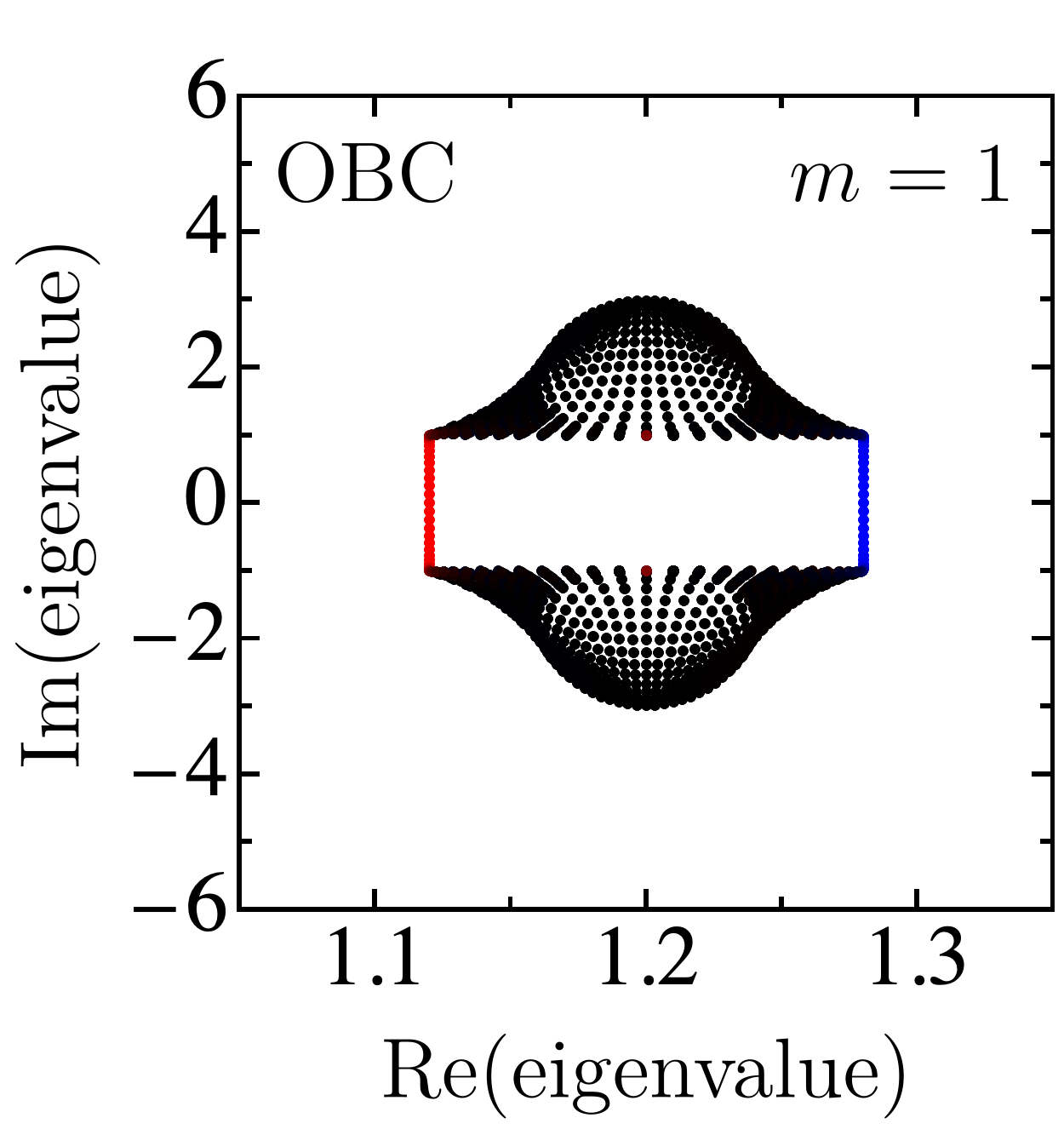}}&\hspace*{0.2cm}&\raisebox{4.985cm}{(e)}\includegraphics[width=0.275\columnwidth]{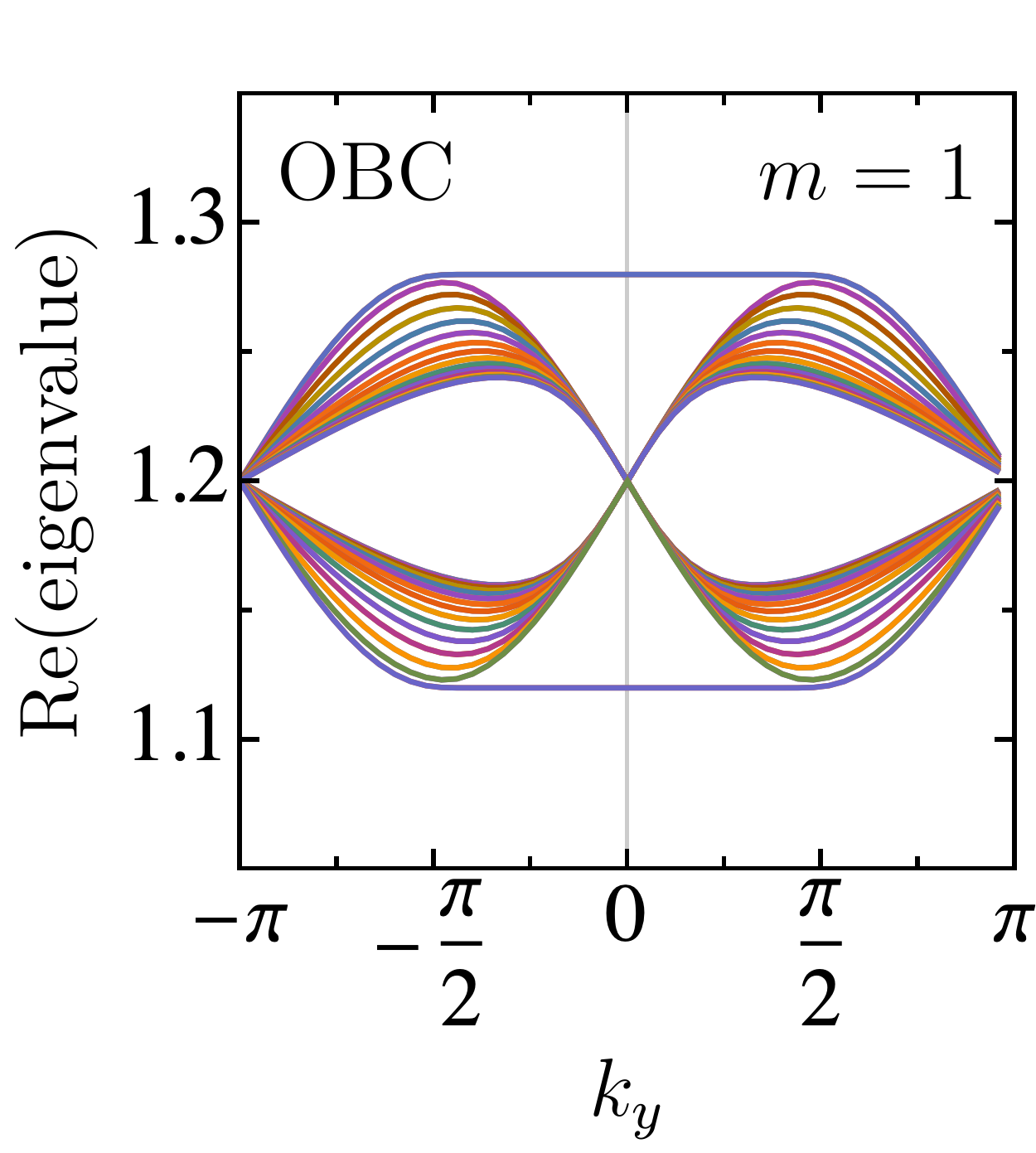}&\hspace*{0.2cm}&\raisebox{4.985cm}{(f)}\includegraphics[width=0.275\columnwidth]{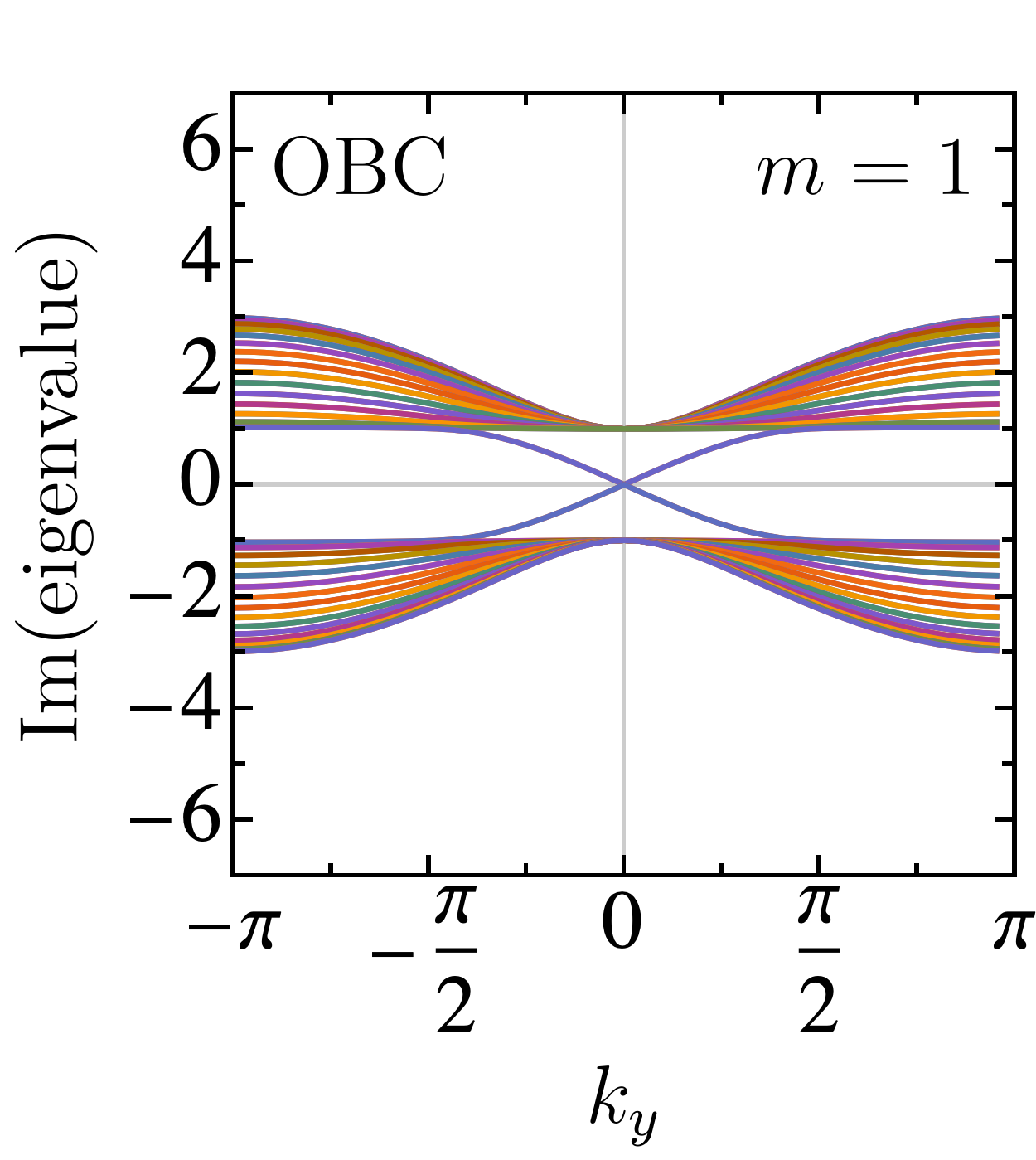}\\
\end{tabular}
\end{center}
\caption{Complex spectrum of the damping matrix $X$ for $m=1$, when $H$ is  in a topological regime with topological edge states and a bulk spectral gap. We compare periodic boundary conditions (PBC, top row) and open boundary conditions (OBC, bottom row). We always choose $\alpha=\beta=1$, $\mu=0$, $\Gamma_0=0.08$ and $\Gamma_z=1.2$, consider 15 sites along $x$, and use periodic boundary conditions along $y$. Panels (a) and (d) depict the complex eigenvalues combining 50 equidistant values of $k_y$ with PBC and OBC, respectively. The color indicates the localization of the corresponding eigenstates: bulk states are shown in black, states at the left (right) edge in red (blue). The dependence of the eigenvalues of $X$ on $k_y$ is illustrated panels (b) (PBC, real part), (c) (PBC, imaginary part),  (e) (OBC, real part), and (f) (OBC, imaginary part).}
\label{fig:damping_spectrum_m1}
\end{figure}

\begin{figure}
\begin{center}
\begin{tabular}{ ccccc }
\raisebox{5.44cm}{(a)}\raisebox{0.33cm}{\includegraphics[width=0.295\columnwidth]{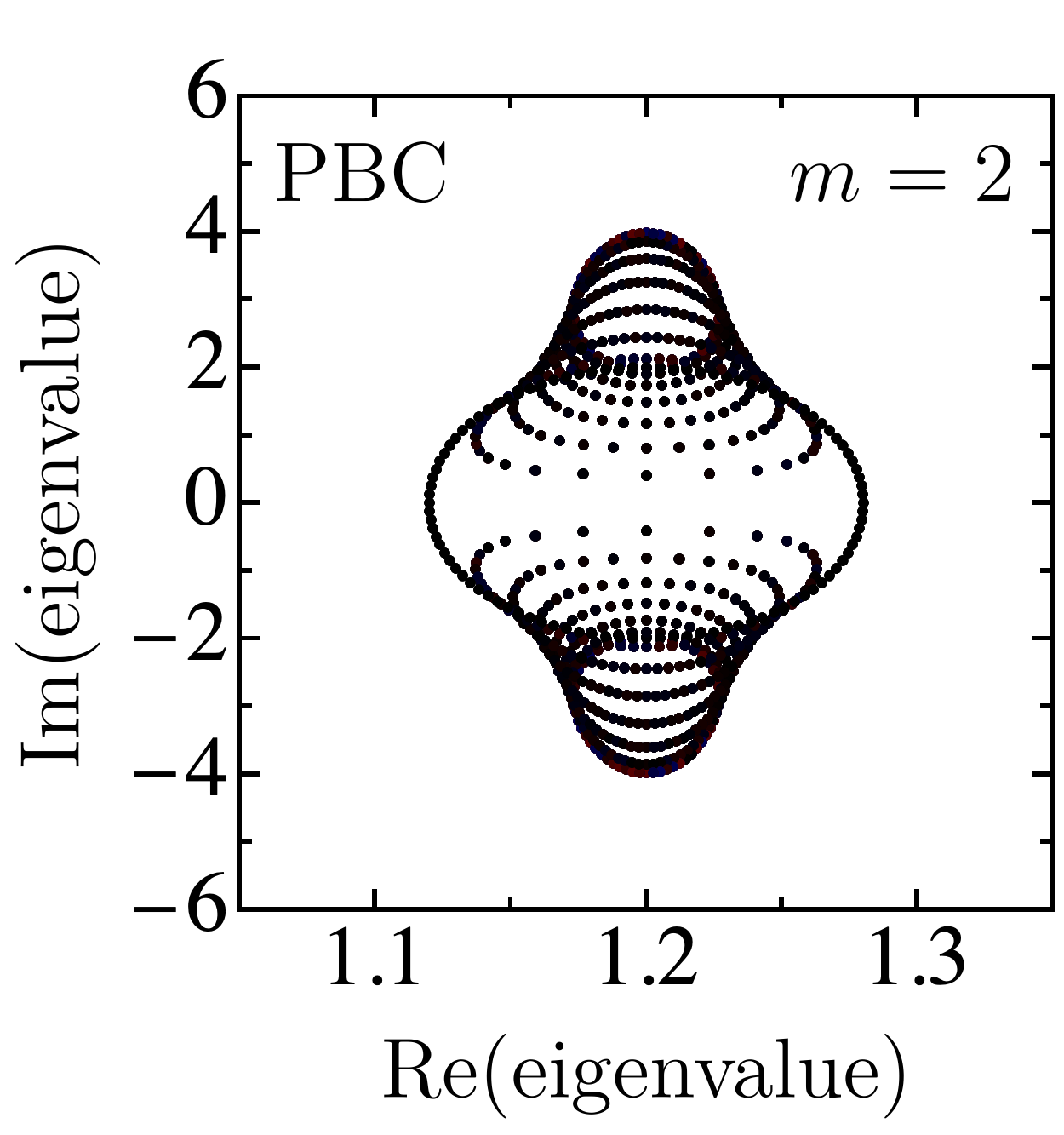}}&\hspace*{0.2cm}&\raisebox{5.44cm}{(b)}\includegraphics[width=0.3\columnwidth]{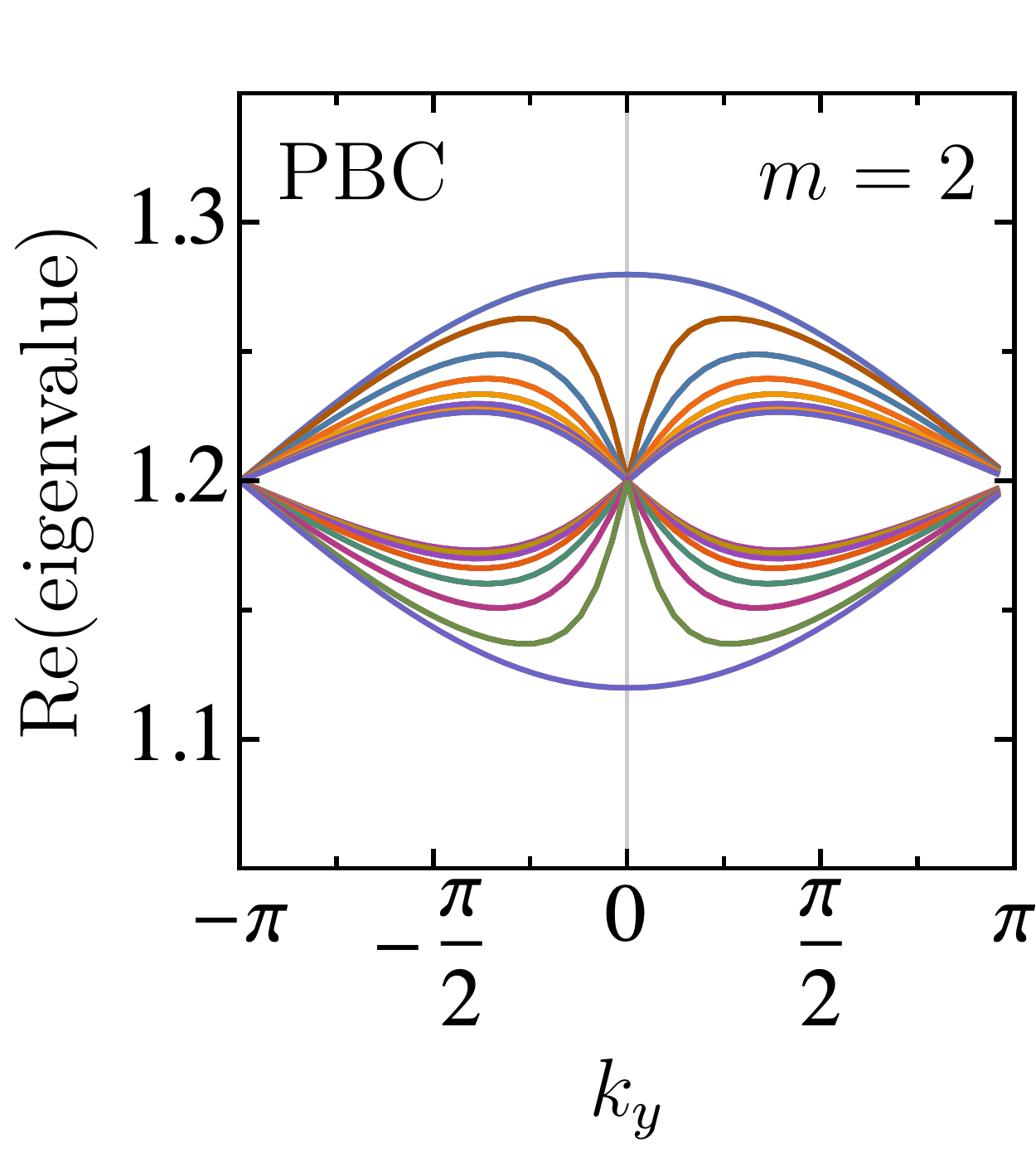}&\hspace*{0.2cm}&\raisebox{5.44cm}{(c)}\includegraphics[width=0.3\columnwidth]{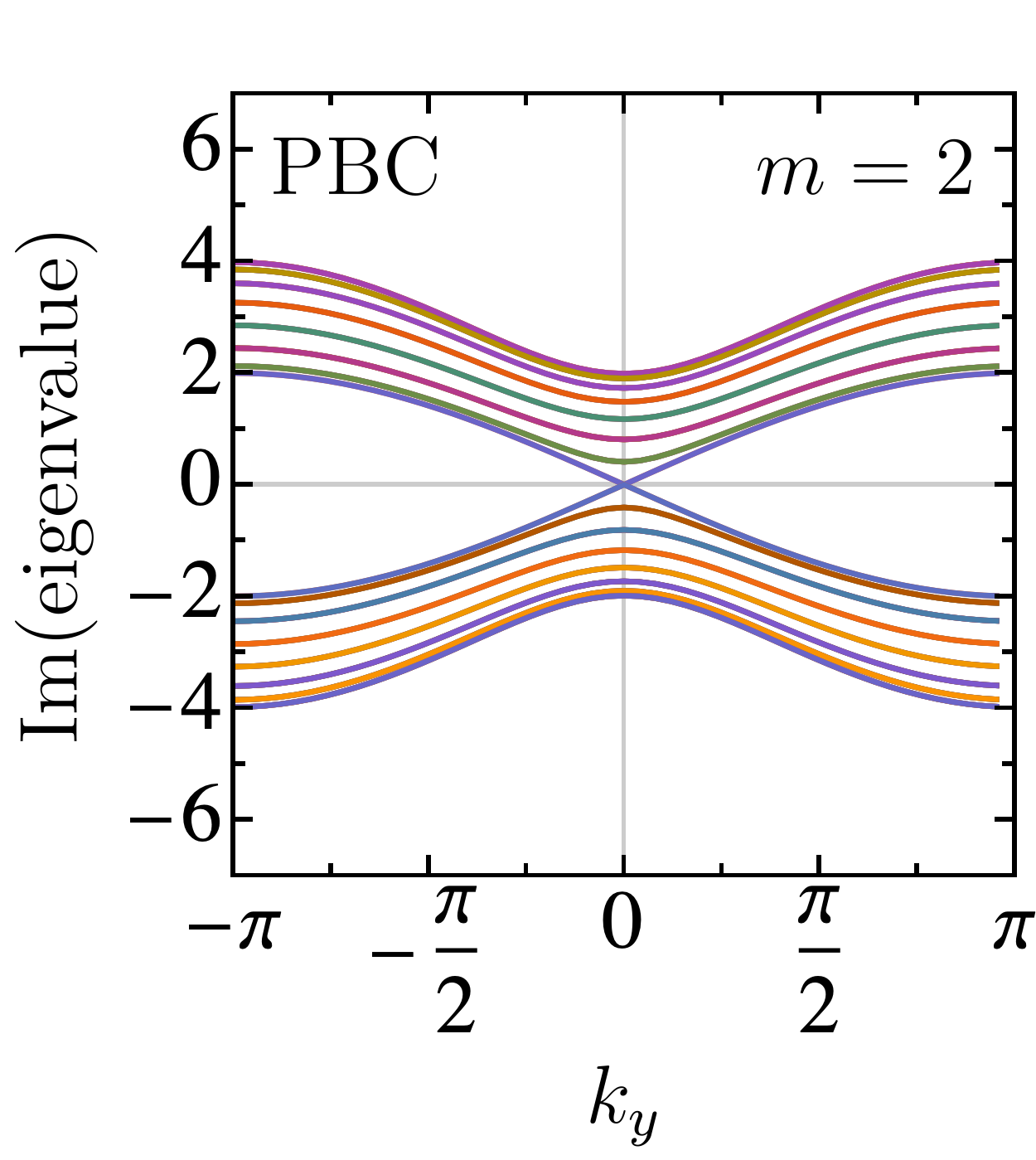}\\
\raisebox{5.44cm}{(d)}\raisebox{0.33cm}{\includegraphics[width=0.295\columnwidth]{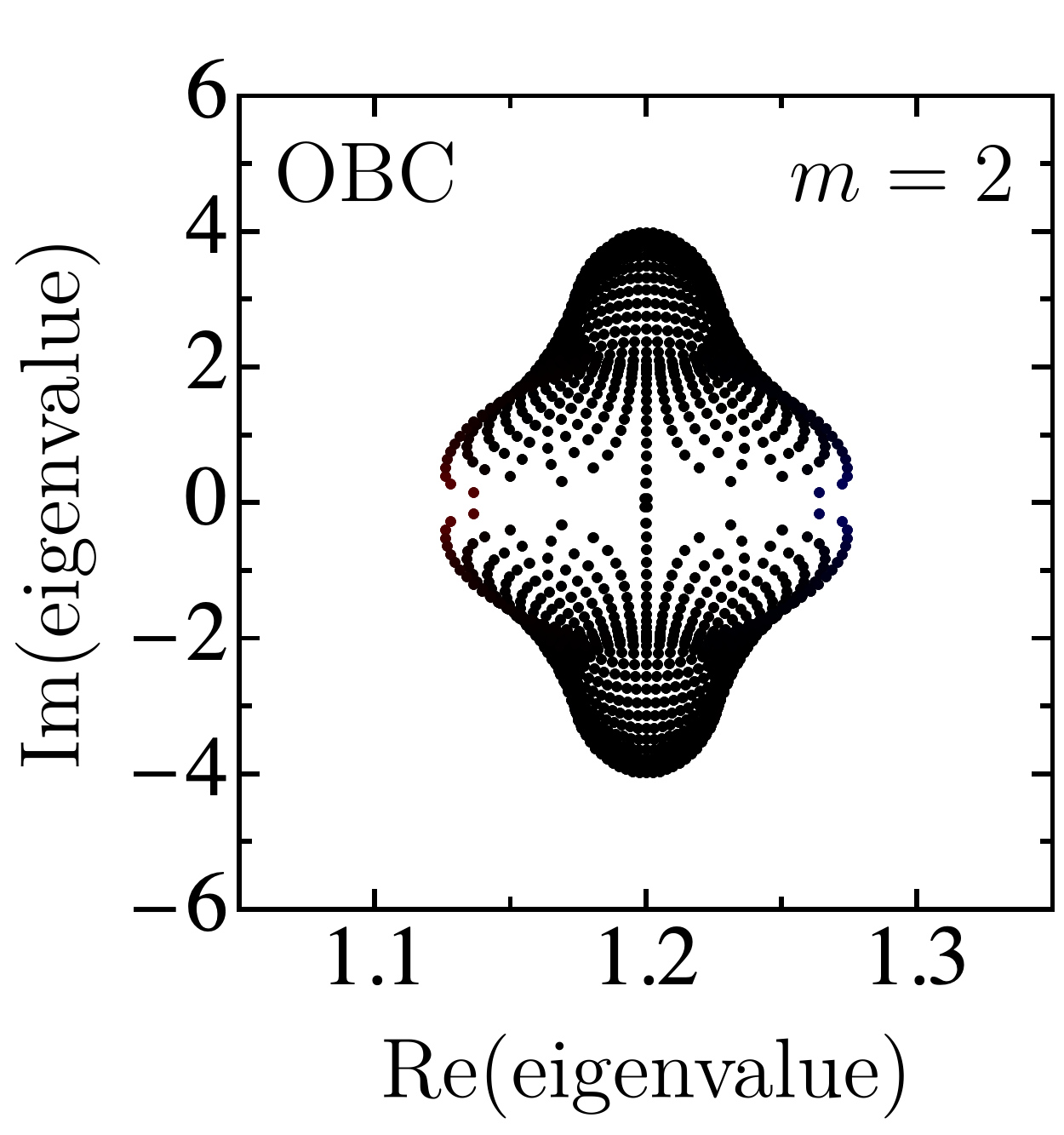}}&\hspace*{0.2cm}&\raisebox{5.44cm}{(e)}\includegraphics[width=0.3\columnwidth]{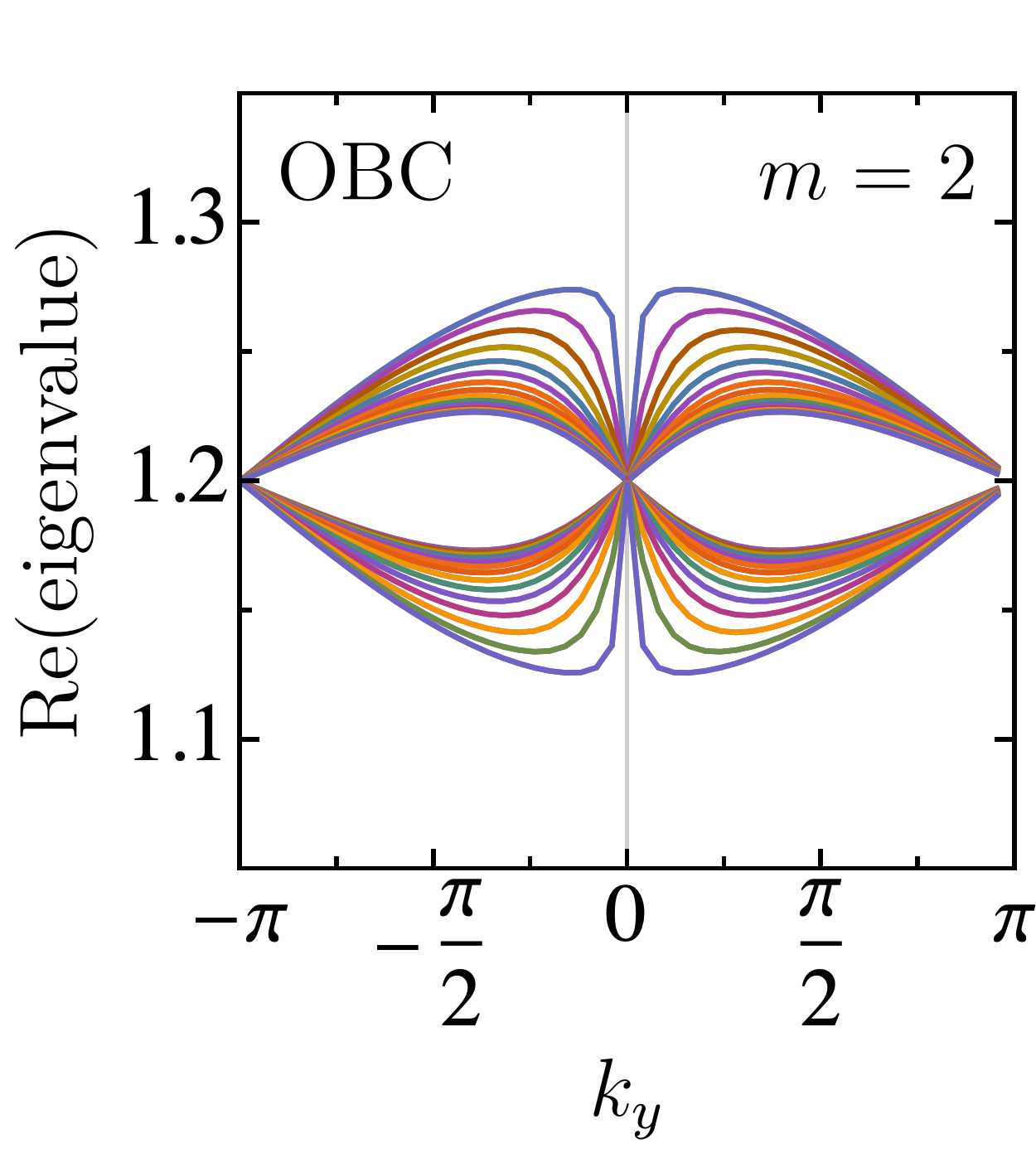}&\hspace*{0.2cm}&\raisebox{5.44cm}{(f)}\includegraphics[width=0.3\columnwidth]{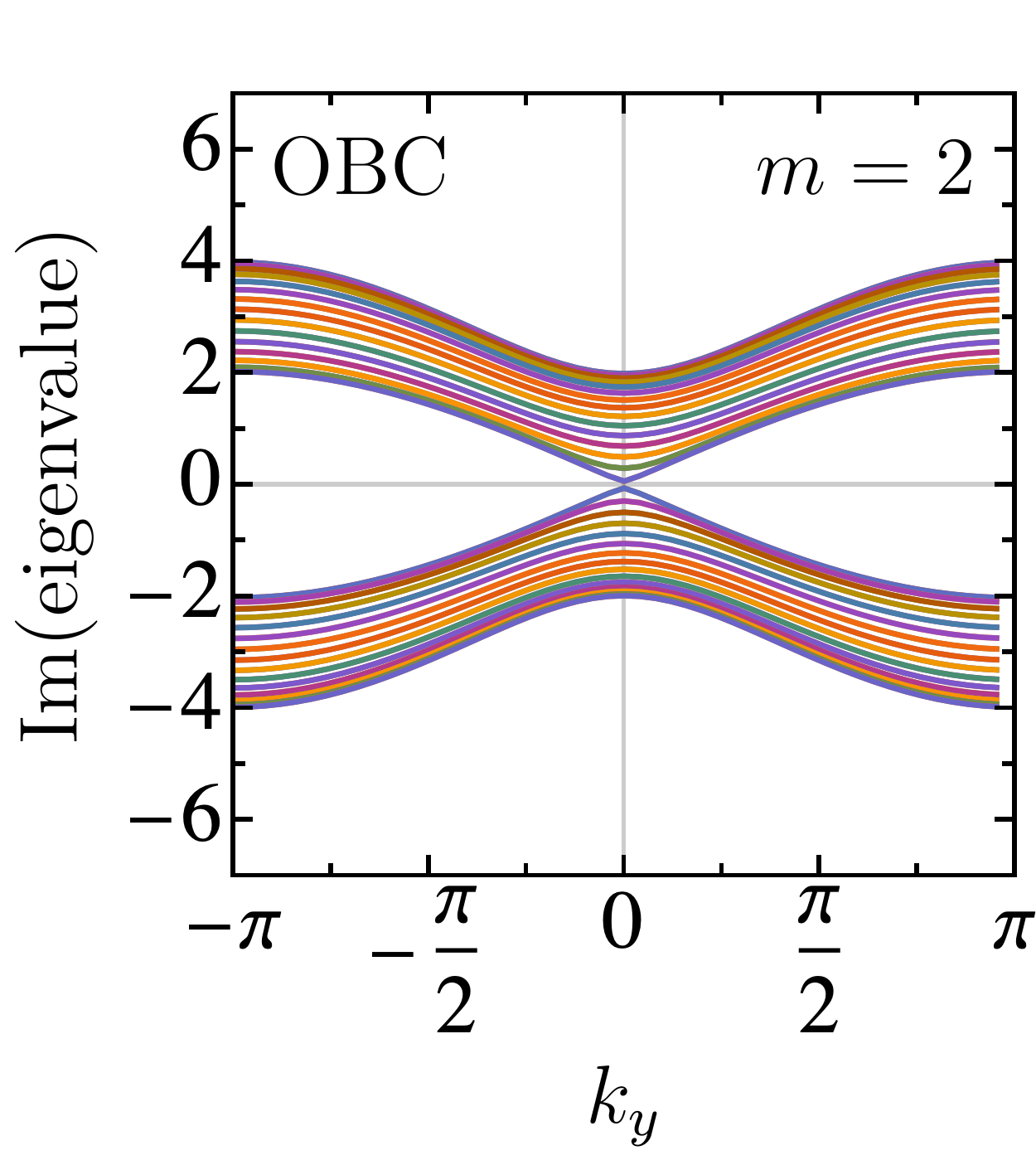}\\
\end{tabular}
\end{center}
\caption{Complex spectrum of the damping matrix $X$ for $m=2$, when $H$ is at a topological transition and has a bulk gap closing without edge states. We compare periodic boundary conditions (PBC, top row) and open boundary conditions (OBC, bottom row). We always choose $\alpha=\beta=1$, $\mu=0$, $\Gamma_0=0.08$ and $\Gamma_z=1.2$, consider 15 sites along $x$, and use periodic boundary conditions along $y$. Panels (a) and (d) depict the complex eigenvalues combining 50 equidistant values of $k_y$ with PBC and OBC, respectively. The color indicates the localization of the corresponding eigenstates: bulk states are shown in black, states at the left (right) edge would be shown in red (blue), but are absent here. The dependence of the eigenvalues of $X$ on $k_y$ is illustrated panels (b) (PBC, real part), (c) (PBC, imaginary part),  (e) (OBC, real part), and (f) (OBC, imaginary part).}
\label{fig:damping_spectrum_m2}
\end{figure}

\begin{figure}
\begin{center}
\begin{tabular}{ ccccc }
\raisebox{5.44cm}{(a)}\raisebox{0.33cm}{\includegraphics[width=0.295\columnwidth]{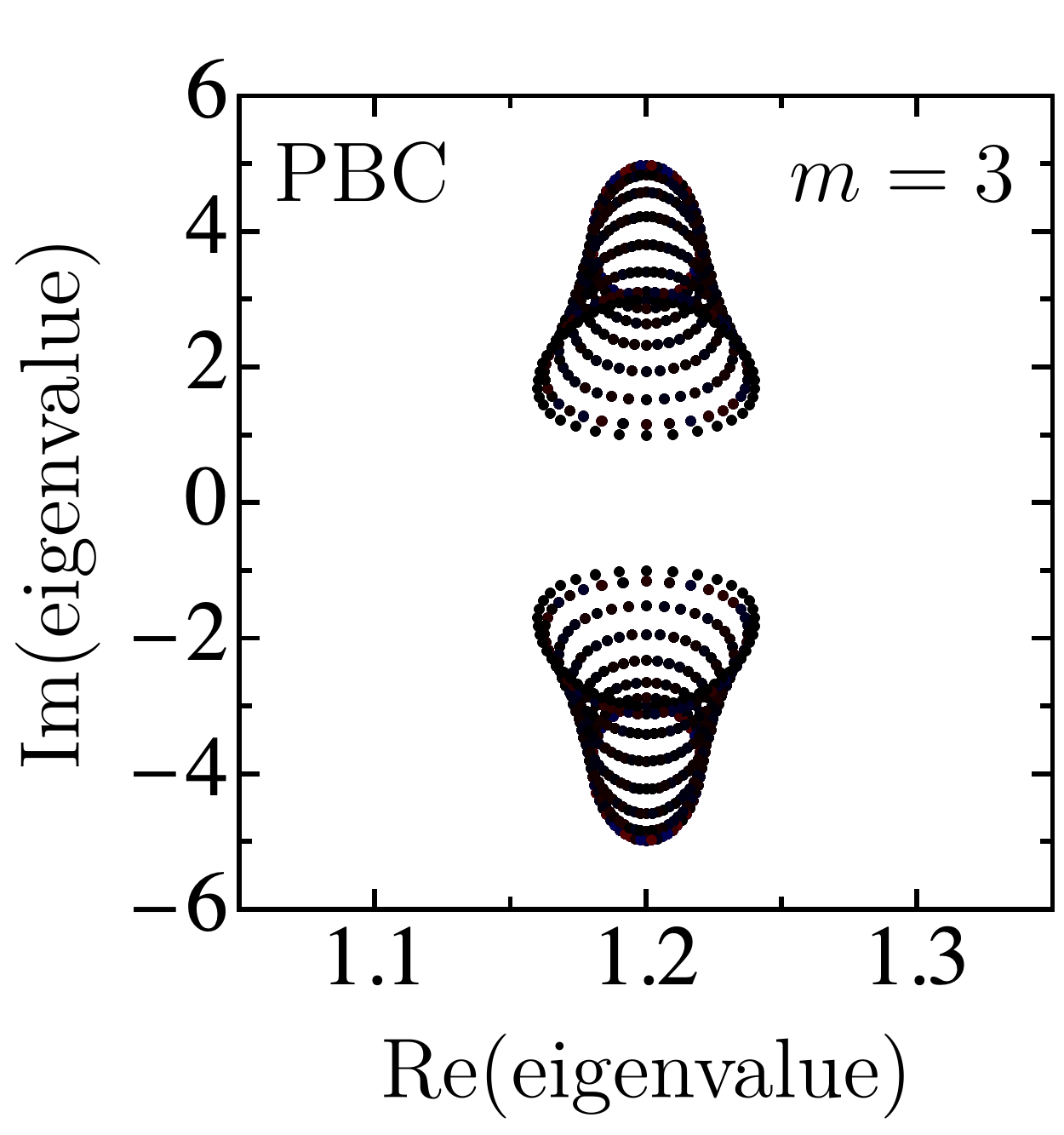}}&\hspace*{0.2cm}&\raisebox{5.44cm}{(b)}\includegraphics[width=0.3\columnwidth]{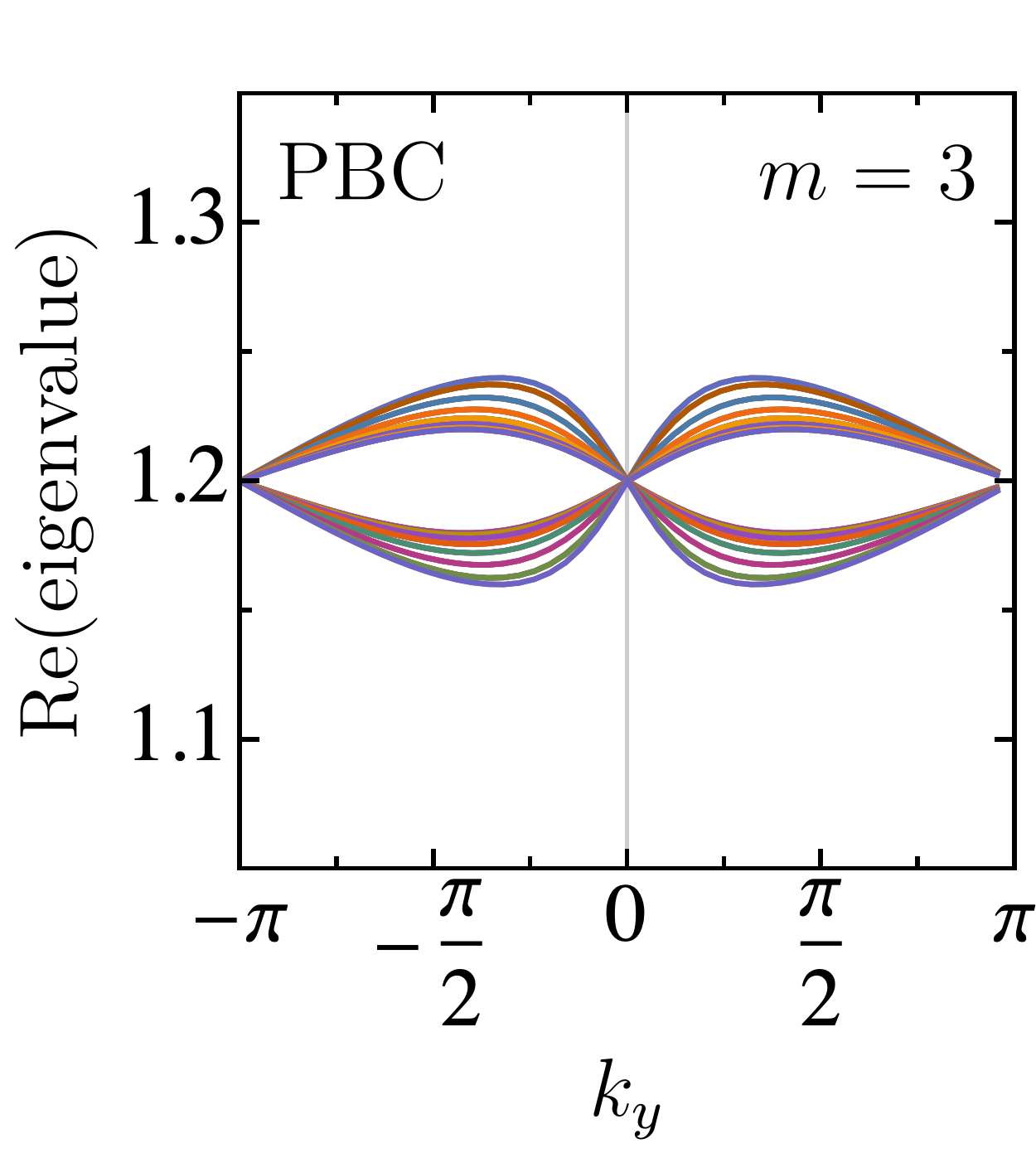}&\hspace*{0.2cm}&\raisebox{5.44cm}{(c)}\includegraphics[width=0.3\columnwidth]{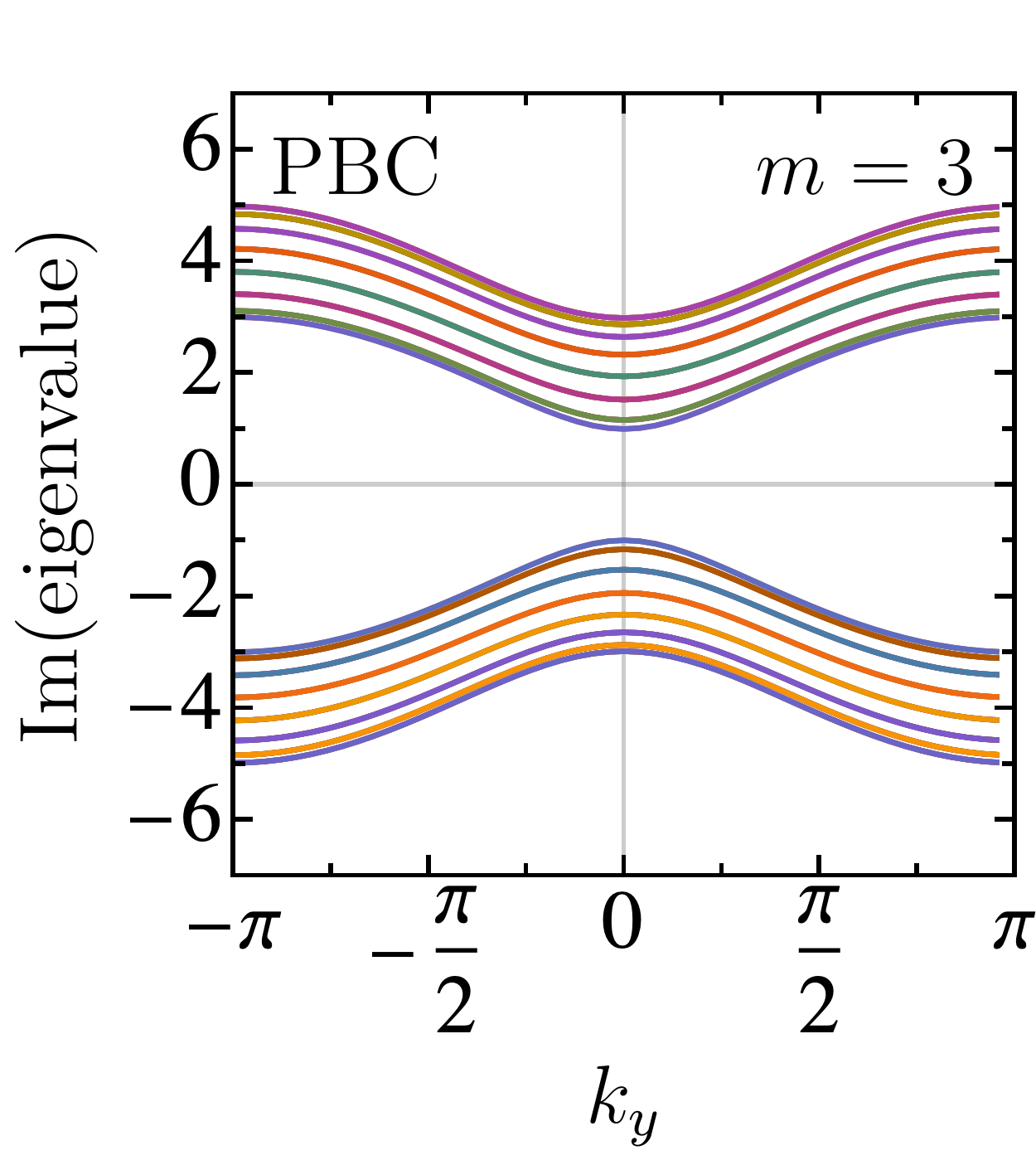}\\
\raisebox{5.44cm}{(d)}\raisebox{0.33cm}{\includegraphics[width=0.295\columnwidth]{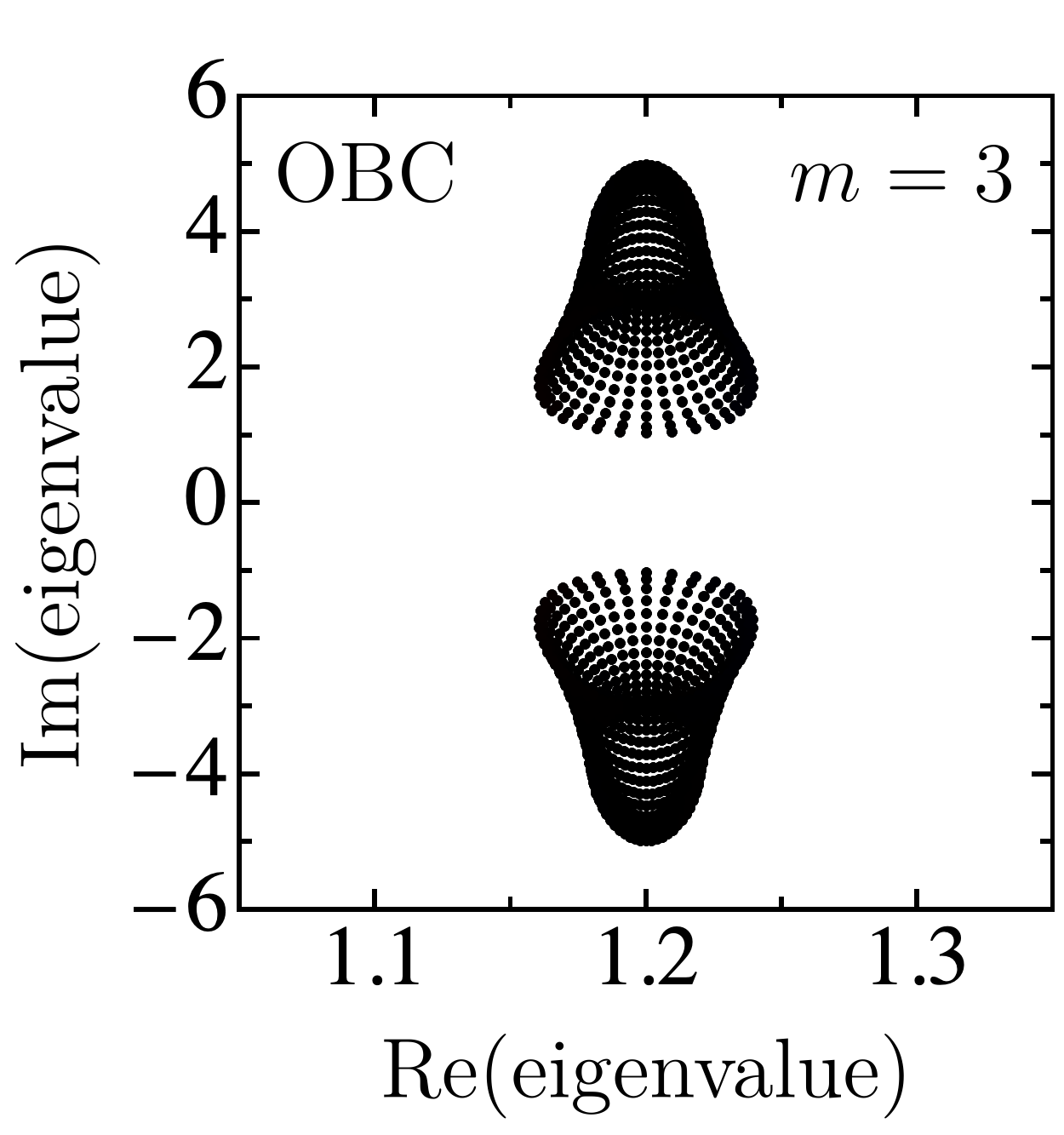}}&\hspace*{0.2cm}&\raisebox{5.44cm}{(e)}\includegraphics[width=0.3\columnwidth]{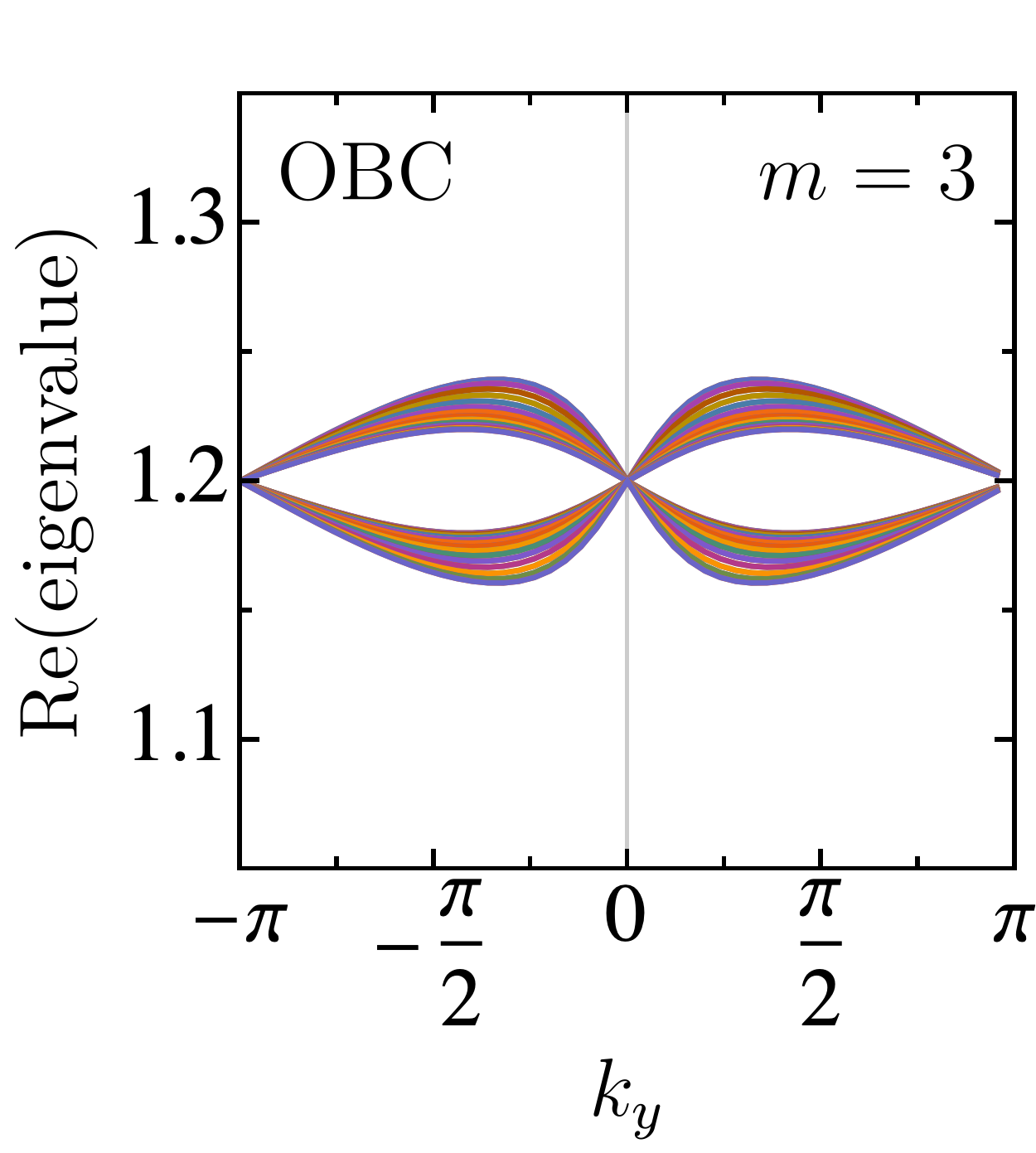}&\hspace*{0.2cm}&\raisebox{5.44cm}{(f)}\includegraphics[width=0.3\columnwidth]{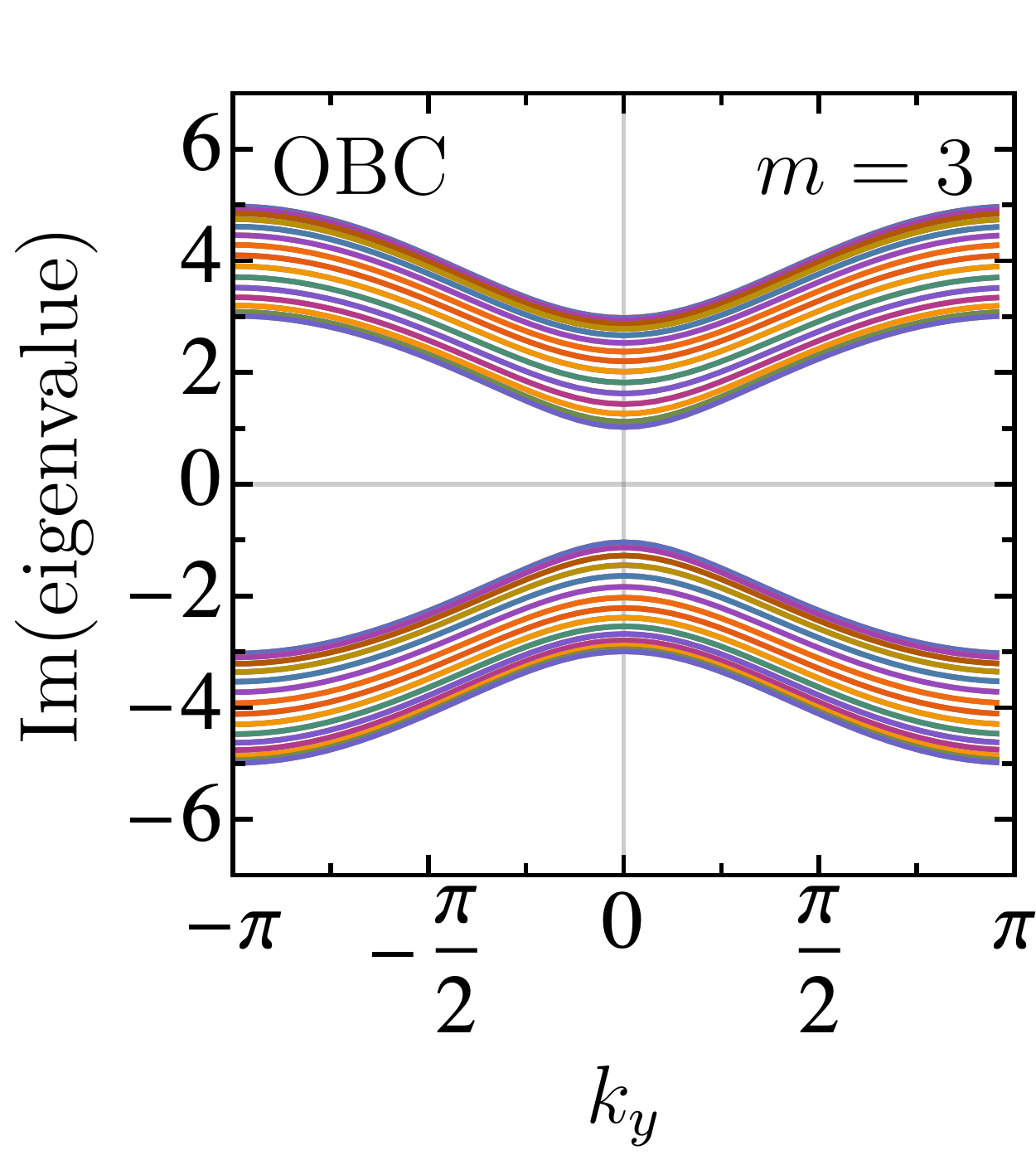}\\
\end{tabular}
\end{center}
\caption{Complex spectrum of the damping matrix $X$ for $m=3$, when $H$ is in a trivial regime without topological edge states, and has no edge states. We compare periodic boundary conditions (PBC, top row) and open boundary conditions (OBC, bottom row). We always choose $\alpha=\beta=1$, $\mu=0$, $\Gamma_0=0.08$ and $\Gamma_z=1.2$, consider 15 sites along $x$, and use periodic boundary conditions along $y$. Panels (a) and (d) depict the complex eigenvalues combining 50 equidistant values of $k_y$ with PBC and OBC, respectively. The color indicates the localization of the corresponding eigenstates: bulk states are shown in black, states at the left (right) edge would be shown in red (blue), but are absent here. The dependence of the eigenvalues of $X$ on $k_y$ is illustrated panels (b) (PBC, real part), (c) (PBC, imaginary part),  (e) (OBC, real part), and (f) (OBC, imaginary part).}
\label{fig:damping_spectrum_m3}
\end{figure}

\section{Presence or absence of edge-selective extremal damping}
\begin{figure}
\centering
\raisebox{4.75cm}{(a)}\includegraphics[height=0.3\columnwidth]{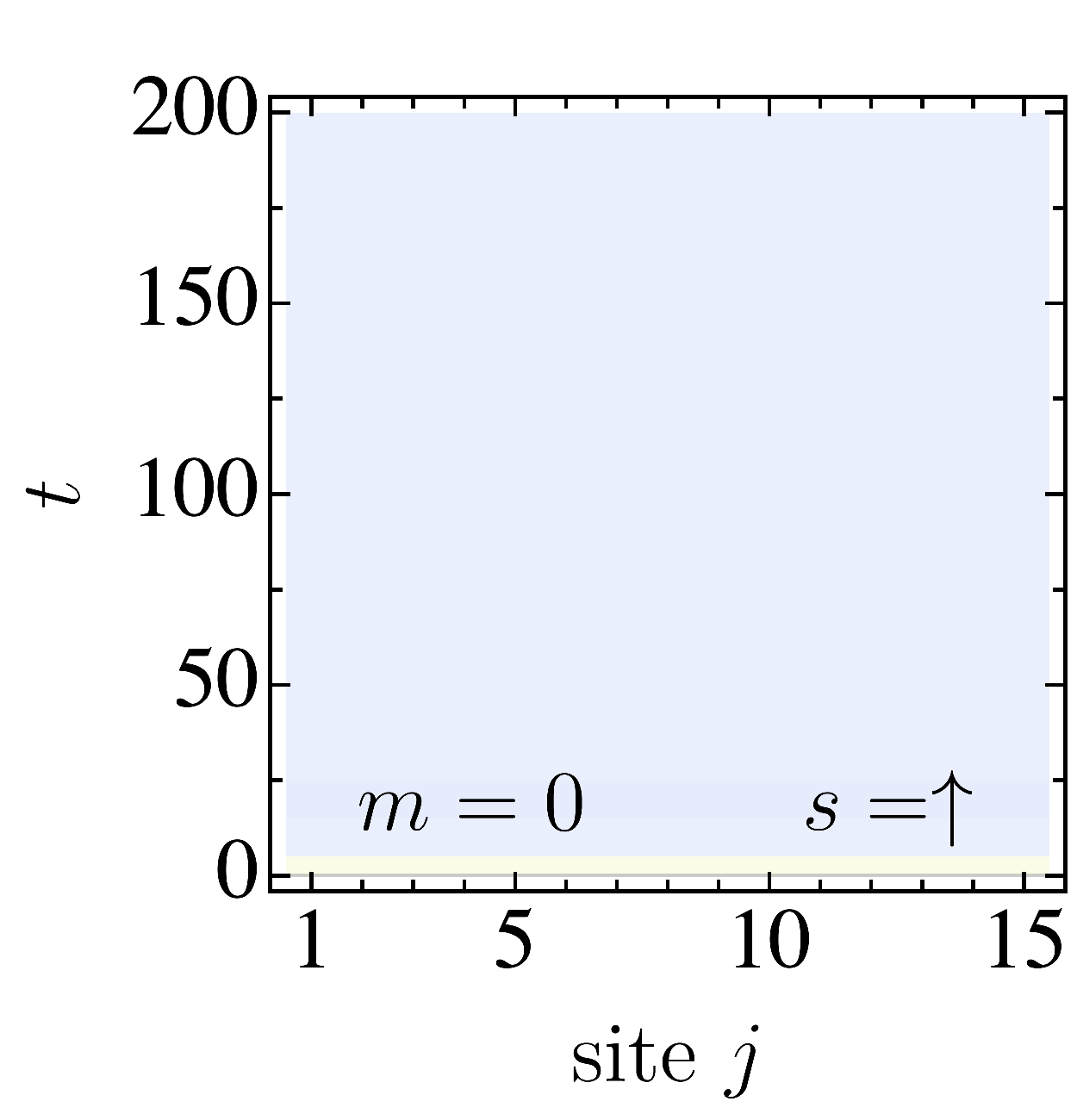}\hspace*{1cm}\raisebox{4.75cm}{(b)}\includegraphics[height=0.3\columnwidth]{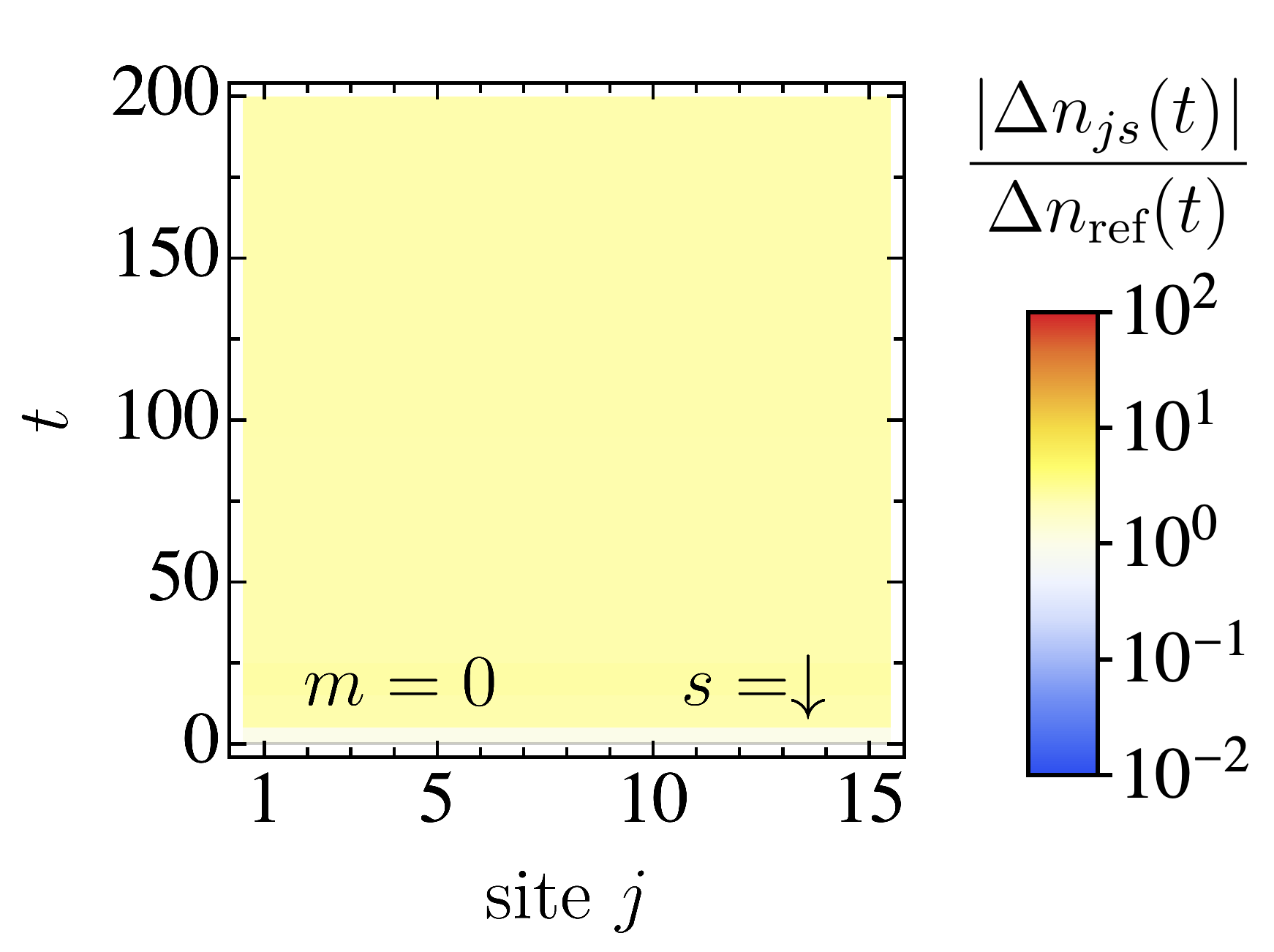}
\caption{Deviation of the electronic density from its steady state value as a function of time $t$ and site number $j$ for spin $s$, $\Delta n_{js}(t)$, in system with 15 sites for $m=0$ and \emph{periodic} boundary conditions along $x$. At $t=0$, the system is initiated in a state in which both spin species (polarized along $z$) are half-filled at every site. Panel (a): time evolution for $s=\uparrow$. Panel (b): time evolution for $s=\downarrow$. We fix $k_y= 1.629$ in both panels. The remaining parameters are $\Gamma_0=0.08$, $\Gamma_z=1.2$, $\alpha=\beta=1$, and $\mu=0$. The reference density deviation $\Delta n_{\text{ref}}(t)$ is the geometric mean of the largest and smallest density deviations from the steady state value at time $t$ over all sites and spins.}
\label{fig:dampingEdgeStatesm0PBC}
\end{figure}

\begin{figure}
\centering
\raisebox{4.75cm}{(a)}\includegraphics[height=0.3\columnwidth]{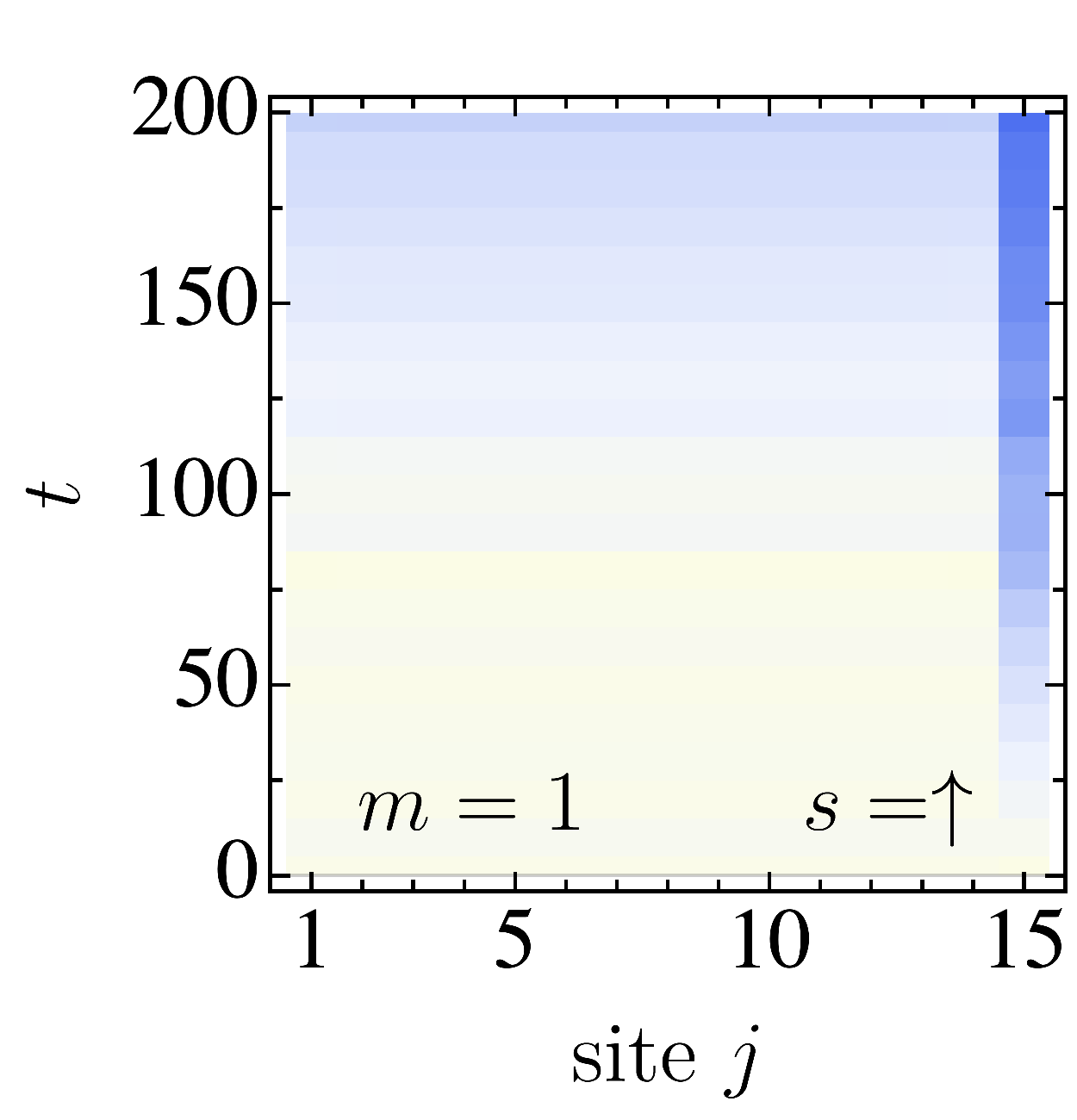}\hspace*{1cm}\raisebox{4.75cm}{(b)}\includegraphics[height=0.3\columnwidth]{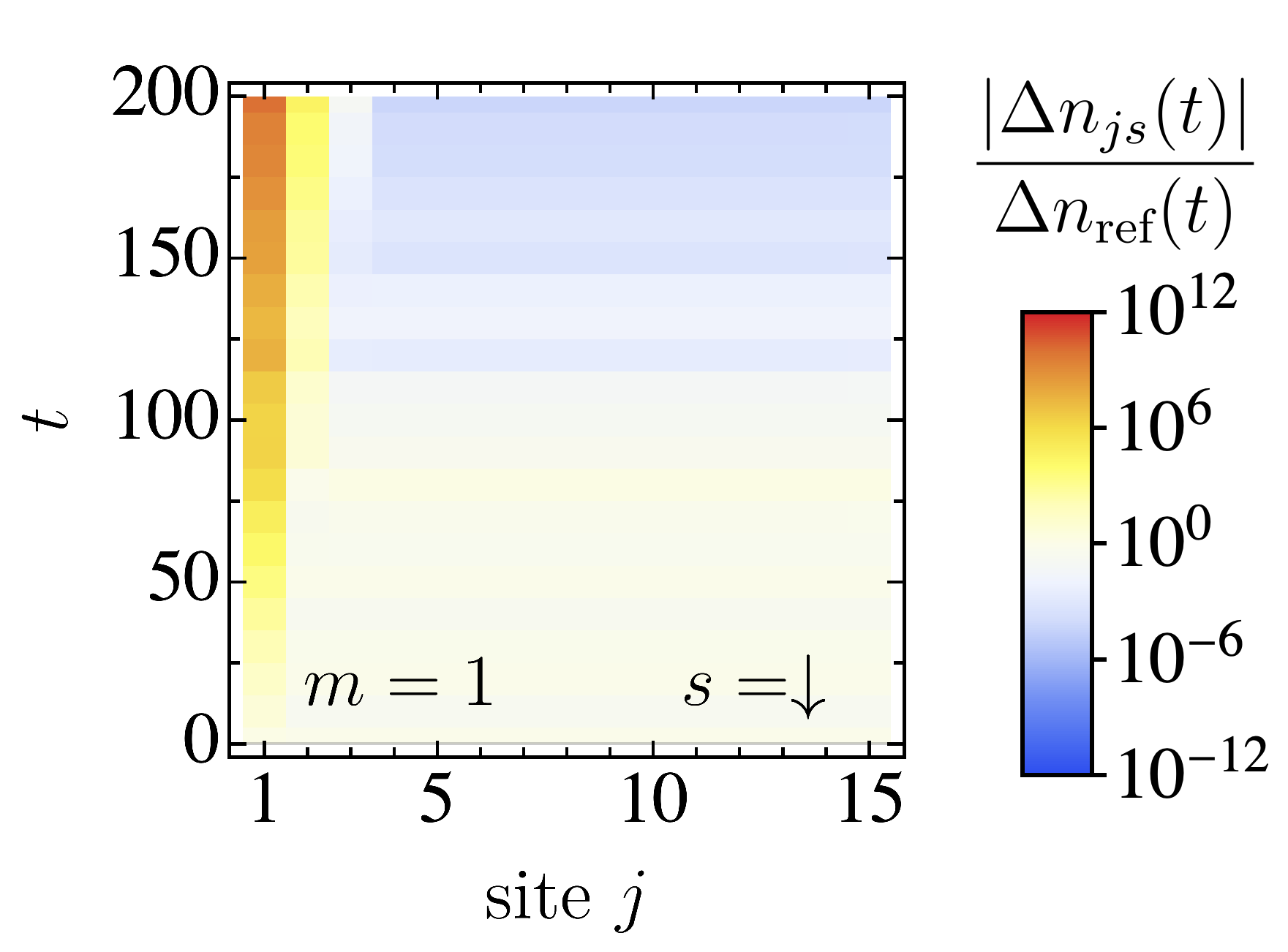}
\caption{Deviation of the electronic density from its steady state value as a function of time $t$ and site number $j$ for spin $s$, $\Delta n_{js}(t)$, in system with 15 sites for $m=1$ and \emph{open} boundary conditions along $x$. At $t=0$, the system is initiated in a state in which both spin species (polarized along $z$) are half-filled at every site. Panel (a): time evolution for $s=\uparrow$. Panel (b): time evolution for $s=\downarrow$. We fix $k_y= -0.05893$ in both panels. The remaining parameters are $\Gamma_0=0.08$, $\Gamma_z=1.2$, $\alpha=\beta=1$, and $\mu=0$. The reference density deviation $\Delta n_{\text{ref}}(t)$ is the geometric mean of the largest and smallest density deviations from the steady state value at time $t$ over all sites and spins.}
\label{fig:dampingEdgeStatesm1OBC}
\end{figure}

\begin{figure}
\centering
\raisebox{4.75cm}{(a)}\includegraphics[height=0.3\columnwidth]{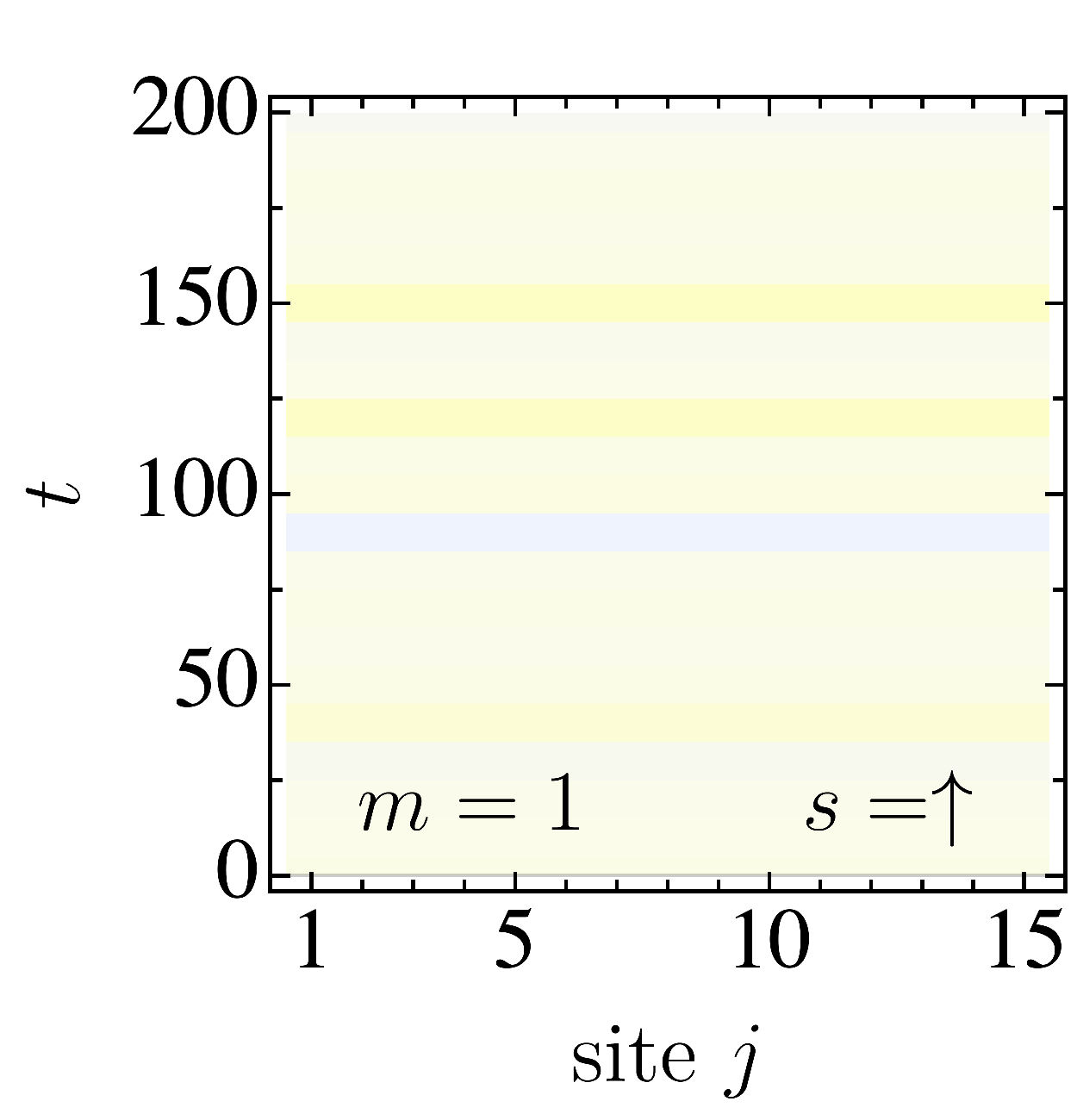}\hspace*{1cm}\raisebox{4.75cm}{(b)}\includegraphics[height=0.3\columnwidth]{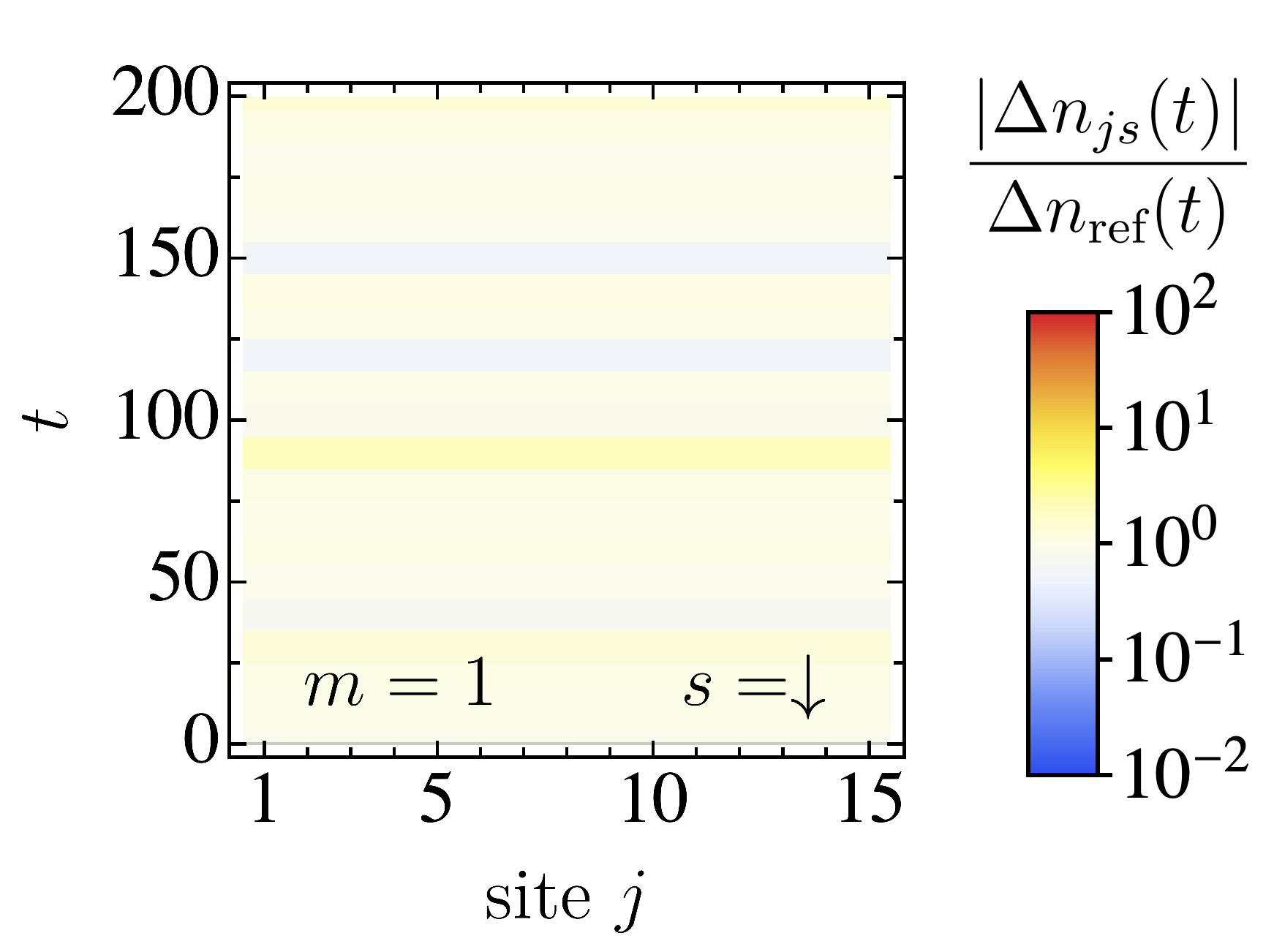}
\caption{Deviation of the electronic density from its steady state value as a function of time $t$ and site number $j$ for spin $s$, $\Delta n_{js}(t)$, in system with 15 sites for $m=1$ and \emph{periodic} boundary conditions along $x$. At $t=0$, the system is initiated in a state in which both spin species (polarized along $z$) are half-filled at every site. Panel (a): time evolution for $s=\uparrow$. Panel (b): time evolution for $s=\downarrow$. We fix $k_y= -0.05893$ in both panels. The remaining parameters are $\Gamma_0=0.08$, $\Gamma_z=1.2$, $\alpha=\beta=1$, and $\mu=0$. The reference density deviation $\Delta n_{\text{ref}}(t)$ is the geometric mean of the largest and smallest density deviations from the steady state value at time $t$ over all sites and spins.}
\label{fig:dampingEdgeStatesm1PBC}
\end{figure}

\begin{figure}
\centering
\raisebox{4.75cm}{(a)}\includegraphics[height=0.3\columnwidth]{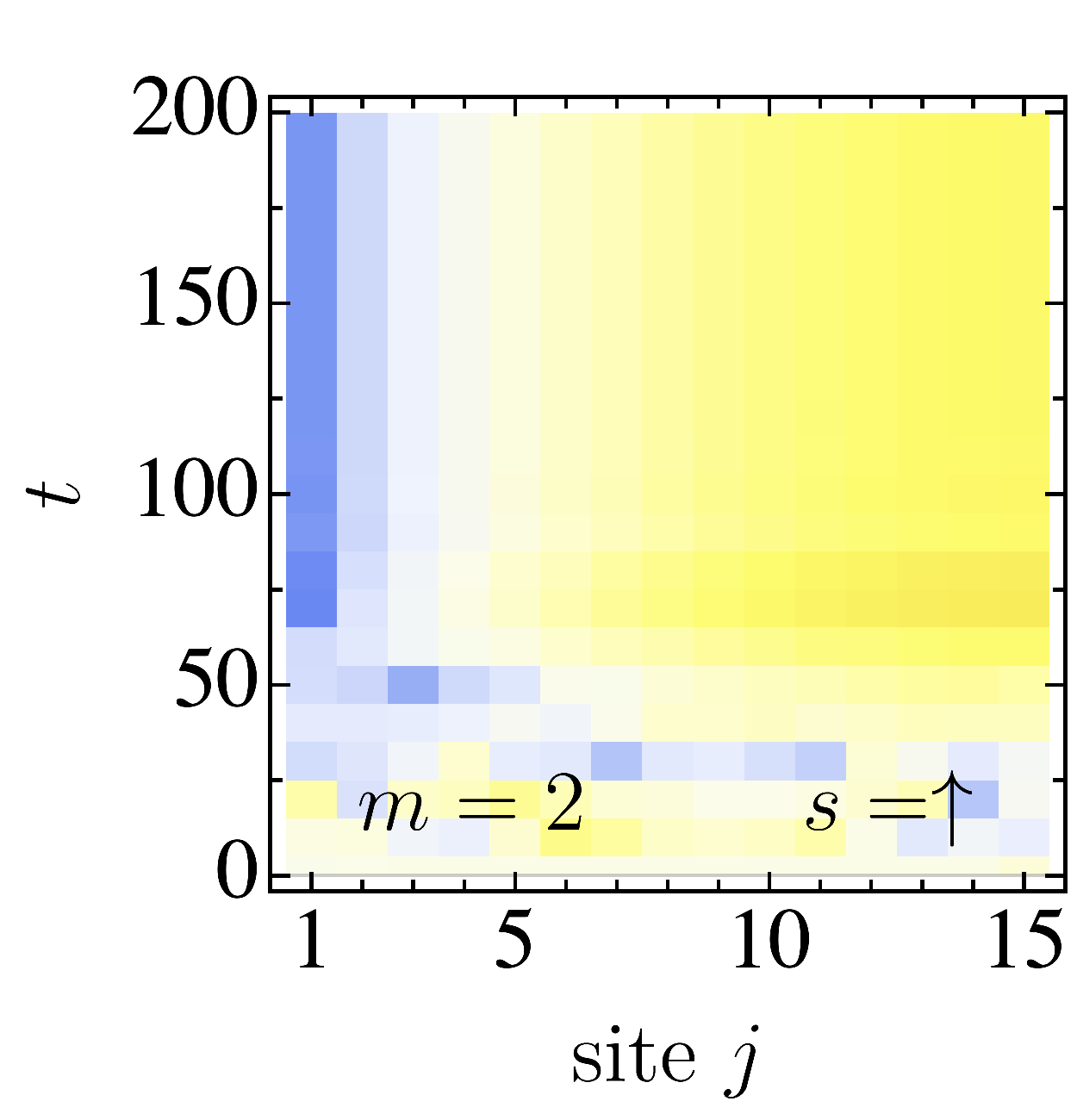}\hspace*{1cm}\raisebox{4.75cm}{(b)}\includegraphics[height=0.3\columnwidth]{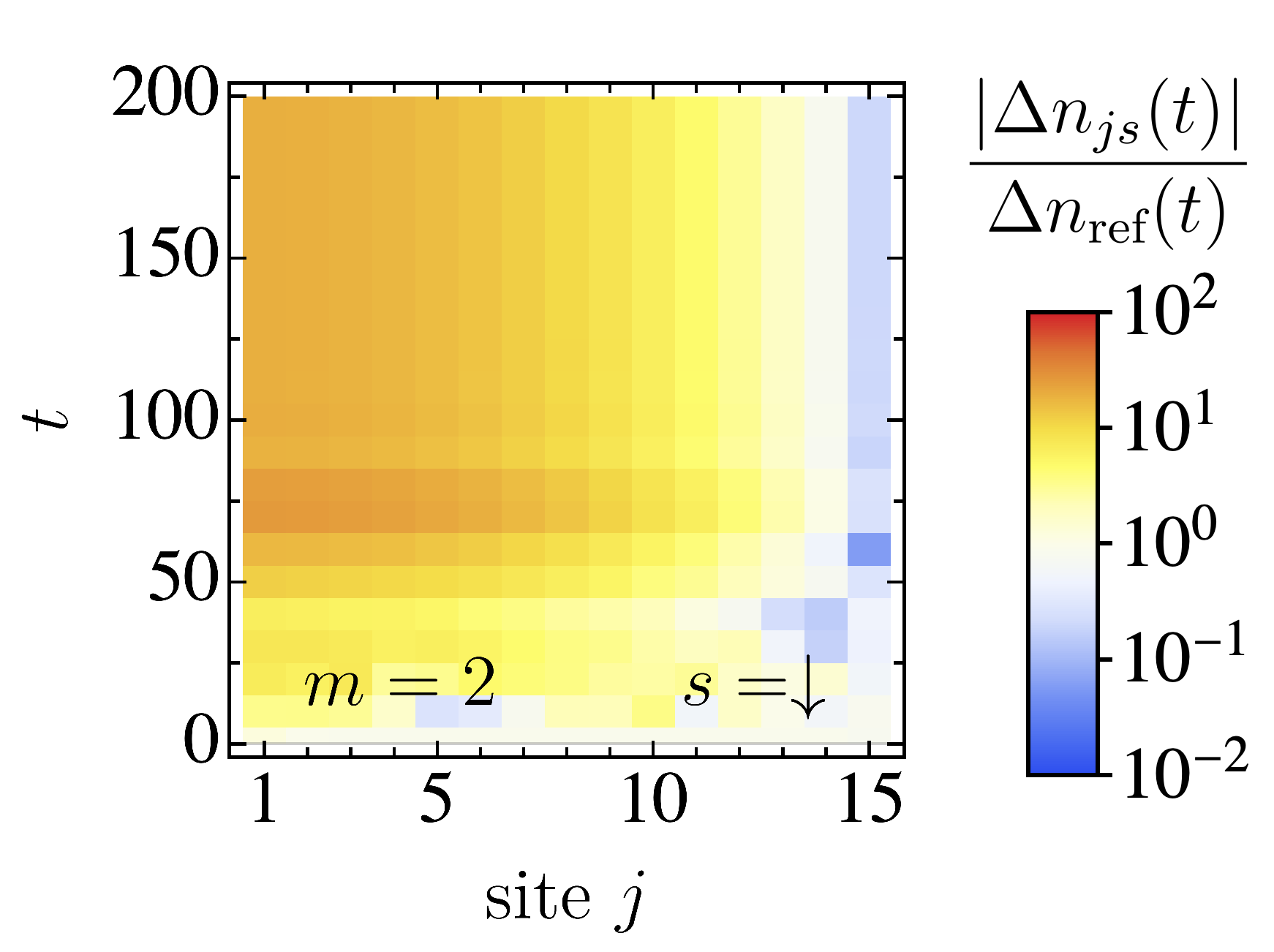}
\caption{Deviation of the electronic density from its steady state value as a function of time $t$ and site number $j$ for spin $s$, $\Delta n_{js}(t)$, in system with 15 sites for $m=2$ and \emph{open} boundary conditions along $x$. At $t=0$, the system is initiated in a state in which both spin species (polarized along $z$) are half-filled at every site. Panel (a): time evolution for $s=\uparrow$. Panel (b): time evolution for $s=\downarrow$. We fix $k_y= -0.05893$ in both panels. The remaining parameters are $\Gamma_0=0.08$, $\Gamma_z=1.2$, $\alpha=\beta=1$, and $\mu=0$. The reference density deviation $\Delta n_{\text{ref}}(t)$ is the geometric mean of the largest and smallest density deviations from the steady state value at time $t$ over all sites and spins.}
\label{fig:dampingEdgeStatesm2OBC}
\end{figure}

\begin{figure}
\centering
\raisebox{4.75cm}{(a)}\includegraphics[height=0.3\columnwidth]{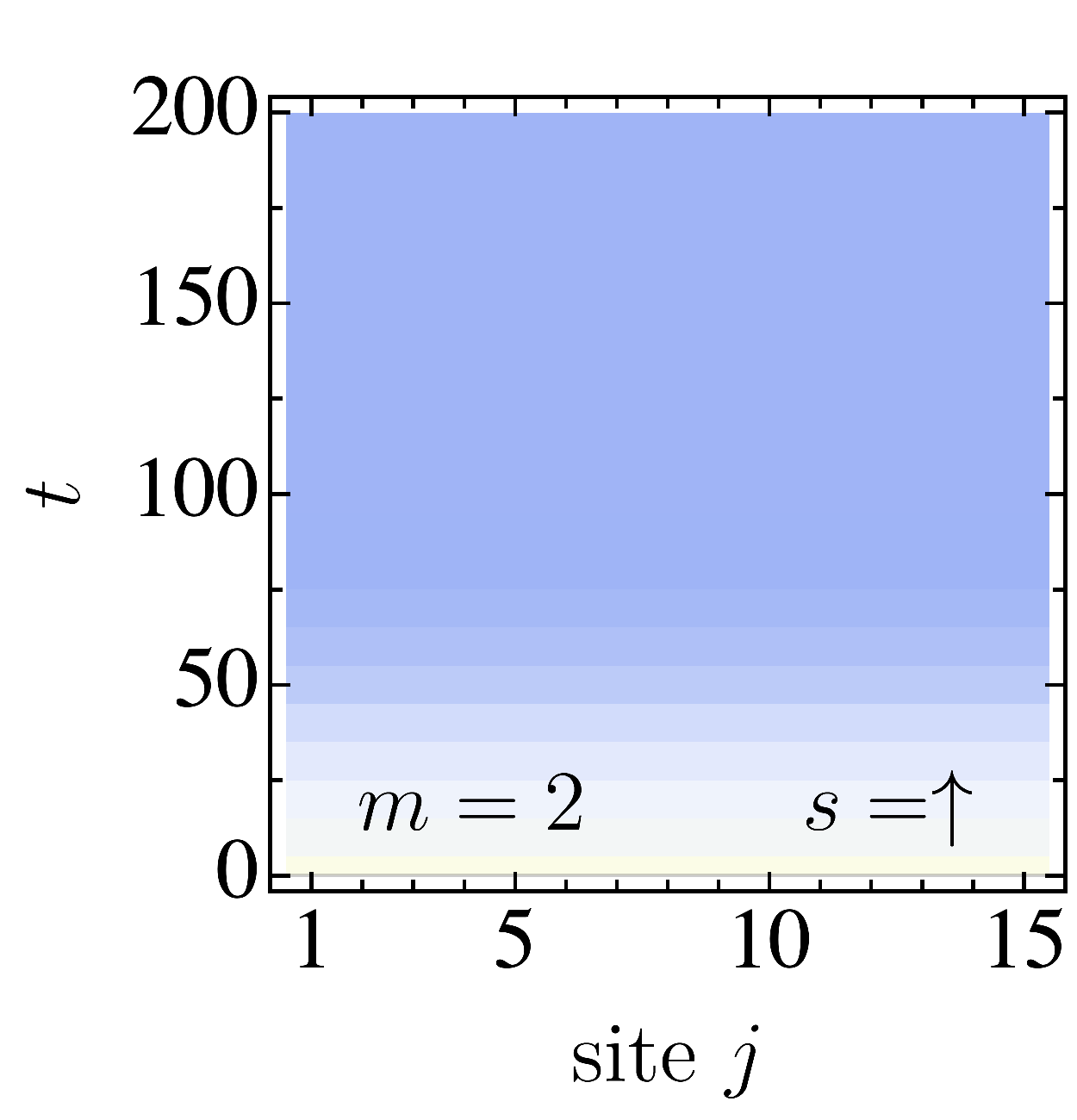}\hspace*{1cm}\raisebox{4.75cm}{(b)}\includegraphics[height=0.3\columnwidth]{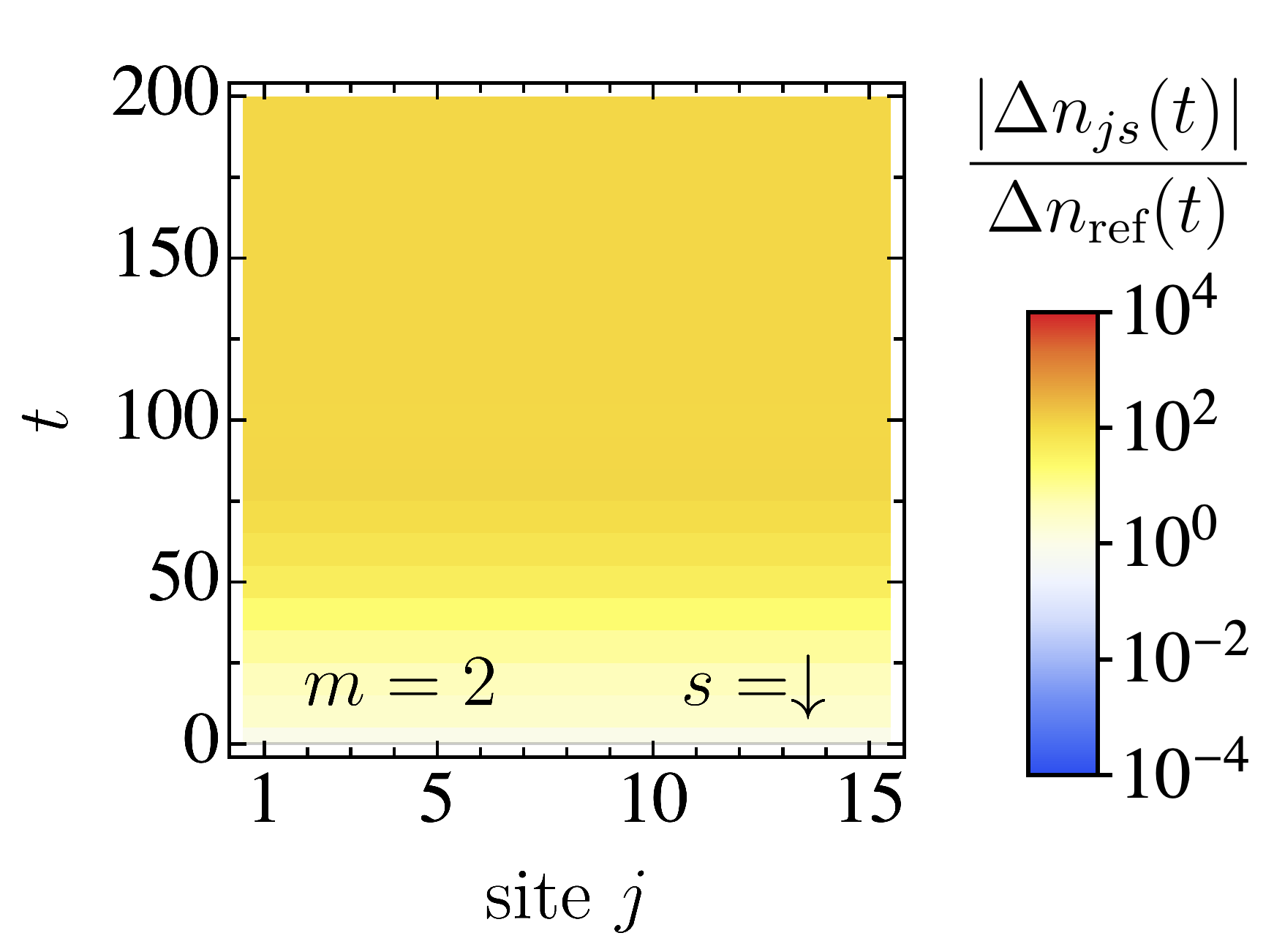}
\caption{Deviation of the electronic density from its steady state value as a function of time $t$ and site number $j$ for spin $s$, $\Delta n_{js}(t)$, in system with 15 sites for $m=2$ and \emph{periodic} boundary conditions along $x$. At $t=0$, the system is initiated in a state in which both spin species (polarized along $z$) are half-filled at every site. Panel (a): time evolution for $s=\uparrow$. Panel (b): time evolution for $s=\downarrow$. We fix $k_y= -0.05893$ in both panels. The remaining parameters are $\Gamma_0=0.08$, $\Gamma_z=1.2$, $\alpha=\beta=1$, and $\mu=0$.The reference density deviation $\Delta n_{\text{ref}}(t)$ is the geometric mean of the largest and smallest density deviations from the steady state value at time $t$ over all sites and spins.}
\label{fig:dampingEdgeStatesm2PBC}
\end{figure}

\begin{figure}
\centering
\raisebox{4.75cm}{(a)}\includegraphics[height=0.3\columnwidth]{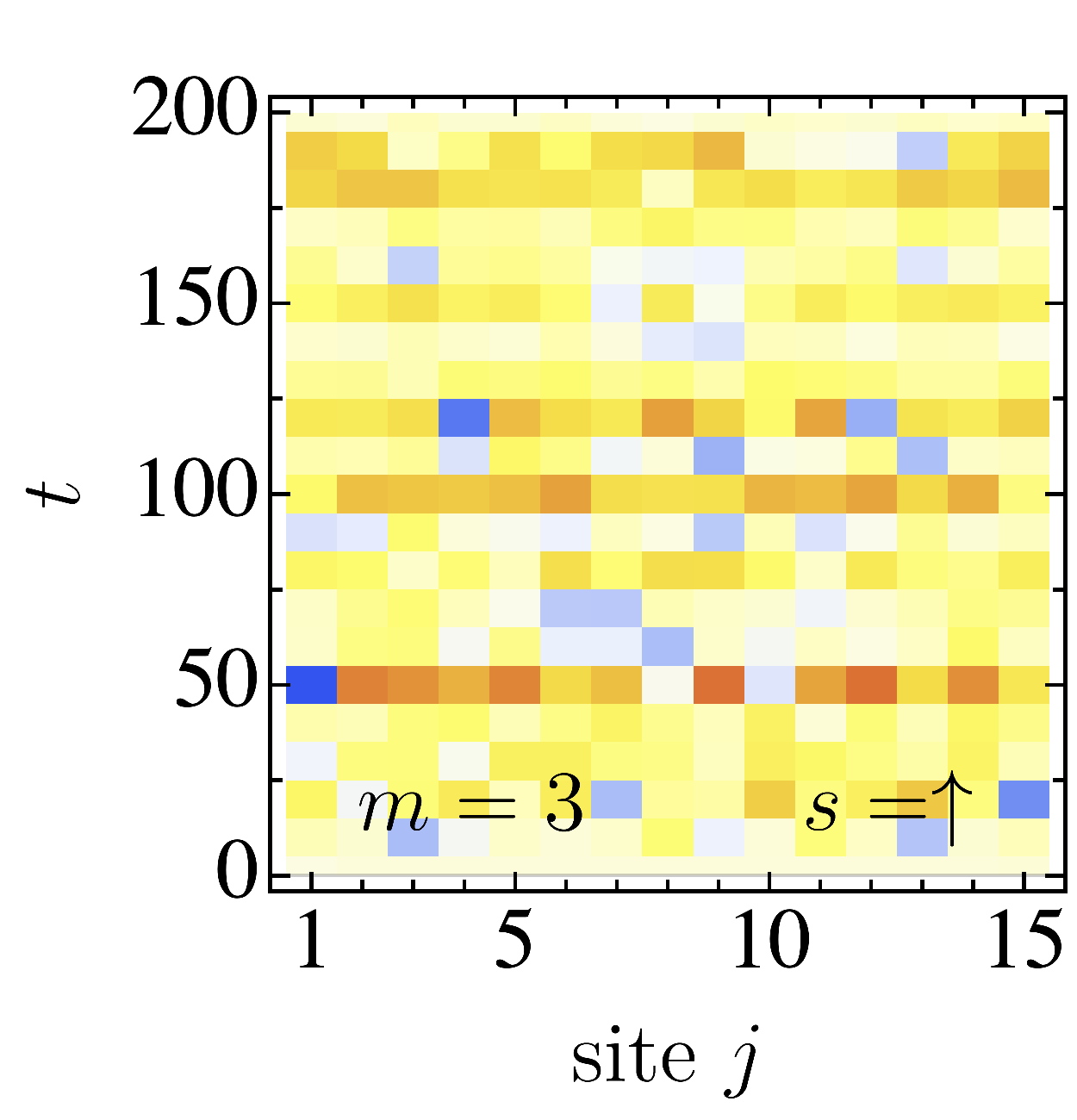}\hspace*{1cm}\raisebox{4.75cm}{(b)}\includegraphics[height=0.3\columnwidth]{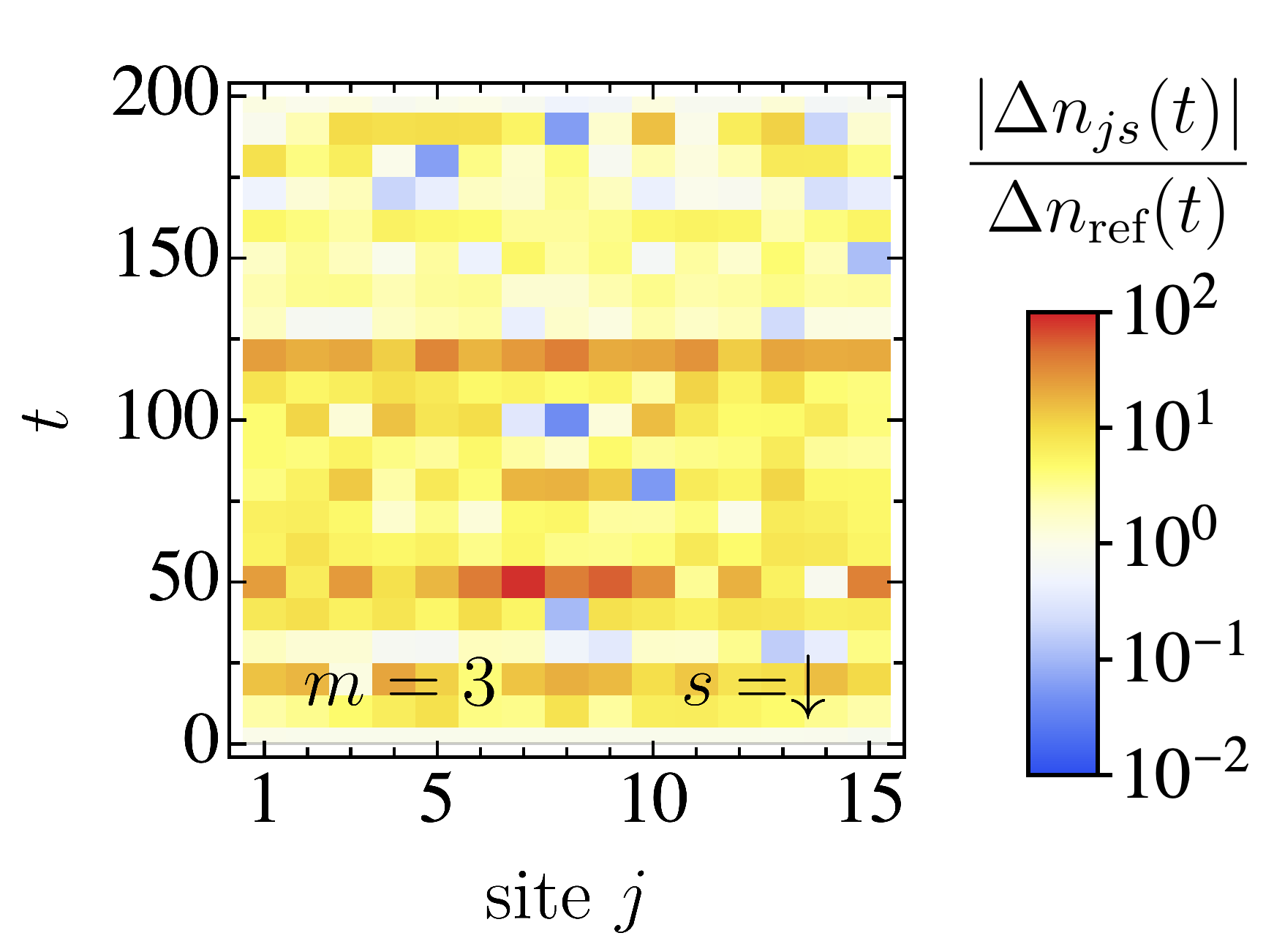}
\caption{Deviation of the electronic density from its steady state value as a function of time $t$ and site number $j$ for spin $s$, $\Delta n_{js}(t)$, in system with 15 sites for $m=3$ and \emph{open} boundary conditions along $x$. At $t=0$, the system is initiated in a state in which both spin species (polarized along $z$) are half-filled at every site. Panel (a): time evolution for $s=\uparrow$. Panel (b): time evolution for $s=\downarrow$. We fix $k_y= -0.05893$ in both panels. The remaining parameters are $\Gamma_0=0.08$, $\Gamma_z=1.2$, $\alpha=\beta=1$, and $\mu=0$. The reference density deviation $\Delta n_{\text{ref}}(t)$ is the geometric mean of the largest and smallest density deviations from the steady state value at time $t$ over all sites and spins.}
\label{fig:dampingNoEdgeStatesm3OBC}
\end{figure}

\begin{figure}
\centering
\raisebox{4.75cm}{(a)}\includegraphics[height=0.3\columnwidth]{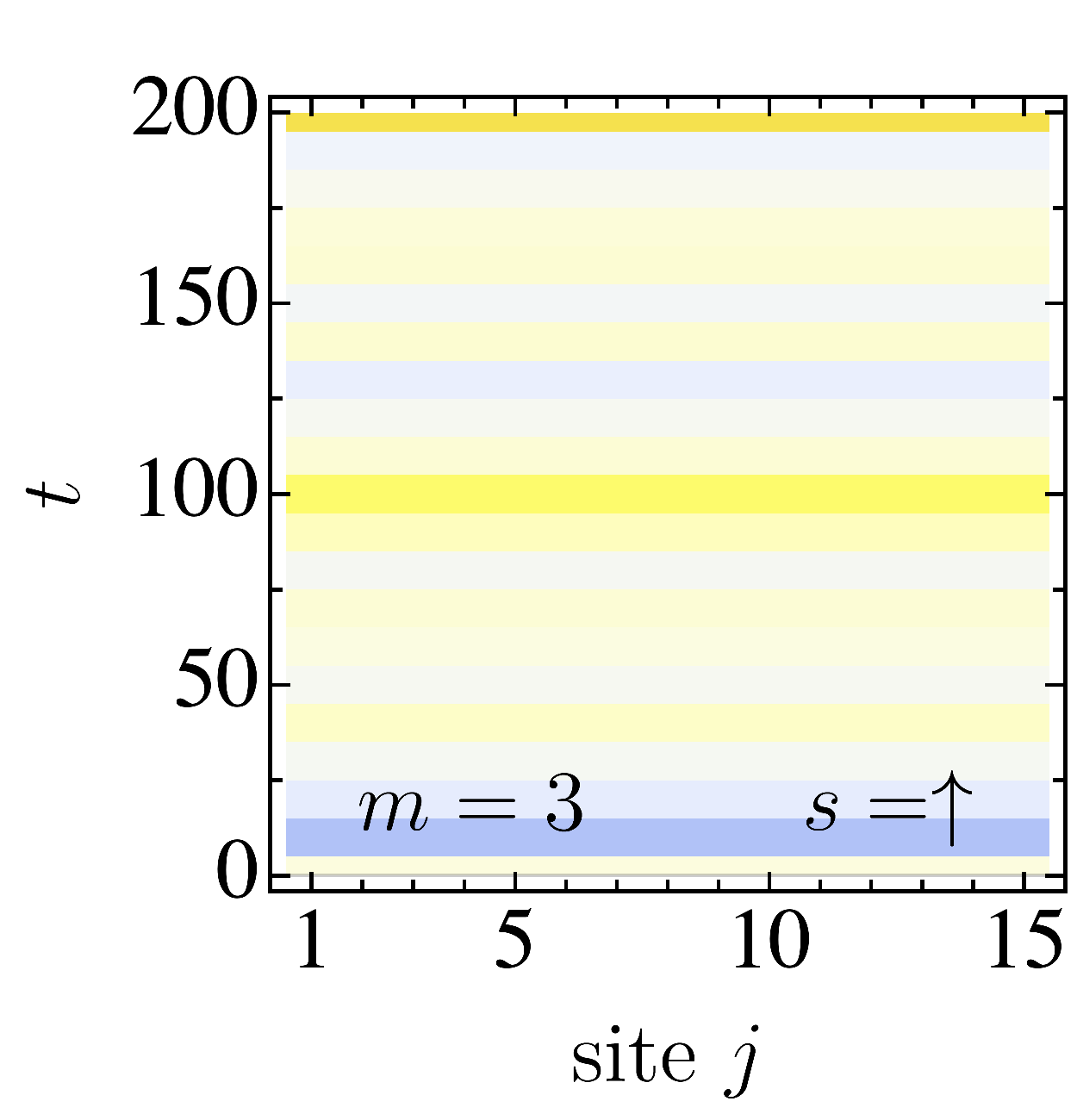}\hspace*{1cm}\raisebox{4.75cm}{(b)}\includegraphics[height=0.3\columnwidth]{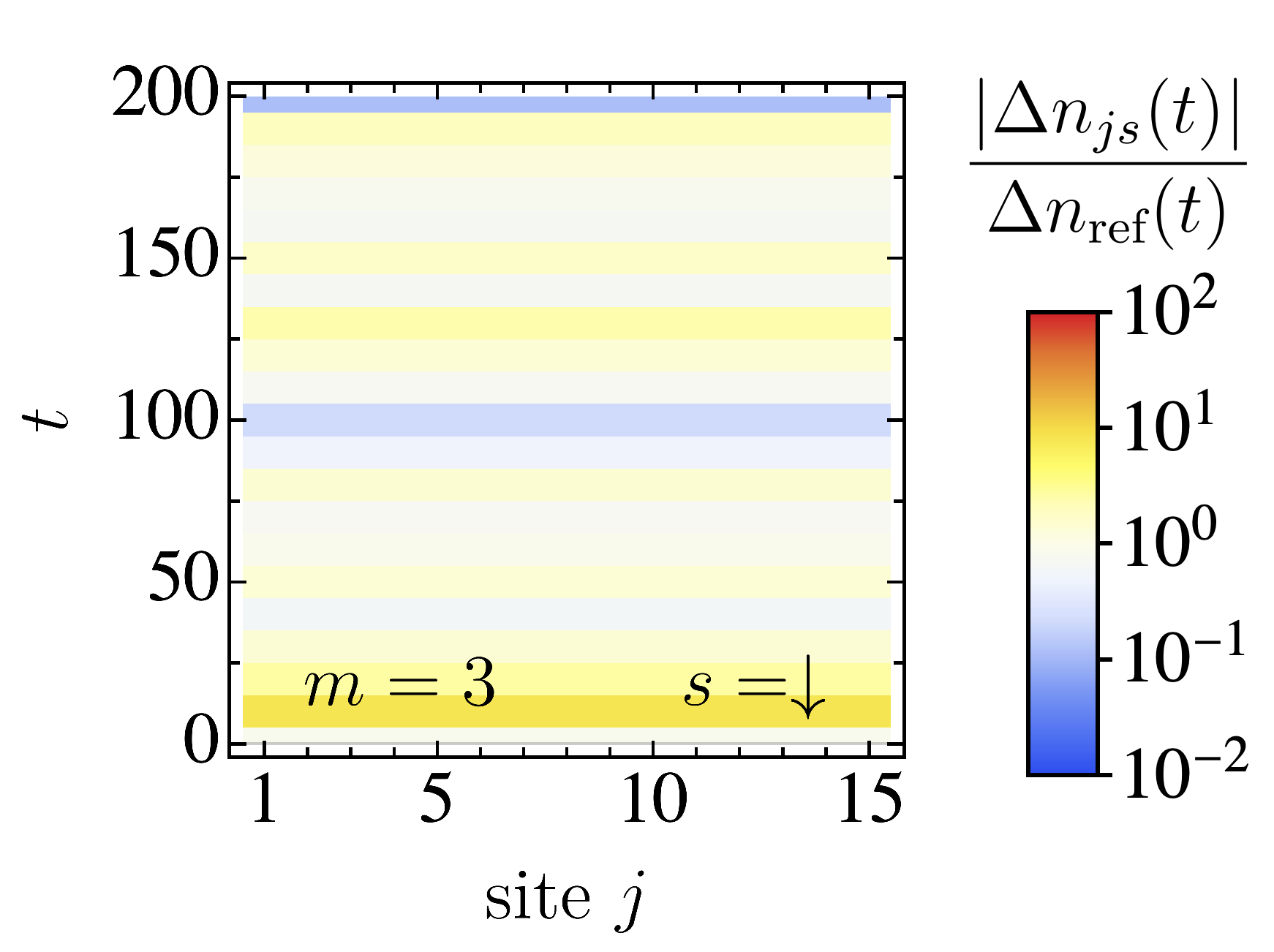}
\caption{Deviation of the electronic density from its steady state value as a function of time $t$ and site number $j$ for spin $s$, $\Delta n_{js}(t)$, in system with 15 sites for $m=3$ and  \emph{periodic} boundary conditions along $x$. At $t=0$, the system is initiated in a state in which both spin species (polarized along $z$) are half-filled at every site. Panel (a): time evolution for $s=\uparrow$. Panel (b): time evolution for $s=\downarrow$. We fix $k_y= -0.05893$ in both panels. The remaining parameters are $\Gamma_0=0.08$, $\Gamma_z=1.2$, $\alpha=\beta=1$, and $\mu=0$. The reference density deviation $\Delta n_{\text{ref}}(t)$ is the geometric mean of the largest and smallest density deviations from the steady state value at time $t$ over all sites and spins.}
\label{fig:dampingNoEdgeStatesm3PBC}
\end{figure}

Fig.~6 of the main text depicts the time evolution of the densities towards their steady-state values for $m=0$ and \emph{open} boundary conditions, i.e.~in a regime in which the damping matrix has extremal edge states. As discussed in the main text, the extremal edge states of the damping matrix entails edge-localized regions of fast(er) and slow(er) decay towards the steady state. Here, we  show further examples of how  the presence (absence) of extremal edge states in the damping matrix entails the presence (absence) of edge-selective extremal damping. Fig.~\ref{fig:dampingEdgeStatesm0PBC} shows this for $m=0$ and periodic boundary conditions (PBC). For $m=1$ (and the other chosen parameters), the damping matrix has a bulk line gap \emph{and} topological edge states. This is shown in Fig.~\ref{fig:dampingEdgeStatesm1OBC} for OBC and Fig.~\ref{fig:dampingEdgeStatesm1PBC} for PBC. For $m=2$ (and the other chosen parameters), the damping matrix has exceptional points \emph{without} topological edge states. As shown in Fig.~\ref{fig:dampingEdgeStatesm2OBC} for OBC and Fig.~\ref{fig:dampingEdgeStatesm2PBC} for PBC, the time evolution is thus not showing exponentially localized and exponentially growing/decreasing damping edge states as in the other plots, but merely a rather constant edge depletion in the case of OBC. For $m=3$, when the damping matrix has no extremal edge states, Fig.~\ref{fig:dampingNoEdgeStatesm3OBC} shows that even with OBC, no edge-localized damping features occur, while Fig.~\ref{fig:dampingNoEdgeStatesm3PBC} shows the same for PBC.


\section{Examples for topological design of damping landscapes}

\subsection{Edge-localized density}
\begin{figure}
\centering
\raisebox{4.75cm}{(a)}\includegraphics[height=0.3\columnwidth]{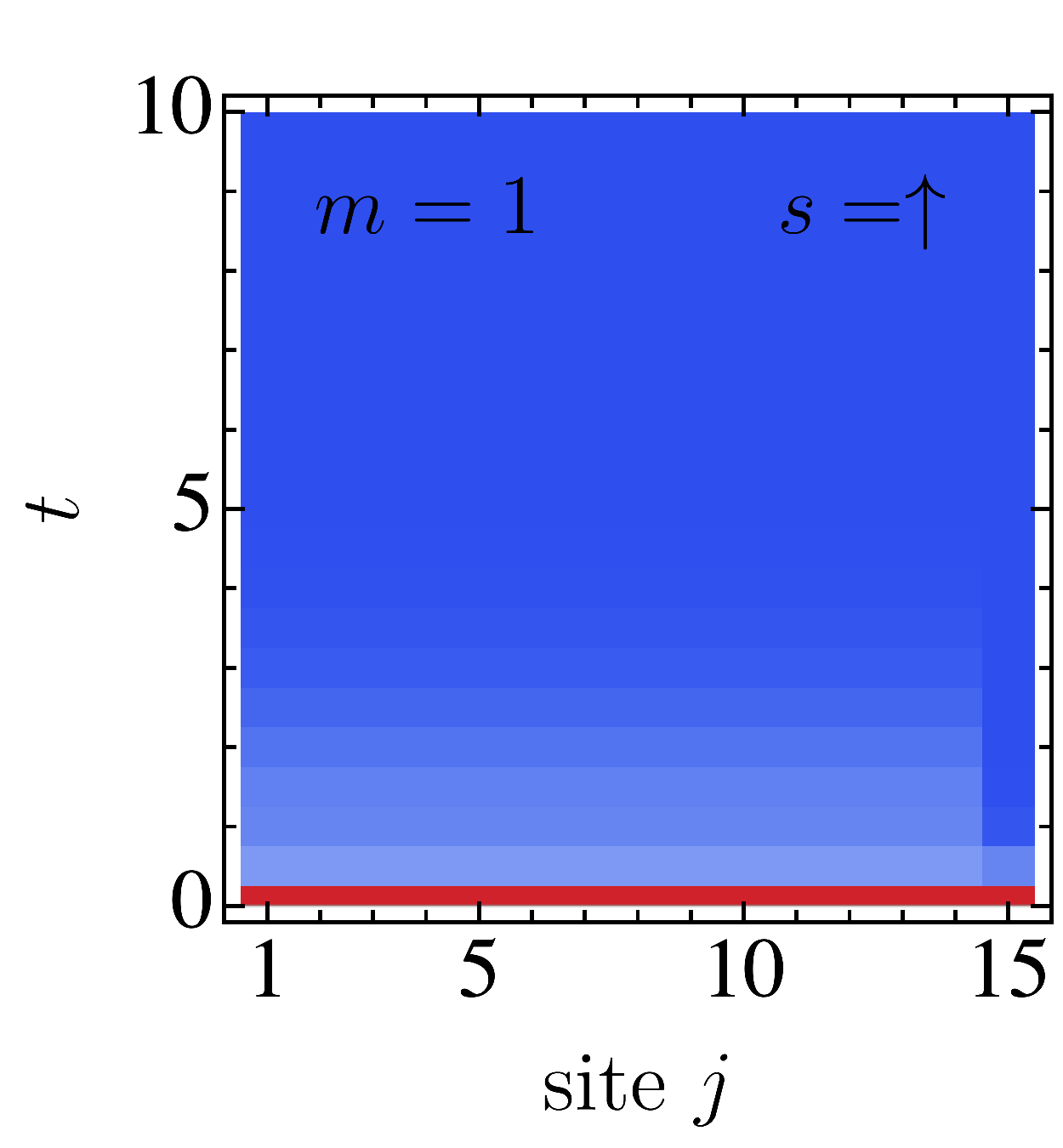}\hspace*{1cm}\raisebox{4.75cm}{(b)}\includegraphics[height=0.3\columnwidth]{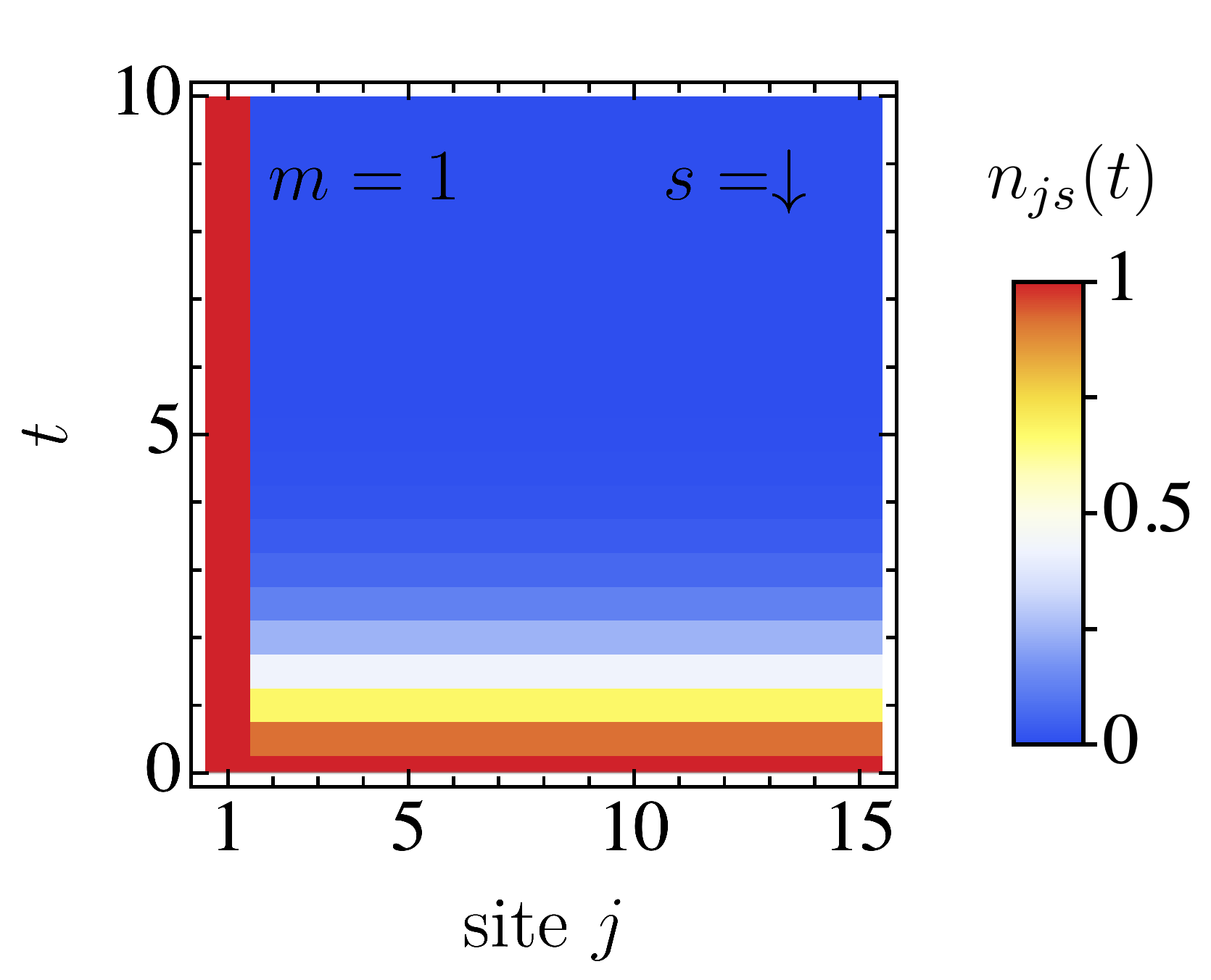}
\caption{Stabilization of an edge-localized stripe of $\downarrow$-spin electrons by topological damping design. We use $\Gamma_0=1$ and $\Gamma_z=0.99999$, as well as $m=1$, $\alpha=\beta=1$, and $\mu=0$. The system has OBC and 15 sites along $x$, PBC along $y$. We fix $k_y=-0.05893$, and initialize the system in a completely filled state at $t=0$. Panel (a) depicts the density of $\uparrow$-spin electrons, panel (b) that of $\downarrow$-spin electrons. Note that in contrast to all other plots of the time evolution of the local densities in the Supplemental Material, the color code here represents the absolute electron density on a linear scale.}
\label{fig:no_damping_edge_state}
\end{figure}

As discussed in the main text, a particularly interesting regime is realized if one species is almost without loss or gain. Here we focus on the case $\Gamma_\uparrow \gg \Gamma_\downarrow\geq0$, i.e.~the regime in which the $\downarrow$-spin electrons are almost without loss. We chose $\Gamma_0=1$ and $\Gamma_z=0.99999$, which means $\Gamma_\uparrow=1.99999$ and $\Gamma_\downarrow=10^{-5}$. Both spin species have loss, such that the asymptotic steady state is an entirely empty system. However, the $\downarrow$-polarized left edge state only reaches this empty state at very large times $\sim 1/\Gamma_\downarrow$. As shown in Fig.~\ref{fig:no_damping_edge_state}, this means that the topological damping edge states quickly stabilize a regime in which the system is essentially empty, except for an exponentially localized stripe of $\downarrow$-spin electrons at the left edge.

\subsection{Oscillating edge damping}
\begin{figure}
\centering
\raisebox{4.75cm}{(a)}\includegraphics[height=0.3\columnwidth]{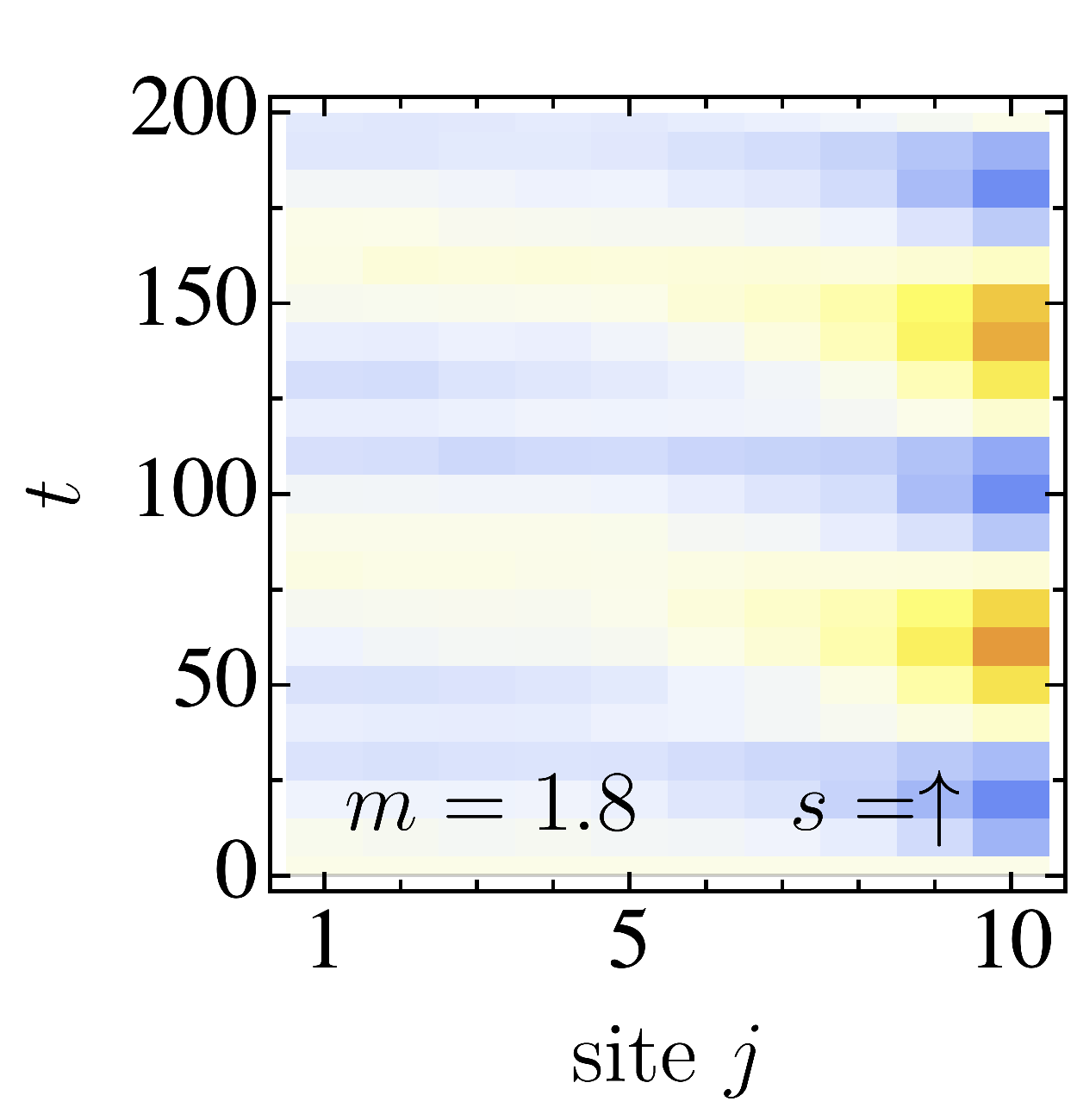}\hspace*{1cm}\raisebox{4.75cm}{(b)}\includegraphics[height=0.3\columnwidth]{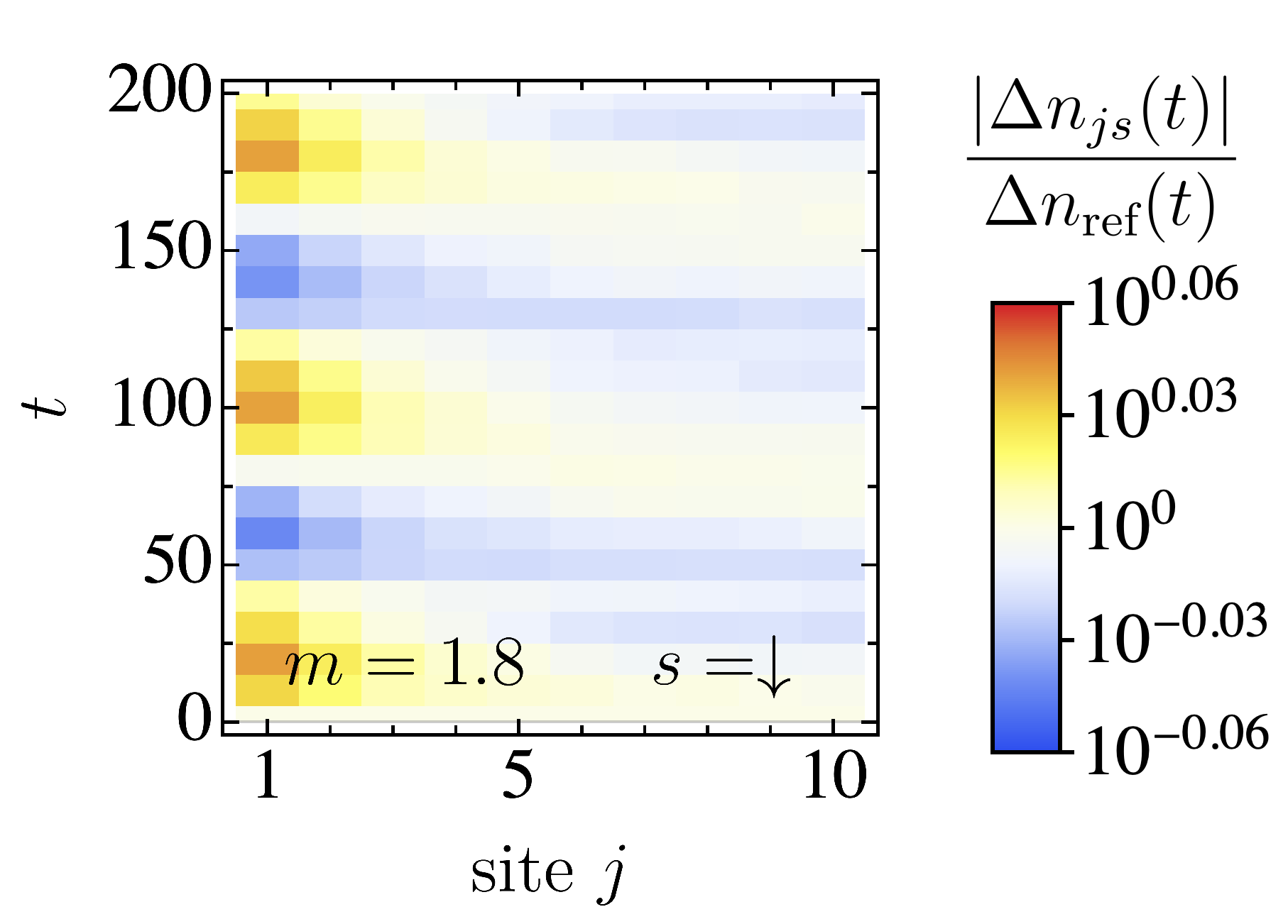}
\caption{Stabilization of an oscillating edge-localized damping pattern by topological damping design. We use $\Gamma_0=1$ and $\Gamma_z=0.01$, as well as $m=1.8$, $\alpha=\beta=1$, and $\mu=0$. The system has OBC and 10 sites along $x$, PBC along $y$. We fix $k_y=0.0001$, and initialize the system at $t=0$ in a state in which both spin species (polarized along $z$) are half-filled at every site. Panel (a) depicts the density of $\uparrow$-spin electrons, panel (b) that of $\downarrow$-spin electrons. The reference density deviation $\Delta n_{\text{ref}}(t)$ is the geometric mean of the largest and smallest density deviations from the steady state value at time $t$ over all sites and spins.}
\label{fig:flashing_edge_state}
\end{figure}

Another example for how edge states of the damping matrix can be used to engineer non-trivial edge-localized damping pattern is shown in Fig.~\ref{fig:flashing_edge_state}, which exhibits a damping pattern that oscillates between the two edges as a function of time. We find that oscillation frequency depends on system size along $x$. The length -dependant oscillatory feature is reminiscent of those found due to the exceptional point occurring at the transition from PT-unbroken to PT-broken regime~\cite{Ruter2010,Zhang2018}. We do find that real energy gap of $iX$ closes, the edge states coalesce and the energy eigenvlaues start to become complex.

\bibliography{Rev_Supplement}